\documentclass[%
 reprint,
superscriptaddress,
showpacs, preprintnumbers,
nofootinbib,
longbibliography,
 amsmath,amssymb,aps,
prd,
floatfix,
]{revtex4-1}

\usepackage[normalem]{ulem} 
\usepackage{soul}
\usepackage[dvipsnames]{xcolor}
\usepackage{graphicx}
\usepackage{dcolumn}
\usepackage{bm}
\usepackage{verbatim}
\usepackage[caption=false]{subfig}
\usepackage{graphicx}
\usepackage{rotating}
\usepackage{relsize}
\usepackage{esdiff}  
\usepackage{booktabs}  
\usepackage[colorlinks=true]{hyperref}
\hypersetup{
    colorlinks,
    linkcolor=blue!50!black,          
    citecolor=blue!50!black,        
    urlcolor=blue!50!black           
}

\def\og{\color{OliveGreen}}

\begin{document}

\title{Inference of neutrino flavor evolution through data assimilation and neural differential equations}

\author{Ermal Rrapaj}
\email{ermalrrapaj@gmail.com}
\affiliation{Department of Physics, University of California, Berkeley, CA 94720, USA}
\affiliation{School of Physics and Astronomy, University of Minnesota, Minneapolis, MN 55455, USA}
\author{Amol V. Patwardhan}
\email{apatward@slac.stanford.edu}
\affiliation{Department of Physics, University of California, Berkeley, CA 94720, USA}
\affiliation{Institute for Nuclear Theory, University of Washington, Seattle, WA 98115, USA}
\affiliation{SLAC National Accelerator Laboratory, 2575 Sand Hill Road, Menlo Park, CA, 94025}
\author{Eve Armstrong}
\email{evearmstrong.physics@gmail.com}
\affiliation{Department of Physics, New York Institute of Technology, New York, NY 10023, USA}
\affiliation{Department of Astrophysics, American Museum of Natural History, New York, NY 10024, USA}
\author{George M. Fuller}
\email{gfuller@ucsd.edu}
\affiliation{Department of Physics, University of California, San Diego, La Jolla, CA 92093-0319, USA}

\date{\today}

\begin{abstract}
 The evolution of neutrino flavor in dense environments such as core-collapse supernovae and binary compact object mergers constitutes an important and unsolved problem. Its solution has potential implications for the dynamics and heavy-element nucleosynthesis in these environments. In this paper, we build upon recent work to explore inference-based techniques for the estimation of model parameters and neutrino flavor evolution histories. We combine data assimilation, ordinary differential equation solvers, and neural networks to craft an inference approach tailored for non-linear dynamical systems.  Using this architecture, and a simple two-neutrino-beam, two-flavor model, we compare the performances of nine different optimization algorithms and expand upon previous assessments of the efficacy of inference for tackling problems in flavor evolution. We find that employing this new architecture, together with evolutionary optimization algorithms, accurately captures flavor histories in  the small-scale model and allows us to quickly explore both model parameters and initial flavor content. In future work we plan to extend these inference techniques to large numbers of neutrinos.   
\end{abstract}

\preprint{SLAC-PUB-17563}

\maketitle 
\noindent
\section{Introduction}

 Core-collapse supernovae and binary compact object mergers are extreme physical environments with the potential to serve as valuable laboratories at the intersection of particle theory, dense matter physics, and high-energy astrophysics. Many of the important physical phenomena in these environments, such as shock propagation, bulk matter outflows, and the synthesis of heavy-elements are driven in part by interactions between nuclear matter and the accompanying prodigious flux of emitted neutrinos~\cite{Burrows1990,Mirizzi:2015,Pinedo2017,Muller:2019}.

In these situations, the flavor evolution of the neutrinos is a complicated, nonlinear problem, wherein the flavor histories of neutrinos with different energies and trajectories are coupled to one another. This has been shown to lead to various collective flavor oscillation phenomena~\cite{Duan:2006an, Duan:2005cp, Duan:2006jv, Hannestad:2006nj, Raffelt:2007cb, Raffelt:2007xt, Dasgupta:2007ws, Johns:2017oky, Cherry:2013mv, Zaizen:2019ufj, Duan:2009cd, Duan:2010bg, Mirizzi:2015eza, Chakraborty:2016yeg}. In particular, in the last few years, it has been demonstrated that relaxing certain assumptions regarding spatial and temporal symmetries in the neutrino flavor field can lead to flavor instabilities not previously identified, and which have not been well studied~\cite{2011PhRvD..84e3013B, Raffelt:2013rqa, Mirizzi:2013wda, Duan:2014gfa, 2015PhLB..751...43A, 2016JCAP...01..028C, 2009PhRvD..79j5003S, 2016PhRvL.116h1101S, 2015PhRvD..92l5030D,  Izaguirre:2016gsx, Dasgupta:2016dbv, Capozzi:2017gqd, 2018PhRvD..97b3017D, 2019PhLB..790..545A, 2019ApJ...883...80S, 2019PhRvL.122i1101C, 2019PhRvD..99j3011D, 2019ApJ...886..139N, 2019ApJ...886..139N, Johns:2019izj, Chakraborty:2019wxe, Morinaga:2018aug, Abbar:2018shq, Cherry:2019vkv, Morinaga:2019wsv, Bhattacharyya:2020dhu, Abbar:2020fcl, Capozzi:2020syn} (see also the recent review in Ref.~\cite{Tamborra:2020cul} and references therein).

Since the flavor evolution of neutrinos is so inextricably linked to the transport of energy and lepton number in these environments, it is important to identify the initial conditions and physical regimes under which these noted flavor-field instabilities and collective phenomena can manifest themselves.  Meanwhile, the next generation of terrestrial detectors such as DUNE~\cite{Abi:2020} and Hyper-Kamiokande~\cite{Abe:2018} could potentially provide a detection of a large number ($\sim \mathcal{O}(10^3\text{--}10^4)$) of neutrinos from Galactic core-collapse supernovae events.  Thus it is pertinent to ask what such a detection could reveal about neutrino properties, as well as the physics of the supernova environment itself.

The last decades have seen the rapid development of machine learning (ML) algorithms that utilize ``big data" to solve difficult problems such as image recognition~\cite{He2016,Zoph2018,Traore2018} and natural language processing~\cite{Yong2018, Otter2018}.  The training of most ML algorithms requires large amounts of data, in part because initial conditions are often not assumed to be well known.  Not surprisingly, scientific fields that typically produce large data sets have leveraged these technological advances~\cite{Tang2019,Zemouri2019}.  Other domains of science and engineering, however, are characterized by sparse data---sparsity that precludes the application of such learning algorithms to problems in these fields.  Instead, these fields have long established research traditions which have led to the development of predictive models. The study of neutrino physics falls into this latter category.

At first glance, then, there appears to be a dichotomy in approach: data-driven machine learning on one hand, and theoretical models on the other.  If no  information about initial conditions---or for that matter any model knowledge---is available but large amounts of data are, the first approach seems very reasonable.  When dealing with sparse data, however,  knowledge of the model and initial conditions is invaluable.  In the physical sciences and engineering fields characterized by  sparse data and highly developed theoretical frameworks, functions chosen to represent any unknown parts of a model are typically dictated by the researchers' experience and intuition. This educated guesswork might not account for all functional forms that may appropriately represent the data. This situation calls for a less biased---and more general---means of model estimation.

Recently, in calculations of neutrino flavor transformation, there have been attempts~\cite{Armstrong:2017,Armstrong:2020} to combine the best from both approaches noted above, by utilizing an inference procedure known as statistical data assimilation (SDA). In an SDA procedure, real or simulated data are considered together with assumed theoretical constraints, to complete a model with one or more unknown parameters. \lq\lq Completing the model\rq\rq\ here refers to the task of simultaneously determining the evolution histories of state variables, as well as the values of unknown parameters, subject to data and physical constraints.
In~\cite{Armstrong:2017,Armstrong:2020} the authors used inference to predict the flavor transformation histories of two mono-energetic neutrino beams coherently interacting with each other and with a matter background potential. Simulated \lq\lq measurements\rq\rq\ of neutrino flavor at the endpoint of the evolution, as well as theoretical constraints at other points along the trajectory, were utilized to compute the evolution histories by approximating the value of the neutrino beam flavor on a grid of uniformly-spaced radial locations.   Several distinct functional forms for the unknown parameters were tested. Also in~\cite{Armstrong:2020} the effects of incrementally adding theoretical constraints to improve the inference efficacy were analyzed.

This work explores new directions in applying SDA for constraining neutrino flavor evolution histories inside core-collapse supernovae. There are two key differences with respect to previous work. Firstly, we   eliminate uniformity in grid discretization.  Specifically, we include an adaptive step differential equation solver in the neural architecture~\cite{Chen2018}, and we employ a deep neural network that replaces any guesswork,  on the part of the researcher, of the functional forms of unknown parameters.  The chosen neural architecture is a universal function approximator~\cite{HORNIK1989}, which removes the need for guessing functional forms of unknown parameters.  The solver  precludes any possible model error associated with grid discretization.  Secondly, we perform the calculations using nine distinct optimization algorithms, for a comparative analysis of performance.
 
Here we study a two-neutrino-beam, two-flavor  system, with one beam initially electron flavor and the other $x$ flavor.  Given the initial flavor configuration, this toy problem can be evolved from the supernova \lq\lq source\rq\rq\ to the \lq\lq detector\rq\rq\ without significant computational difficulties~\cite{Keister:2014ufa}. Our focus, however, is not the forward evolution, but rather the estimation of model parameters responsible for a given detection. 
 
Our results show that inference, even given such an apparently-simple setup, is  an intricate task. We find that, for the specific small-scale model examined in this paper, evolutionary algorithms perform  well  in both the estimation of unknown parameters and initial conditions. However, this may change in more complicated setups and with large number of unknown parameters and a hybrid approach of consecutive optimizations, for instance, evolutionary algorithms followed by gradient and hessian based methods, might be more appropriate.  In what follows, we will discuss the potential advantages of adaptive-step-size solvers.

In Sec.~\ref{sec:opt}, we describe the algorithmic setup and specifics of the problem.  Then we perform two sets of inference experiments. In the first set (Secs.~\ref{sec:exp1} and \ref{sec:exp2}), we estimate unknown parameters governing the matter potential, assuming the initial conditions on flavor to be known.  In the second set (Secs.~
\ref{sec:exp3} and \ref{sec:exp4}), we instead estimate the initial conditions, assuming the matter potential to be known. In Sec.~\ref{sec:considerations} we make further considerations regarding the comparison of different models and estimation of uncertainties. We present conclusions in Sec.~\ref{sec:conclusion}.

\section{Inverse problems and neural network optimization}
\label{sec:opt}

\subsection{General Framework}
\label{subsec:GF}

Data assimilation is an inverse problem formulation~\cite{Tarantola2005}: a procedure whereby information in measurements is used to complete a model of the system from which the measurements were obtained.  For our purposes, the model $\bm{F}$ is written as a set of ordinary differential equations that evolve in affine parameter $r$ as:
\begin{equation}
  \diff{\bm{u}}{r} = \bm{F}(\bm{u},r,\bm{\theta})
  \label{eq:Feq}
\end{equation}
where the vector $\bm{u}$ is the observable being modeled, with initial value $\bm{u}_0$.  The affine parametrization $r$ may be, for example, time or distance. Any unknown parameter that influences the forward problem (differential equation) is denoted by $\bm{\theta}$. An observation $\bm{u}_D$ is made at a detector location $R$ and one seeks to estimate $\bm{\theta}$
that best fits this observation. This is achieved through minimization of the cost function:
\begin{equation}
 \text{Cost}(\bm{\theta})=\left[\bm{u}_D - \bm{u}_{\theta}(R)\right]^T \bm{W} \left[\bm{u}_D - \bm{u}_{\theta}(R)\right]
 \label{eq:cost}
\end{equation}
where $\bm{u}_{\theta}(R)$ is the prediction from $\bm{F}$.  In practice, we may only observe a subset of the components of the  vector $\bm{u}_D$, while through $\bm{F}$ all components are evolved in order to predict the final values at the detector. The sparse matrix $\bm{W}$ is introduced to select these components when computing the cost function. 

\begin{figure}[ht]
 \includegraphics[scale=0.45]{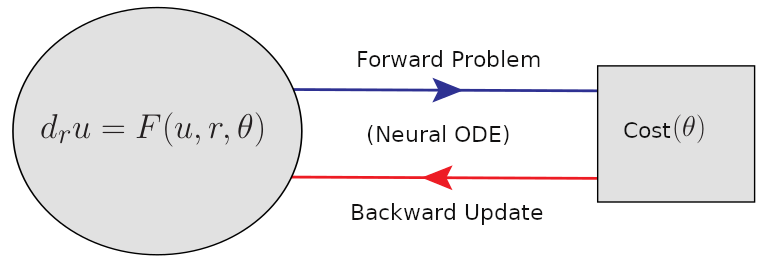}
 \caption{Data assimilation as a neural network architecture.  The forward problem is represented by a set of first-order ordinary differential equations, $\bm{F}(\bm{u},r,\bm{\theta})$, that depend on the unknown model parameters $\bm{\theta}$. Cost$(\bm{\theta})$ is the objective function whose minimum denotes the optimal solution of the inverse problem.
 Blue arrows denote the forward process of prediction, and red arrows denote the backward pass for error correction.}
 \label{fig:setup}
\end{figure}

In \cite{Armstrong:2017,Armstrong:2020}, the cost function was comprised of three parts: model error, measurement error, and physical constraint terms. The model error,  in addition to permitting uncertainty in the model parameter estimates, included terms related to the uniform discretization of the domain and finite difference approximation of the derivative.  In addition, as the optimization algorithm took the grid points to be independent, we had imposed co-variation of the model coordinates into the cost function as an equality constraint.  Here, we explicitly include only the measurement term in the cost function, as the neural architecture ensures that the other terms are automatically satisfied; this will be explained in detail below.  In \cite{Armstrong:2020}, we also considered $\bm{u}_0$ to be an input in the form of a measurement. Here, in the first two experiments we follow the same assumption, but without explicitly including $\bm{u}_0$ as part of the cost function (the cost function implicitly depends on $\bm{u}_0$, via the dynamical equations).  In the last two experiments, we assume all model parameters are known, and we instead optimize the cost function by varying the initial conditions $\bm{u}_0$. 

In recent decades, machine learning has been used to provide solutions to ordinary differential equations~\cite{vanMilligen1995,Lagaris1998,Berg2018,Magill2018,RAISSI2019}. As the focus of this work is the inverse problem, the system of differential equations is a building block of our setup.  As such, our goal is not to approximate the ODE solution through a neural architecture or discretized grid, but rather to understand which parameters in the ODE definition lead to a solution that best matches observations.  Hence, we use the existing vast and established literature on solving ODEs through forward integration in the vein of incorporating model knowledge with machine learning.  This is the motivation for using the recently developed neural ODE~\cite{Chen2018} network for data assimilation.  This network will automatically incorporate our model knowledge of the dynamical system, through Eq.~(\ref{eq:Feq}).

We include an adaptive step solver method in the neural architecture, removing the need for domain discretization and errors induced by such discretization.  Specifically, we employ the Radau method~\cite{Hairier1999}.  The forward-problem arrow in Fig.~\ref{fig:setup} refers to this part of the architecture.  The solution $\bm{u}_{\theta}(r)$ then satisfies all the physical constraints associated with Eq.~(\ref{eq:Feq}).  Consequently, no model and no constraint terms are needed in the cost function.  As such, the cost function in Eq.~(\ref{eq:cost}) contains only the measurement term.
We have verified that the errors associated with model and constraint terms are within numerical precision ($\lesssim 10^{-16}$).  An additional feature of this setup is the reduction of the number of unknown parameters that require optimization.  Obtaining adequate resolution in the previous setup required a rather large number of grid points.  In this new architecture, the points in the domain are automatically chosen by the adaptive Radau method to solve the forward problem to machine precision. 

 At first glance, the choice to incorporate the ODE solver into the neural architecture may seem puzzling.  The most general definition of a neural layer is a differentiable function with tensor input and output.  This is the basis of a general differentiable programming architecture.  Traditionally, neural layers are superpositions of simple primitive functions, but they need not be.  Differential equation solvers naturally fit this framework, as an ODE solver has an input vector $\bm{u}(r_n)$ that outputs a new vector $\bm{u}(r_{n+1})$, where the points ($r_n$ and $r_{n+1}$) and the separation between those points are determined adaptively.  In order for the solver to be a neural layer, the output of the ODE solver must be differentiable with respect to the unknown parameters.
This is achieved through automatic differentiation~\cite{Revels2016,Baydin2018} and adjoint sensitivity analysis~\cite{SANDU2003,Ca02006}.  Automatic differentiation in computer science encompasses a set of techniques for converting a program into a sequence of primitive operations that have specified routines for computing derivatives.  It is efficient, in that there exists a linear time cost in computing values, and it is numerically stable.  Adjoint sensitivity analysis allows for the automatic differentiation of ordinary differential equations.
The code for this work was written in the Julia programming language~\cite{Bezanson2017},  and we used the DiffEqFlux package~\cite{Rackauckas2017,Innes2018,Rackauckas2019}.  While the interested reader may delve deeper into these interesting topics, for the purposes of this work, it suffices to state that the current architecture allows us to compute the Jacobian and Hessian matrices of the dynamical model ($\bigtriangledown_{\bm{\theta}}\text{Cost}(\bm{\theta})$ and $\partial_{\theta^i}\partial_{\theta^j}\text{Cost}(\bm{\theta})$).

Once the forward problem is implemented, one performs the minimization of the cost function.  In Fig.~\ref{fig:setup} this step is represented by the backward update arrow. Typically, in deep learning, the minimization is performed through stochastic gradient descent (SGD)~\cite{Bottou1998} (which requires the computation of the gradients mentioned). 
In practice, the step size of the parameter update is a hyper parameter of the training procedure, which must be tuned to achieve convergence. Many improvements on SGD have been developed through decades. For instance, AdaGrad~\cite{Duchi2011} and Adam~\cite{Kingma2015} are algorithms widely used in deep learning. LBFGS~\cite{Robert2002}, which also approximates the Hessian of the cost function, is a commonly used alternative to SGD.   

The focus of this work is global optimization (finding the optimal value in the entire region of interest), and the algorithms mentioned above work locally, informed by gradients of the cost function. 
Hence, we  also employed a number of global algorithms in our analysis. A list of algorithms used in this work, along with their classification, is given below:
\begin{itemize}
\item Monte Carlo based methods (gradient free):
``Simulated Annealing'' (SAMIN)~\cite{Goffe1994,Goffe1996} --- based on Metropolis-Hastings algorithm to generate samples from a thermodynamic system,
 \item Evolutionary algorithms (gradient free):
 \begin{enumerate}
  \item ``Improved Stochastic Ranking Evolution Strategy'' (ISRES)~\cite{Runarsson2000,Runarsson2005} --- based on a combination of a mutation rule (with a log-normal step-size update and exponential smoothing) and differential variation (update rule similar to Nelder–Mead~\cite{Nelder1965}) ,
  \item ``Adaptive Particle Swarm Algorithm'' (APS)~\cite{Zhan2009} --- improve global coverage and convergence by switching between four evolutionary states: exploration, exploitation, convergence, and jumping out.
 \end{enumerate}
 \item Jacobian and Hessian based methods: 
 \begin{enumerate}
  \item ``Interior Point Optimizer'' (IPOPT)~\cite{Wachter2006} --- a primal-dual interior point method which uses line searches based on filter methods. IPOPT is designed to exploit $1^{\text{st}}$ and $2^{\text{nd}}$ derivative information if provided. If no Hessians are available, IPOPT will approximate them using a quasi-Newton methods, specifically a BFGS update.
  
  \item ``Newton method with Trusted Region Hessian'' (NTR)~\cite{NoceWrig06} --- quadratic approximation of the objective function by means of the hessian with steps restricted to be within a `trusted' region where the approximation is believed to be valid.
 \end{enumerate}
 \item Combination of global and local optimization: 
 \begin{enumerate}
    \item  ``Stochastic Global Optimization'' (STOGO)~\cite{Madsen2002} ---  systematically divide the search space (which must be bound-constrained) into smaller hyper-rectangles via a branch-and-bound technique, and searching them by a gradient-based local-search algorithm.
     \item `Multi-Level Single-Linkage'' (MLSL)~\cite{Kan1987,Kucherenko2005} --- global optimization by a sequence of local optimizations from random starting points,
 in conjunction with local optimizations algorithms
 \begin{itemize}
  \item BOBYQA~\cite{Powell2009} --- (gradient free) bound-constrained optimization using an iteratively constructed quadratic approximation for the objective function,
  \item ``Method of Moving Asymptotes'' (MMA)~\cite{Svanberg2002} --- local, convex and separable approximation of the objective function from the gradient,
  \item LBFGS~\cite{Nocedal1980,Liu1989} ---  quasi-Newton method that approximates the Broyden–Fletcher–Goldfarb–Shanno algorithm (BFGS)~\cite{Luenberger2015} using a limited amount of computer memory.
 \end{itemize}
 \end{enumerate}
 
\end{itemize}

These algorithms cover a wide range of methodologies.  For instance, multi-level algorithms (MLSL) have been used for over three decades in optimization, and nowadays are part of many statistical programming languages.  Simulated annealing is another widely used algorithm, with over four decades of applications.  In addition to the familiar Newton's method and IPOPT, we have also included evolutionary algorithms like ISRES and APS which are inspired by biological evolution~\cite{Vikhar2016}. Clearly, optimization is a rather fascinating and varied field.  
The numerical implementation can be found in NLopt~\cite{Johnson} and Optim~\cite{mogensen2018optim} packages. In all experiments covered in this work, we set the maximal number of iterations for the optimization procedure to 1000.

\subsection{Specifics of the problem}
\label{subsec:SOP}
The neutrino flavor evolution problem has been explained in detail in~\cite{Armstrong:2020}.
Here we summarize the system of differential equations,
\begin{equation} \label{eq:model}
  {\bm{F}}_i=\diff{{\bm{P}}_{i}}{r} = \left(\Delta_i {\bm{B}} + V(r) \hat{z} 
  +\mu(r) \sum_{j\neq i} {\bm{P}}_j \right) \times {\bm{P}}_i
\end{equation}
Here, $\Delta_{i} = \delta m^2/(2E_{i})$ are the vacuum oscillation frequencies of neutrinos with energies $E_i$. The mass-squared differences in vacuum are $\delta m^2$. The unit vector representing neutrino flavor mixing in vacuum is ${\bm{B}}=\sin(2 \alpha) \hat{x} -\cos(2 \alpha) \hat{z}$, where $\alpha$ is the mixing angle between the flavor and mass eigenstates. The functions $V(r)$ and $\mu(r)$ are the potentials arising from neutrino-matter and neutrino-neutrino interactions, respectively. The \lq\lq polarization vectors\rq\rq\ $\bm{P}_i$, which contain information about the flavor composition of the neutrinos, play the role of the state variable $\bm{u}$ from Eq.~(\ref{eq:Feq}), and the only components that are measured at the detector are the $P_z$ of each neutrino.  Anti-neutrinos are not considered at present, and will be included in future work. The generalization to include anti-neutrinos is straightforward, and simply involves allowing for negative oscillation frequencies $\Delta_i$ in Eq.~(\ref{eq:model}).

In our model, we take the neutrino-neutrino potential to be,
\begin{equation}
\mu(r) = \frac{\mu_0}{(r+\delta_0)^4}.
\end{equation}
 This choice is consistent with how this coupling strength varies in the neutrino bulb model calculations employing the single-angle approximation. In our SDA experiments, $\mu_0$ is taken to be a constant with a known value and $\delta_0=10^{-3}$ is added to avoid any numerical singularities at $r_{0}=0$.  The matter potential $V(r)$ is chosen to be
\begin{equation}
 V(r) = \frac{V_0}{(r+\delta_0)^3}.
 \label{eq:vr0}
\end{equation}
In the first two experiments, $V_0$ is treated as an unknown parameter that we optimize. To generate the simulated \lq\lq detector data\rq\rq\ ($\bm{u}_D$ in Eq.~(\ref{eq:cost})), we use a constant value $V_0 = \tilde{V}_0$ of the matter potential coefficient, given in table~\ref{table:Known}.
As a thought experiment and proof of concept, we study a system of two neutrino beams, with all the parameters used for simulated data generation displayed in table~\ref{table:Known}. In future work we intend to study much larger systems.

\setlength{\tabcolsep}{5pt}
\begin{table}[ht]
\small
\centering
\begin{tabular}{|l |c | c |c|} \toprule
\hline
 \textit{Parameter} & \textit{Value} & \textit{Initial polarization} & \textit{Value} \\\midrule \hline
 $\Delta_1$ & 30 &  $P_{1, z}(r_0)$ & -1.0\\ 
 $\Delta_2$ & 55 & $P_{2, z}(r_0)$ & \ 1.0 \\ \cline{3-4}
 $\mu_0$ & 10.0 & \textit{Final polarization} & \textit{Value} \\ \cline{3-4}
 $\tilde{V}_0$ & 50.0 & $P_{1, z}(R)$ & 0.20575 \\
 $\alpha$ & 0.15 & $P_{2, z}(R)$ & -0.96750 \\
 $r_{0} $ & 0 & & \\
 $R$ & 5 & & \\
 \bottomrule \hline
\end{tabular}
\caption{\textbf{Model parameters used for generating the simulated `detector' data.}  $\Delta_i$ are the vacuum oscillation frequencies of the neutrinos, and ($\mu_0, \tilde{V}_0$) are the multiplicative factors governing the neutrino-neutrino coupling potential $\mu(r)$ and matter potential $V(r)$.  Parameter $\alpha$ is the mixing angle in vacuum. Neutrino 1 is initially $x$ flavor and neutrino 2 is initially electron flavor,  as is reflected in the respective initial $P_z$ values.} 
\label{table:Known}
\end{table}

\begin{figure}[ht]
 \includegraphics[scale=0.45]{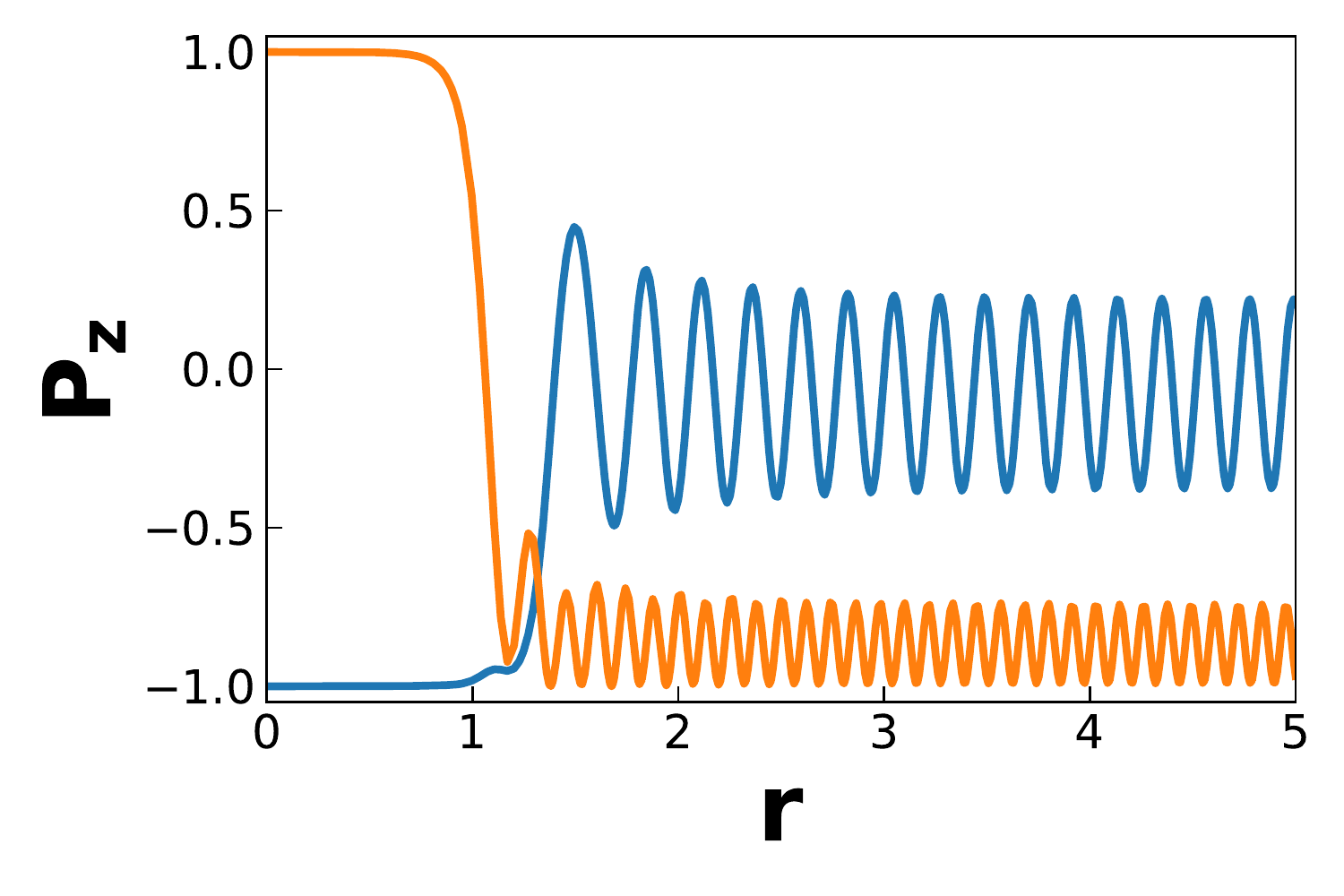}
 \caption{ The solution to the forward problem, in terms of the $z$ components of the polarization vectors, ${\bf P_z}$, of the  two neutrino beams, as functions of the affine parameter ${\bf r}$. The parameters used in the forward integration are shown in table~\ref{table:Known}. One neutrino beam (orange) is initially electron flavor, and the other (blue) is initially x flavor. In this toy problem, we assume the detector to be at location $r=5$.  For solving the inverse problem, the only information available to the procedure is the polarization measured at this location.}
 \label{fig:polarizations}
\end{figure}

Fig.~\ref{fig:polarizations} displays the  solution to the forward problem---represented here by the $z$ components of the polarization vectors as functions of $r$, for the model parameters of table~\ref{table:Known}. The two neutrino beams are initially in pure flavor states (electron and $x$). At some intermediate distance, there occurs a large flavor transformation, and the two neutrino beams interchange flavors.  As expected, with increasing distance, both matter and neutrino potentials become less relevant and we can observe vacuum oscillations.

\section{Matter potential coupling as an unknown constant}
\label{sec:exp1}
\begin{figure}[ht]
 \includegraphics[scale=0.4]{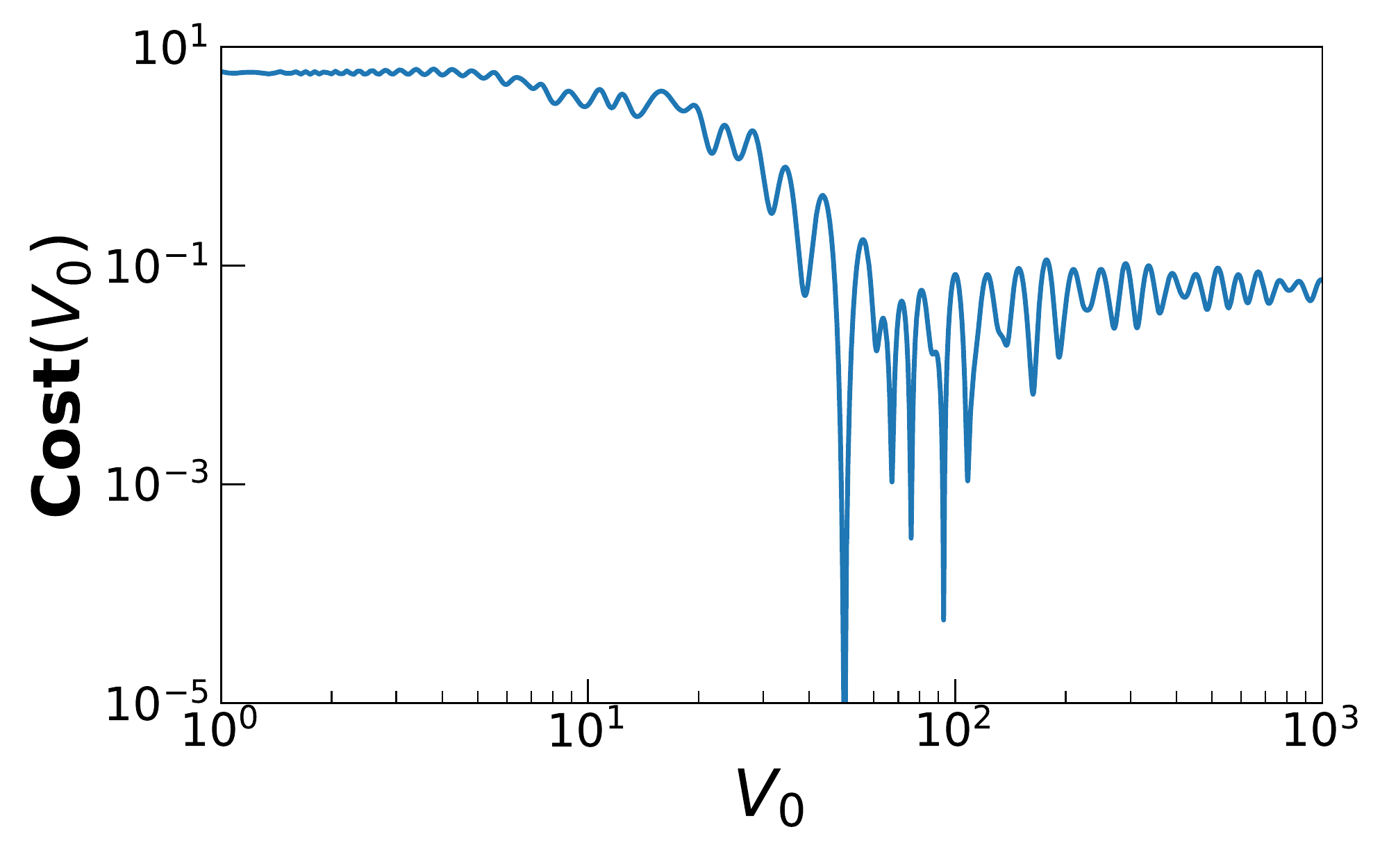}
 \caption{Cost function dependence on the free parameter $V_0$, where the simulated \lq\lq detector\rq\rq\ data was obtained through forward integration using the parameter values of table~\ref{table:Known}. The global minimum is obtained at $V_0=\tilde{V}_0$ (see table~\ref{table:Known}), but there are many local minima present.   Thus, even with only a single unknown parameter, this problem presents a very difficult challenge for the optimization.}
 \label{fig:lossf}
\end{figure}

In this section, we assume the matter potential coupling $V_0$ is an unknown parameter and ask whether the $z$ components of the polarization vectors of the two neutrino beams at the detector provide sufficient information to infer the true value $\tilde{V}_0$ provided in Table~\ref{table:Known}. As this is a small system, and there is only one unknown parameter, we can plot the dependence of the cost function  on the unknown parameter using repeated forward integration, as shown in Fig.~\ref{fig:lossf}.  That is, the forward code was run several times with different parameter values $V_0$, and the corresponding values of $P_z$ at the endpoint in each case were compared with the true values (that is, with $V_0 = \tilde{V}_0$) to generate the cost function using Eq.~(\ref{eq:cost}). In practice, physical systems contain many more particles, and more unknowns, rendering it infeasible to create the plot analogous to Fig.~\ref{fig:lossf}.  We are considering a \lq toy\rq\ problem, however, as a proof of concept for the approach we propose, and for the relative ease with which we may examine figures such as those displayed in this work.

\begin{figure}[ht]
 \includegraphics[scale=0.4]{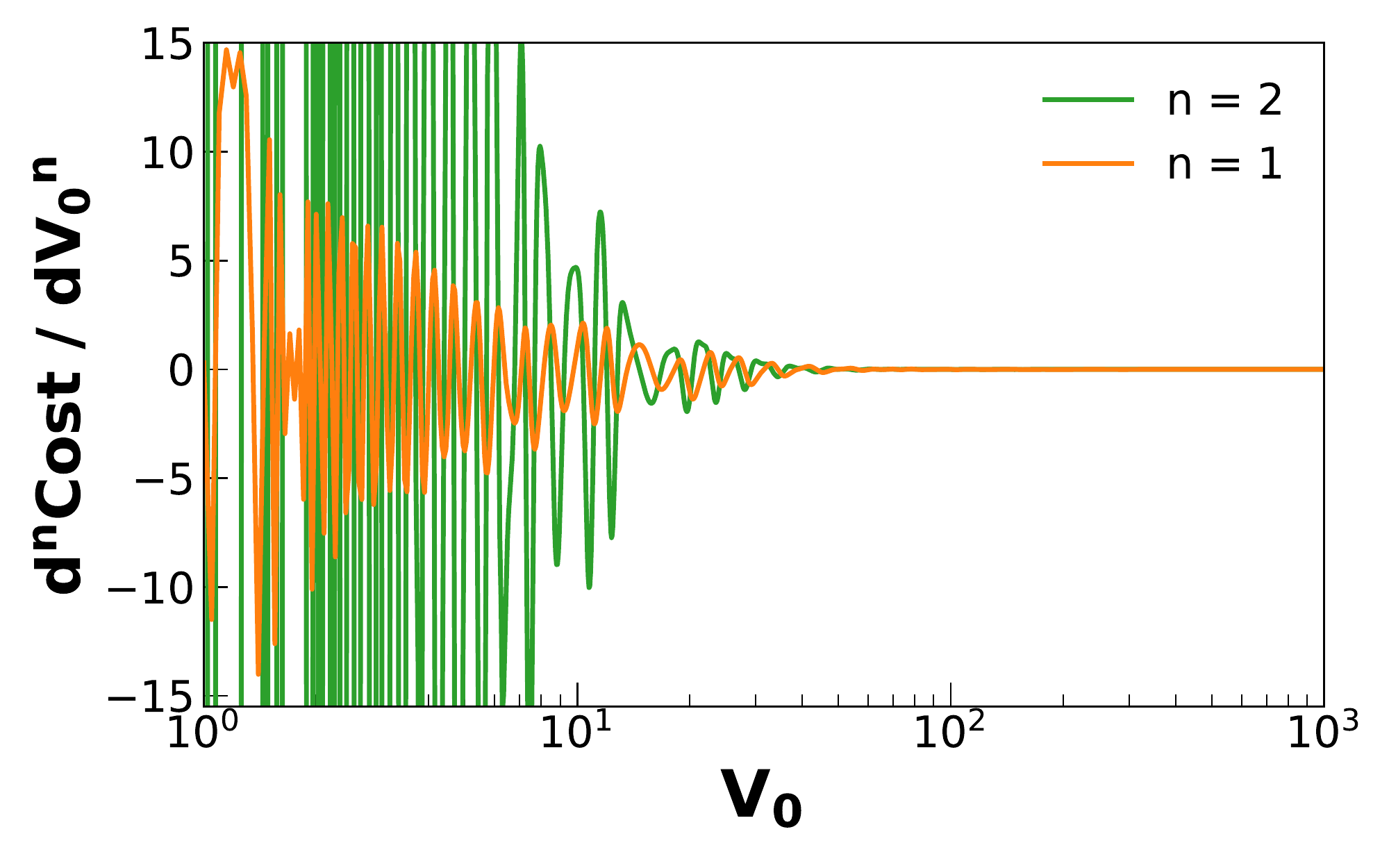}
 \caption{Gradient (orange) and Hessian (green) of cost function with respect to the free parameter $\theta=V_0$. Ideally, one expects that these derivatives will be large at locations far from the global optimum, and that they will gradually go to zero as the minimum is approached. In our setup there is a sudden transition from large oscillations to tiny ones near $V_0 = \tilde{V}_0$. For small $V_0$, gradient based methods will have very large parameter updates in either direction, and for large $V_0$ the updates will be very small.}
 \label{fig:lossdf}
\end{figure}

From Fig.~\ref{fig:lossf} we see that at $V_0 = \tilde{V}_0$ the cost function attains its global minimal value as expected.  In addition, though, there are many local minima present, which will make it difficult for any local optimization algorithm to find the correct value $\tilde{V}_0$ if the initial guess is in the vicinity of a  different local minimum. In addition, in the parameter range with small $V_0$ values, the cost function changes very quickly between large and small values. To understand how gradient and Hessian based optimizers would perform, in Fig.~\ref{fig:lossdf} we plot the first and second derivatives of the cost as function of $V_0$.
As the figures show, there are large fluctuations in the derivatives of the cost function at small values of $V_0$, and many stable local minima at large $V_0$.   On the one hand, if the initial guess is small, gradient-based optimization will change the value of the guess drastically, and on the other hand, if the initial guess is large, the changes will be minuscule. This does not bode well for gradient- (and Hessian-) based optimization. Even with just two neutrino beams, the problem is rather complicated.  Hence our choice for global optimization and the wide range of optimization algorithms that we test. 

\begin{figure}[ht]
\subfloat[][]{\includegraphics[scale=0.25]{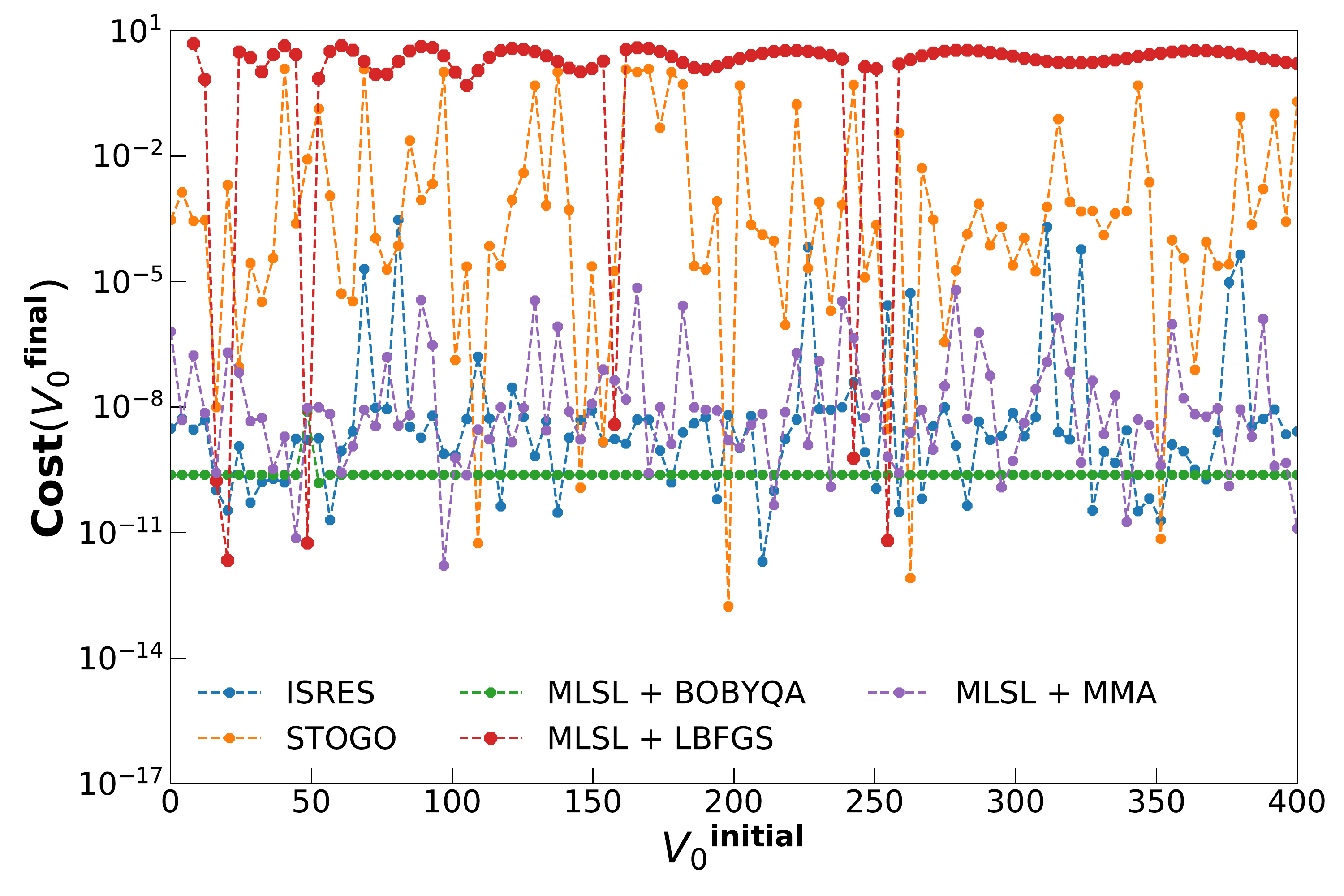}\label{fig:tr1}}\\
\subfloat[][]{\includegraphics[scale=0.25]{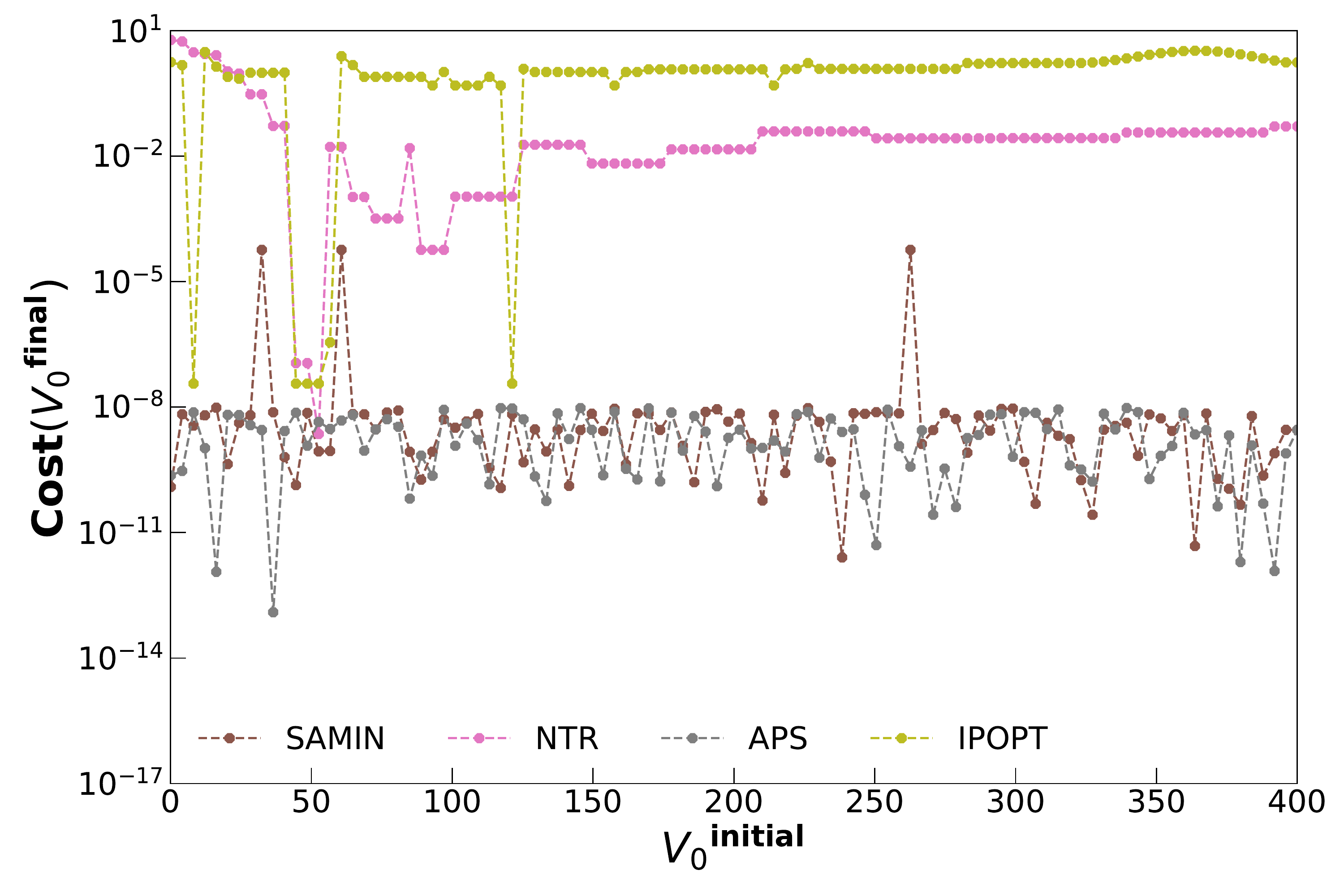}\label{fig:tr2}}\\
 \caption{The cost function at the final inferred values of $V_0$  (i.e., at the end of the optimization procedure) as function of the initial  guess $V_0^{\text{initial}}$, for each  of the nine optimization algorithm. Ideally this value should be $0$, and in practice, the smaller it is the better the optimization. For visual clarity, the nine algorithms are plotted  in two groups \ref{fig:tr1} and \ref{fig:tr2}.  In this set of experiments, the optimization procedure is instructed to regard $V_0$ as a constant unknown parameter.}
 \label{fig:train1}
\end{figure}

\begin{figure}[ht]
 \includegraphics[scale=0.28]{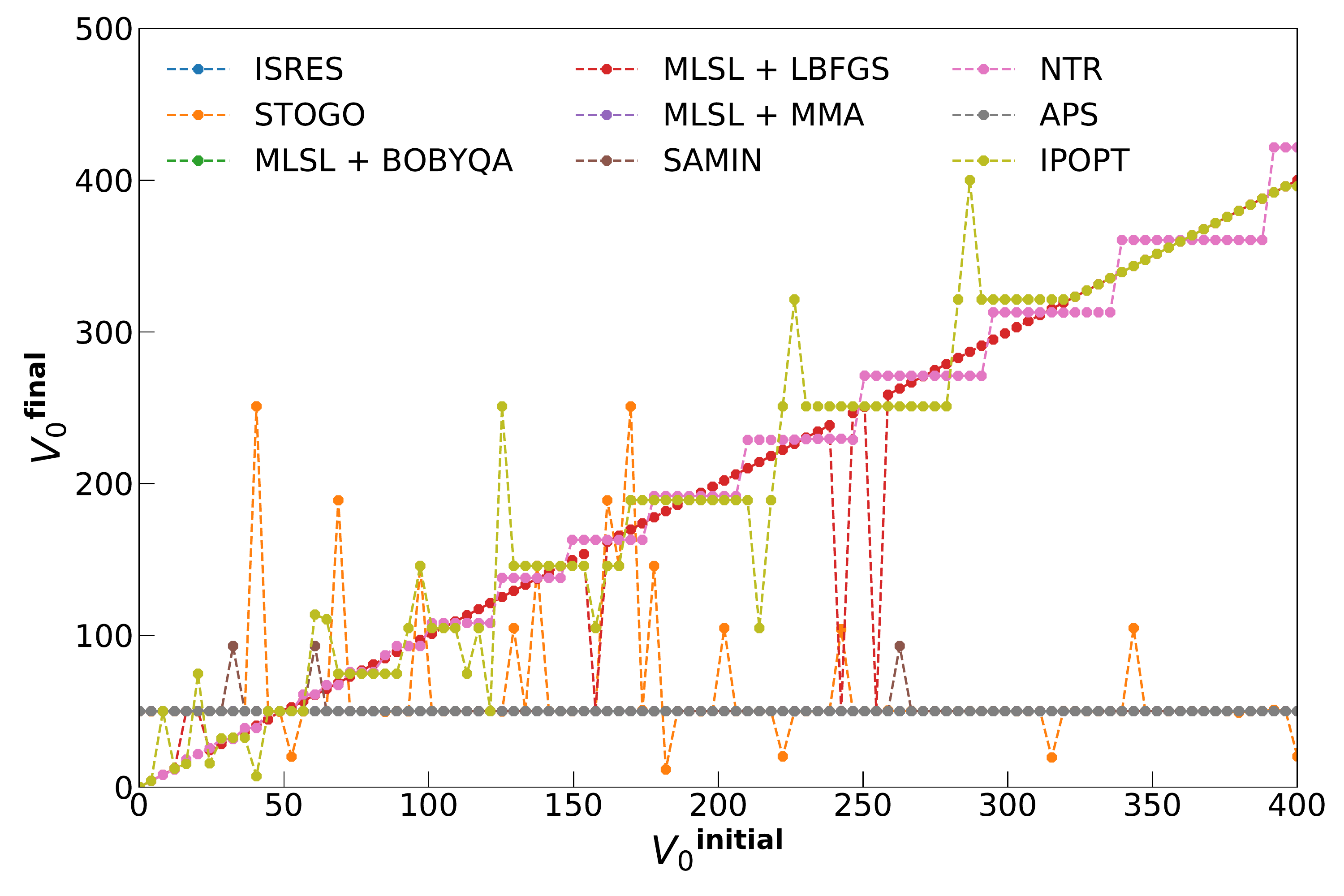}
 \caption{The final inferred value of $V_0$ as function of the initial guess $V_0^{\text{initial}}$, for each optimization algorithms. For a converged optimization this value should be $\tilde{V}_0=50$, independent of the initial guess (see table~\ref{table:Known}).  These results are for the same set of experiments as those in Fig.~\ref{fig:train1}.}
 \label{fig:train2}
\end{figure}

In Fig.~\ref{fig:train1} we plot the cost function at the end of the iterations for each algorithm as a function of the initial guess for $V_0$. We sampled uniformly 100 initial values in the range $[0,400]$. 
Not surprisingly, gradient and Hessian based algorithms have a final cost value much larger than the rest. To verify that small cost indeed translates to convergence to the global minimum, we also plot the final inferred values of the unknown parameter as function of the initial guess in Fig.~\ref{fig:train2}. 

As can be seen from both figures, most gradient-based methods have difficulties in converging to the optimal value, while gradient-free methods perform rather well. In particular, with the Jacobian/Hessian based methods like IPOPT and NTR, the final inferred value of $V_0$ is positively correlated with the initial guess, suggesting that the optimization procedure simply finds a local minimum close to the initial guess. This outcome agrees with expectations laid out by Figs.~\ref{fig:lossf} and~\ref{fig:lossdf}. In addition, the combination of a global (MLSL) and local method (MMA) seems successful.

\section{Matter Potential Coupling as an unspecified function of position}
\label{sec:exp2}

In the calculations presented in this section, we treat the numerator in Eq.~(\ref{eq:vr0}), $V_0$, as a function of the affine parameter $r$. Note that the simulated \lq\lq detector data\rq\rq\ ($\bm{u}_D$) used is identical to the data used in the previous section.  That is, it was generated using a constant matter potential coefficient $\tilde{V}_0$.  Instead of selecting a specific functional form for this dependence, we represent the numerator by a two-layer feed-forward neural network, $V_0(r)=|\mathcal{N_A}(\bm{\theta},r)|$. Each layer has five neurons. The first layer has a hyperbolic tangent activation function, and the second is linear; for a total of 16 parameters denoted by $\bm{\theta}$. The depth, width, and activation functions of the neural architecture are hyper parameters.  We choose these hyper parameters strictly, as our goal here is to merely explore whether a neural architecture can provide us with reasonable estimates for $V_0$ and the architecture chosen can represent a wide range of functions. Since only positive values are physically meaningful, the matter coupling parameter is taken to be the absolute value of the output of the neural architecture. We opted to perform global optimization in the range $\pm 10^3$ for each of the parameters. For each method we sampled uniformly 40 initial parameter sets for each optimization.  

\begin{figure}[ht]
 \includegraphics[scale=0.3]{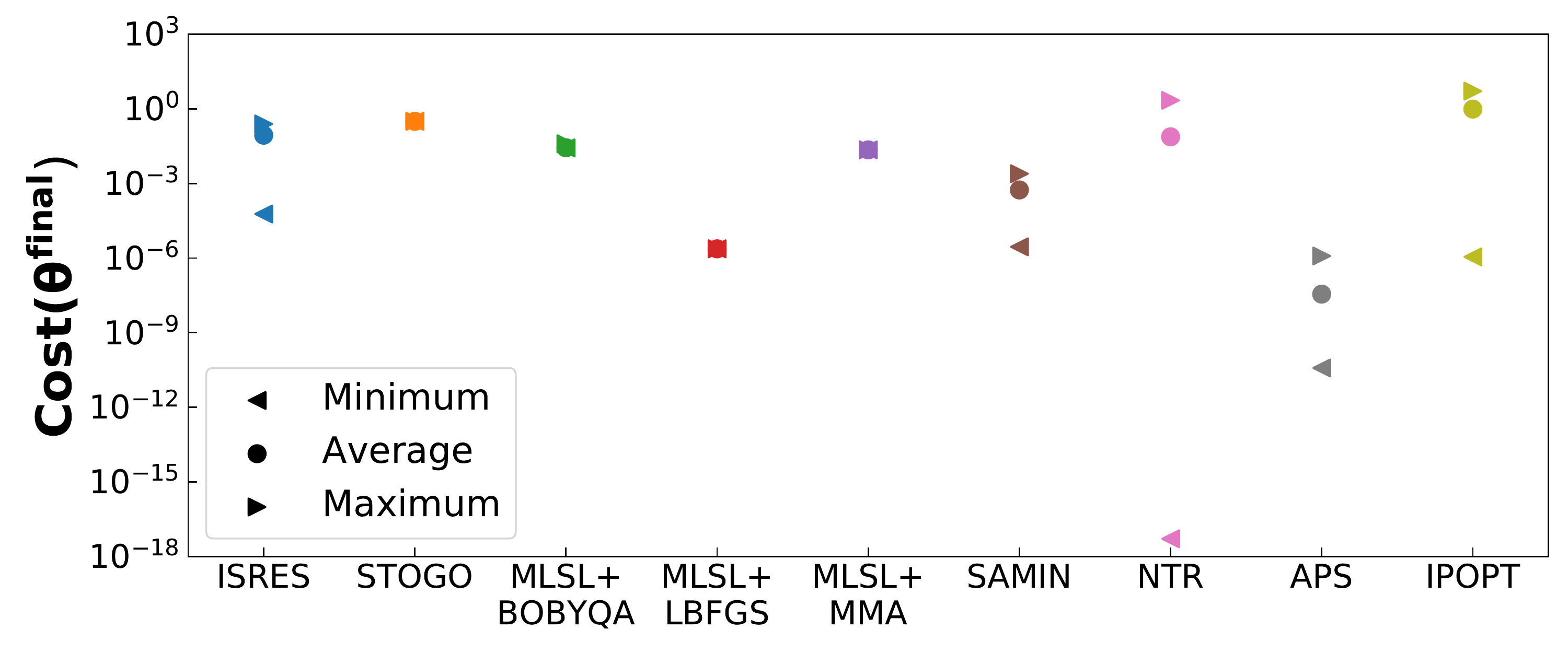}
 \caption{The minimum, maximum and average values of the cost function at the final values of $\bm{\theta}$  (i.e., at the end of the optimization procedure), for each optimization algorithm. Ideally, all these three values should be $0$. The spread shows how dependent the optimization algorithms are on the initial guess of the unknown parameters.  In this set of experiments, the optimization procedure treats $V_0$ as a generic function, encoded in terms of a two-layer feed-forward neural network, that depends on unknown parameters $\bm\theta$.}
 \label{fig:nntheta}
\end{figure}

In Fig.~\ref{fig:nntheta} we show the maximal, minimal and average values of the cost function at the end of the each optimization procedure. APS stands out from the rest: it performs quite well for all initial guesses and provides overall small cost values.  While NTR achieves a near-zero cost value for a particular initial guess, its results are quite spread and have a strong dependence on initial conditions.  Generally, dependence on the initial guess is to be expected.  If a guess happens to be close to the optimal result, one would expect the optimization procedure to produce a final cost value close to zero.  On the other hand, if the initial guess is quite far from the optimal value, the optimization might converge to local minimal nearby.  An additional complication arises from the possibility of degeneracies as the number of unknown parameters increases.  \lq\lq Degeneracies\rq\rq\ here refers to the possibility of multiple solutions $|\mathcal{N_A}(\bm{\theta},r)|$ that yield the same values of $P_z(r)$ at the endpoint.

\begin{figure}[ht]
 \includegraphics[scale=0.35]{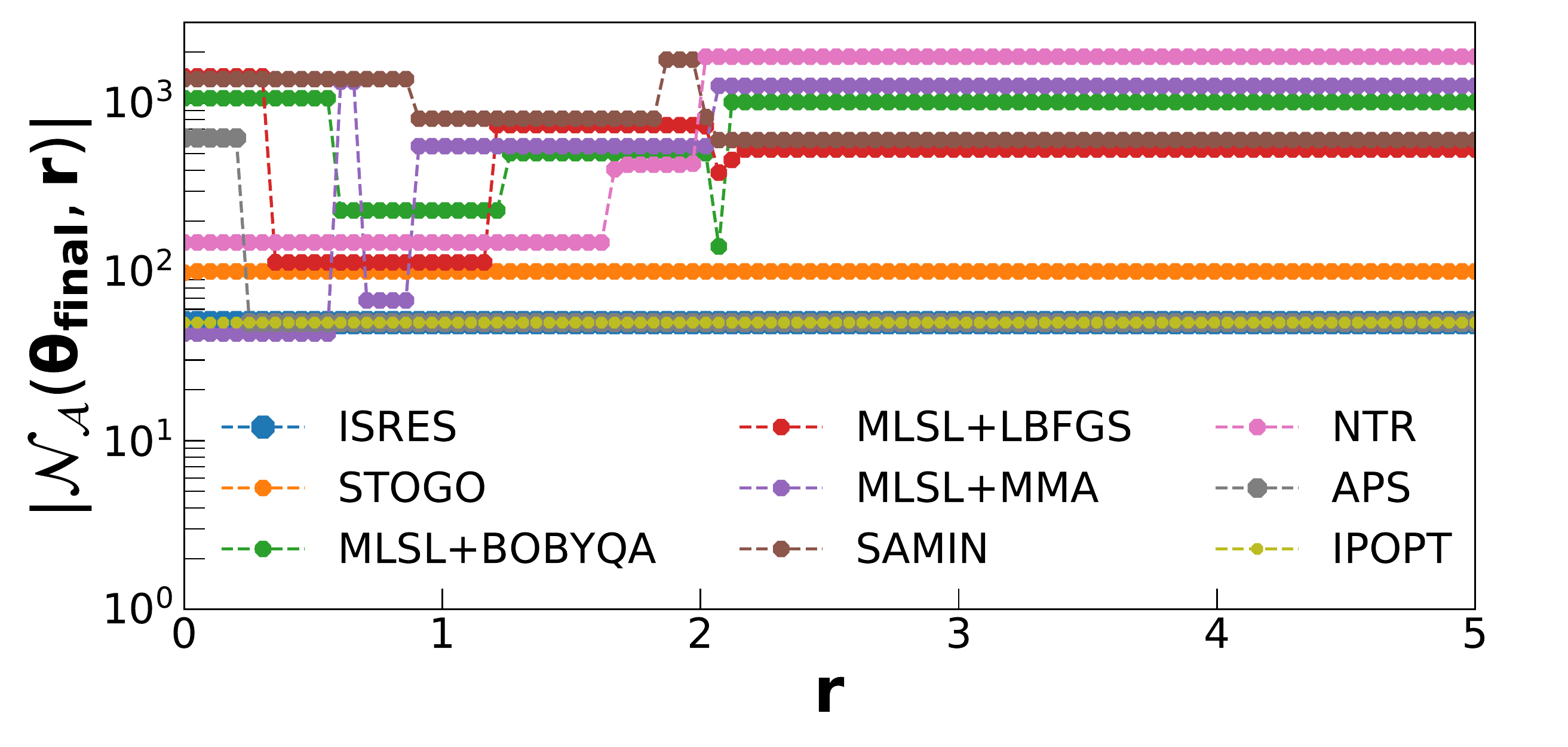}
 \caption{Optimal matter potential coupling {\og $V_0$} found by each method as function of the affine parameter $r$. Some algorithms find the coupling to be a constant close to the value in table~\ref{table:Known}, while others show sharp matter profile changes. This shows that various matter profiles, despite being quite different, lead to very similar values of the polarization $z$-components detected  at $r=5$.  This is for the same set of experiments described in the caption of Fig.~\ref{fig:nntheta}.}
 \label{fig:nn}
\end{figure}

To illustrate the behavior behind this remark, we plot the numerator $|\mathcal{N_A}(\bm{\theta},r)|$ for each method using the parameters that produced the smallest cost value in Fig.~\ref{fig:nn}. A priori, we know that a constant function of $r$, namely $V_0(r) = \tilde{V}_0$, is one possible optimal solution.

As the plot shows, both ISRES and IPOPT converge to a constant function close to $\tilde{V}_0$. APS shows a sharp transition from a region of high density to the optimal $\tilde{V}_0$. A rather interesting result is displayed by NTR, where the matter density profile experiences two sharp transitions and yet the cost value is small ($\approx 10^{-18}$). Other methods that display sharp transitions are MLSL+LFBGS and SAMIN, which perform slightly better than ISRES. In fact, many methods show sharp changes in the matter profile. For the model chosen, the matter density profile is inversely proportional to $r^3$, so the sharp changes in the numerator amount to small changes in the profile itself.  In an actual core-collapse supernova environment, such sharp transitions may represent, for instance, a dense matter outflow in a lower density background, or alternatively the presence of a shock during the supernova explosion. 
In this manner, allowing for a variable numerator represented by the neural architecture can lead us to discovering other possible matter profiles consistent with the same detector measurement.

\section{Partially unspecified initial conditions}
\label{sec:exp3}
In this section we study the influence of initial conditions on the inference of the neutrino flavor composition at the detector. 
As an illustration, we assume that Neutrino 1, which in the original setup was taken to be initially in the $x$ flavor, decouples earlier and can potentially oscillate in flavor before the second neutrino is emitted.  By assuming the matter and neutrino potentials of table~\ref{table:Known}, and given detector measurements at $R=5$ in Fig.~\ref{fig:polarizations} (i.e., the same simulated detector data as in the previous experiments), we investigate whether we can infer the initial polarization of this neutrino. The initial flavor polarization is normalized, so it can expressed by two free parameters, the azimuthal and polar angles in flavor space; that is,
 $u^{(1)}_0=\{\cos(\theta_{1_A}) \sin(\theta_{1_P}), \sin(\theta_{1_A}) \sin(\theta_{1_P}), \cos(\theta_{1_P}) \}$,  where $\theta_{1_A}\ \epsilon \ [0,2\pi],\ \theta_{1_P}\ \epsilon\ [0,\pi]$. Here, $\bm{\theta}=\{\theta_{1_A},\theta_{1_P}\}$ are the unknown parameters to optimize. As we assume coherent evolution, the polarization is normalized throughout the evolution.
\begin{figure}[ht]
 \includegraphics[scale=0.5]{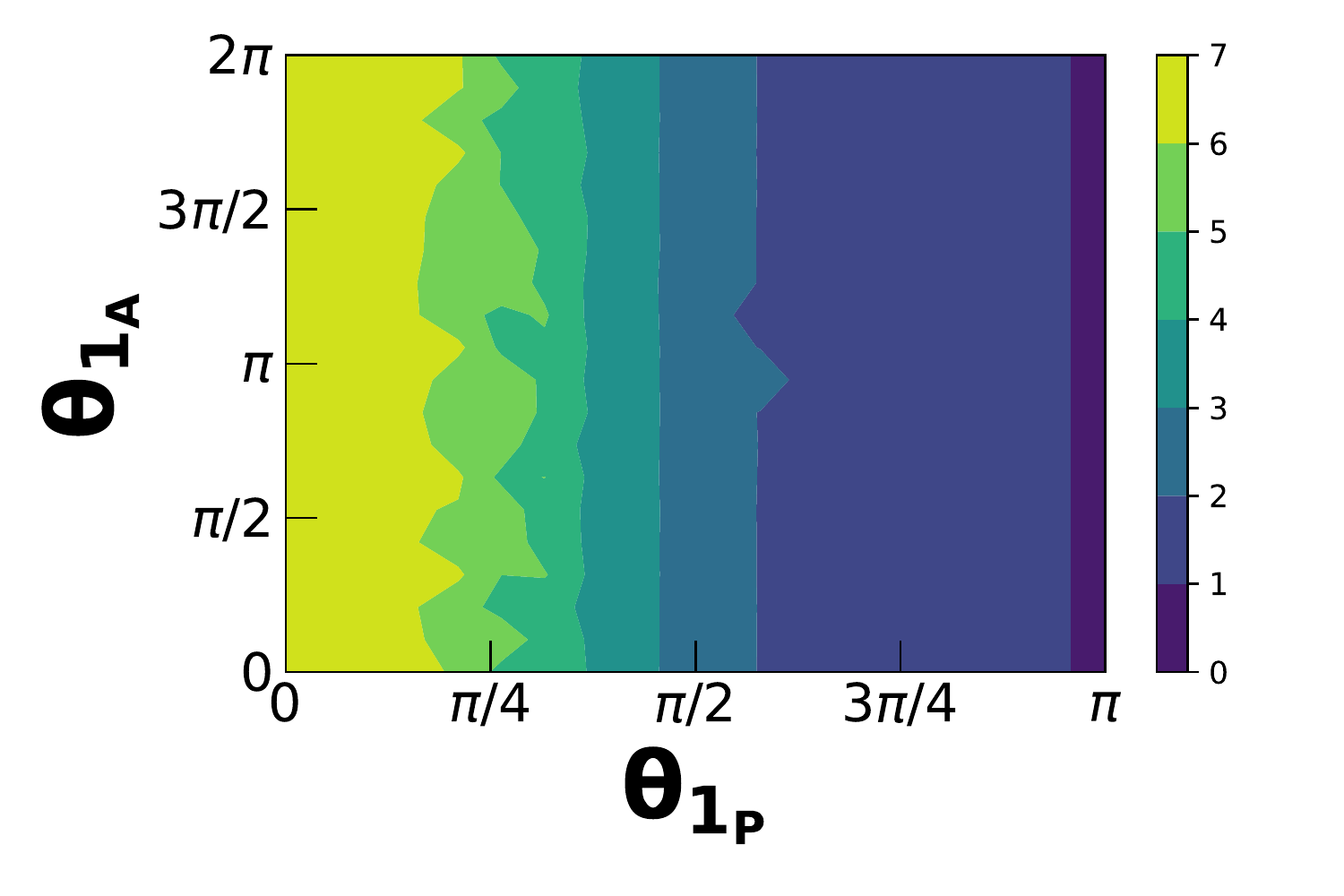}
 \caption{Dependence of the cost function, Cost$(\theta_{1_P}, \theta_{1_A})$, on   the initial orientation in flavor space of the polarization vector of the first neutrino. There is a relatively strong dependence on the polar angle, $\theta_{1_P}$, and weak dependence on the azimuthal angle, $\theta_{1_A}$. This is an indication of degeneracy in parameter space.  Here, the matter potential was taken to be a known constant $\tilde V_0$ (see table~\ref{table:Known}), and only the initial angles $\theta_{1_P}$ and $\theta_{1_A}$ were varied. The simulated \lq\lq detector\rq\rq\ data was generated by using $P_{1,z} = -1$ as the starting point of the forward integration.}
 \label{fig:cu0}
\end{figure}

In Fig.~\ref{fig:cu0} we plot the cost function dependence on these two parameters. Given that the second neutrino is initially of electron flavor, the optimal value for the first neutrino is to be an $x$ flavor (i.e., $P_z = -1$, or equivalently, $\theta_{1_P}=\pi$), as the figure confirms. In addition, we can see that the polar angle plays a major role in determining the value of the cost function. The figure shows the cost value decrease as the polar angle changes from $0$ to $\pi$, which is the optimal value. There is also minor dependence on the azimuthal angle for a fixed polar angle. Overall, this is to be expected, as for $\theta_{1_P}=\pi$, the neutrino is $x$ flavor regardless of $\theta_{1_A}$.

We have performed 100 optimization experiments with different initial guesses of $\bm{\theta}$,  from a uniform grid of initial values within the allowed range.  The statistics for the final values are summarized in table~\ref{table:angles1}. 
\setlength{\tabcolsep}{5pt}
\begin{table}[ht]
\small
\centering
\begin{tabular}{|l|c |c | c |} \toprule
\hline
 \textit{Method}& $\left(\mu_{\theta_P},\mu_{\theta_A}\right)/\pi$ & $\left(\sigma_{\theta_P},\sigma_{\theta_A}\right)/\pi$ \\\midrule \hline
  ISRES & $(0.996,1.05)$ & $(0.001,0.57)$ \\
  STOGO & $(0.87,1.52)$ & $(\approx 0,\approx 0)$ \\
  MLSL + BOBQYA & $(1.0,1.4)$ &  $(\approx 0, \approx 0)$ \\ 
 MLSL + LBFGS & $(0.484,1.0)$ & $(0.31,0.64)$ \\
 MLSL + MMA  & $(0.997,1.011)$ & $(0.00013,\approx 0)$ \\
    SAMIN & $(0.997,0.996)$ & $(0.002,0.615)$ \\
    NTR & $(0.47,0.98)$ & $(0.31,0.65)$ \\
      APS & $(1.0,1.0)$ & $(\approx 0 ,0.9)$\\
      IPOPT & $(0.9,1.1)$ & $(0.2 ,0.4)$\\ \hline
     \textit{Optimal Values}& $(1,-)$ & $(0,-)$\\
 \bottomrule \hline
\end{tabular}
\caption{Sample average and standard deviation of the  inferred angles in flavor space, for the initial polarization of the first neutrino at the end of the each optimization procedure. The optimal values shown at the end of the table; there is no preferred azimuthal angle.  In this set of experiments, the matter coupling parameter $V_0$ is taken to be a known constant, and the initial polar and azimuthal angles, $\theta_{1_P}$ and $\theta_{1_A}$, of the polarization vector of the first neutrino are regarded as being unknown parameters that the optimization procedure is tasked with inferring.} 
\label{table:angles1}
\end{table}
APS and MLSL + BOBQYA reach the optimal value of $\theta_{1_P}$ for most of the initial guesses, as shown by the tiny variances in the table.  Many other algorithms (ISRES, MLSL + MMA, SAMIN, IPOPT) converge quite close to the optimal value and have small variances. All methods converge to large values of the azimuthal angle. To understand this behavior, we computed the gradients of the cost function and found one major gradient flow toward $\theta_{1_P}=\pi$, as expected.  We also found a rather small flow toward $\theta_{1_A}=2\pi$.  But interestingly, the average inferred values of the azimuthal angle seem to be clustering around $\pi$ rather than $2\pi$. This is an unexpected result,  and a priori hard to guess, as one would need to solve to the flavor evolution equations for all initial conditions to notice this secondary flow.

\begin{figure}[ht]
 \includegraphics[scale=0.3]{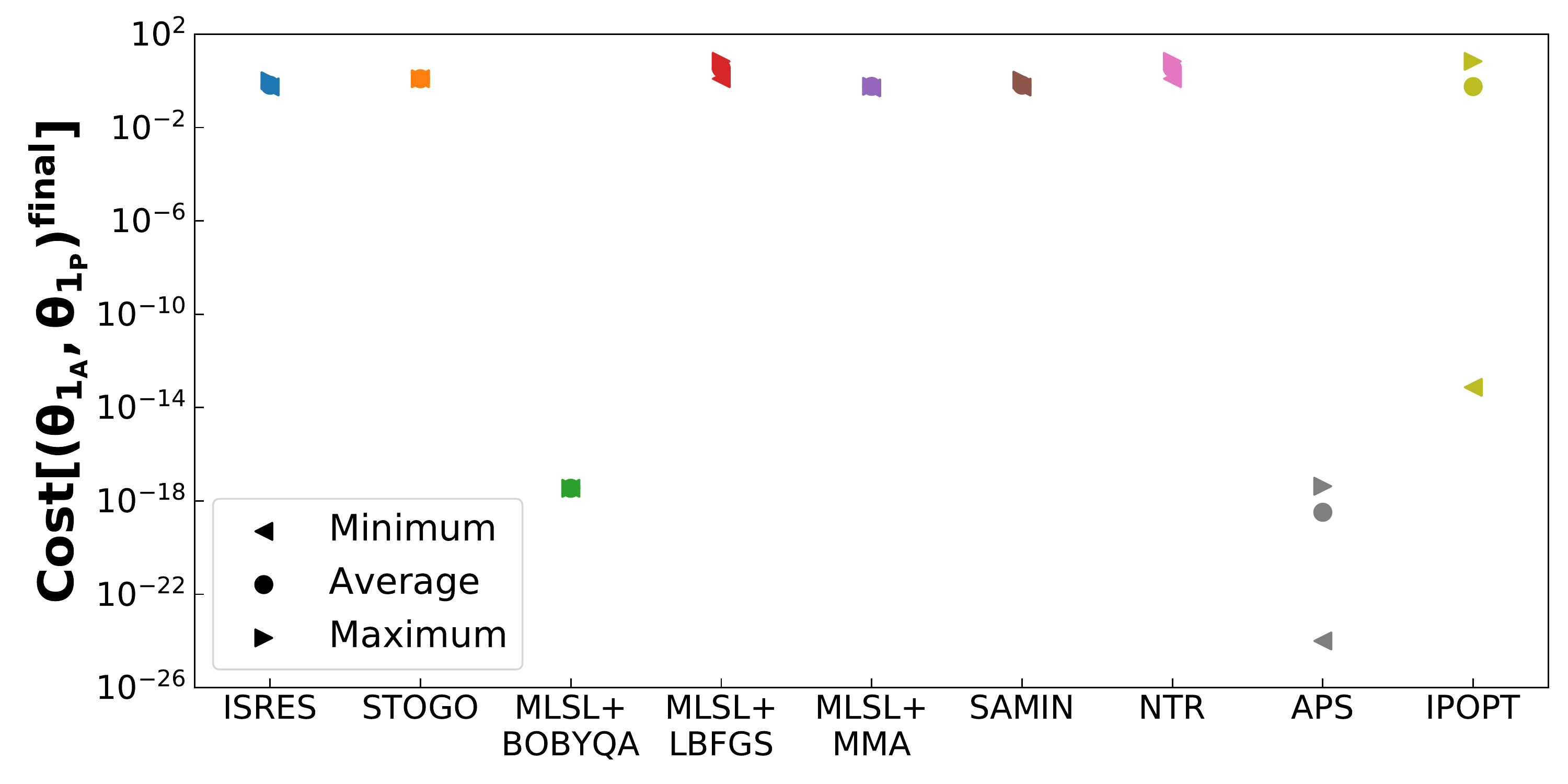}
 \caption{The minimum, maximum and average values of the cost function at the final values of $\bm{\theta}$  (i.e., at the end of the optimization procedure), for each optimization algorithm. Ideally, all these three values should be $0$.  For each algorithm, the spread  of cost function values shows how dependent the optimization algorithms are on the initial guess of the unknown parameters. These results are for the set of experiments described in the caption of table~\ref{table:angles1}, where the unknown parameters to be optimized are $\bm{\theta}=\{\theta_{1_P},\theta_{1_A}\}$.}
 \label{fig:u01theta}
\end{figure}

In Fig.~\ref{fig:u01theta} we display the maximal, minimal and average cost value obtained from each method. 
APS and MLSL + BOBQYA result in small cost values for all initial guesses, in agreement with table~\ref{table:angles1}. 
Most methods do not show any spread in the final values of the cost function, apart from APS and IPOPT. In these two cases, for some initial conditions, the methods achieve very low cost function values.  
\section{Completely unspecified initial conditions}
\label{sec:exp4}
In this section, we make no assumptions about the initial polarizations. Instead, given the detector data generated by the parameters in table~\ref{table:Known}, we optimize the cost function for the azimuthal and polar angles in flavor space for both neutrinos. We pick 5 uniformly-spaced values for each of the 4 angles for a total of 625 experiments for each optimization method. These are initial guesses of the angles in flavor space. 

\begin{figure}[ht]
 \includegraphics[scale=0.3]{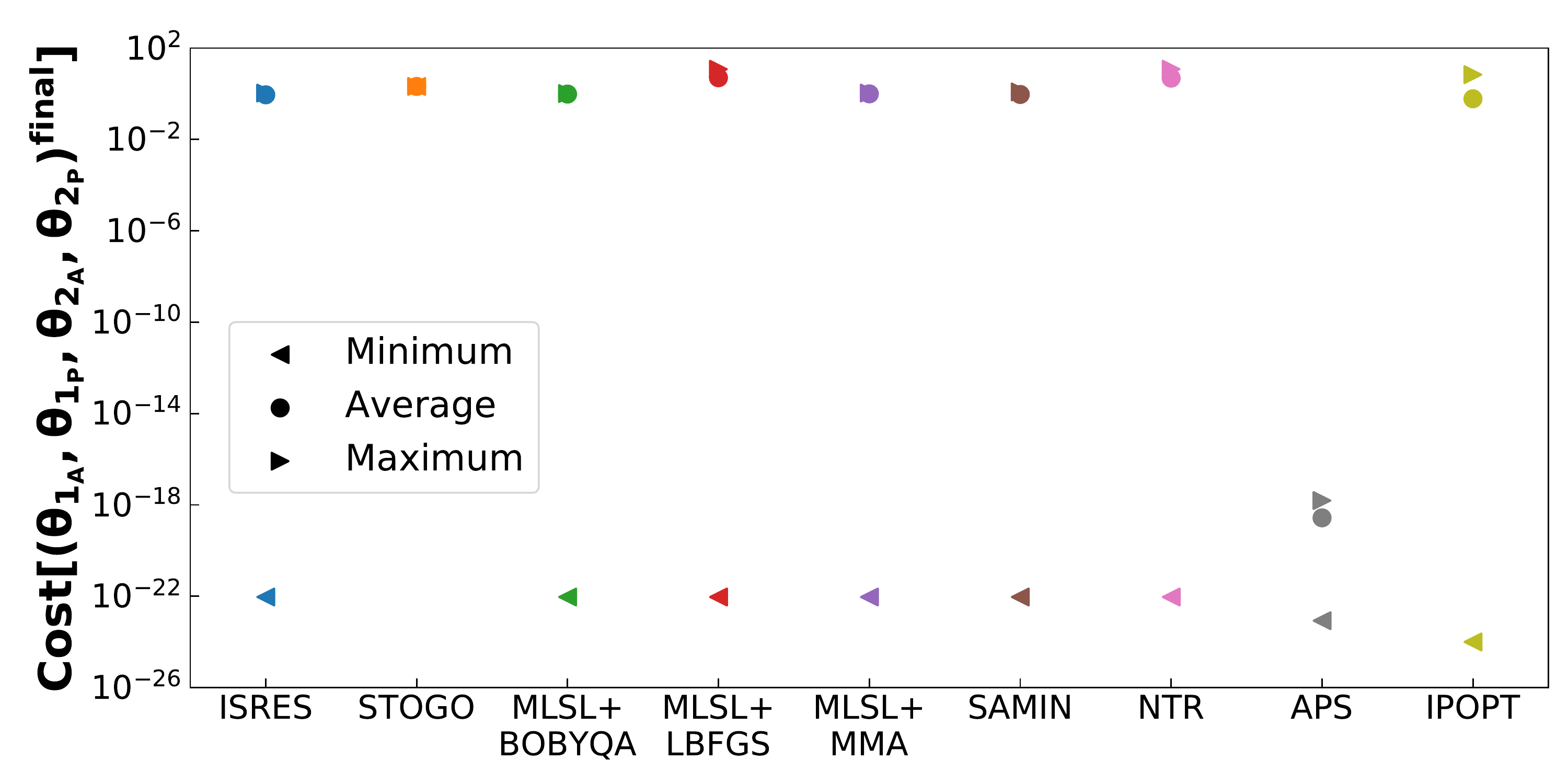}
 \caption{The minimum, maximum and average values of the cost function at the final values of $\bm{\theta}$  (i.e., at the end of the optimization procedure), for each optimization algorithm. Ideally, all these three values should be $0$.  For each algorithm, the spread  of cost function values shows how dependent the optimization algorithms are on the initial guess of the unknown parameters.  These results are for the set of experiments where the matter coupling parameter is taken to be a known constant, and the unknown parameters to be optimized are the polar and azimuthal angles of both the initial neutrino polarization vectors, i.e., $\bm{\theta}=\{\theta_{1_P},\theta_{1_A},\theta_{2_P},\theta_{2_A}\}$.}
 \label{fig:u0boththeta}
\end{figure}

As Fig.~\ref{fig:u0boththeta} shows, most methods converge to sub-optimal solutions. There is convergence for initial conditions close to optimal values, but this does not happen for initial guesses farther away. APS is the only method that performs well for all initial guesses. On the hand, STOGO does not provide a small cost value even for initial guesses close to the optimal one.

As an additional check, in table~\ref{table:angles21} we summarize the statistics for the final values obtained from each method for each of the two neutrinos. The sample standard deviation shown is an additional indication of the dependence on the initial guess. Ideally, this deviation should be zero as the methods should converge to the optimal value regardless of the initial guess for the unknown parameter, but this is not the case in practice.


\setlength{\tabcolsep}{5pt}
\begin{table}[ht]
\small
\centering
\subfloat[Inferred angles for neutrino 1]{
\begin{tabular}{|l|c |c | c |} \toprule
\hline
 \textit{Method}& $\left(\mu_{\theta^{(1)}_P},\mu_{\theta^{(1)}_A}\right)/\pi$ & $\left(\sigma_{\theta^{(1)}_P},\sigma_{\theta^{(1)}_A}\right)/\pi$ \\\midrule \hline
  ISRES &  $(0.89,0.98)$ & $(0.07,0.53)$ \\
  STOGO & $(0.73,1.0)$ & $ (\approx 0,\approx 0) $\\
  MLSL + BOBQYA & $(0.81,1.88)$ &  $(0.04,0.27)$ \\ 
 MLSL + LBFGS & $(0.5,1.0)$ & $(0.35,0.71)$ \\
 MLSL + MMA  & $(0.81,0.14)$ & $(0.21,0.04)$ \\
    SAMIN & $(0.87,0.99)$ & $(0.07,0.59)$ \\
    NTR & $(0.5,1.0)$ & $(0.5,1.0)$ \\
      APS & $(1.0,1.04)$ & $(\approx 0,0.54)$ \\
      IPOPT & $(0.45,0.89)$ & $(0.26, 0.57)$ \\ \hline
      \textit{Optimal Values}& $(1,-)$ & $(0,-)$\\
 \bottomrule \hline
\end{tabular}}

\subfloat[Inferred angles for neutrino 2]{
\begin{tabular}{|l|c |c | c |} \toprule
\hline
 \textit{Method}& $\left(\mu_{\theta^{(2)}_P},\mu_{\theta^{(2)}_A}\right)/\pi$ & $\left(\sigma_{\theta^{(2)}_P},\sigma_{\theta^{(2)}_A}\right)/\pi$ \\\midrule \hline
  ISRES &   $(0.17,0.98)$ & $(0.10,0.53)$ \\
  STOGO & $(0.43,0.99)$ & $(\approx 0,\approx 0)$ \\
  MLSL + BOBQYA & $(0.15,0.18)$ & $(0.03,0.26)$ \\ 
 MLSL + LBFGS & $(0.5,1.0)$ & $(0.35,0.71)$ \\
 MLSL + MMA  & $(0.21,0.33)$ & $(0.04,0.22)$ \\
    SAMIN & $(0.19,1.0)$ & $(0.07,0.59)$ \\
    NTR & $(0.35,0.71)$ & $(0.35,0.71)$ \\
      APS & $(\approx 0,1.03)$ & $(\approx 0,0.64)$ \\
      IPOPT & $(0.19, 0.89)$ & $(0.21, 0.59)$ \\ \hline
      \textit{Optimal Values}& $(0,-)$ & $(0,-)$\\
 \bottomrule \hline
\end{tabular}}
\caption{Sample average and standard deviation of the inferred angles in flavor space, for the initial polarization vectors of the {\og two} neutrinos at the end of the each optimization procedure. The optimal values shown at the end of the table; there is no preferred azimuthal angle. These results are for the set of experiments described in the caption of Fig.~\ref{fig:u0boththeta}.} 
\label{table:angles21}
\end{table}


As the tables show, most methods tend toward a large polar angle for the first neutrino and a small value for the second one. In other words, most methods expect the first neutrino to be mostly $x$ flavor and the second to be mostly electron flavor. This result indicates that, for the two neutrino beam system, the final polarization values can provide information on initial conditions, under the assumption that we know the matter density profile. In addition, we have identified APS as a method that works quite well in understanding the initial flavor composition of the system.


 
\section{Further Considerations}
\label{sec:considerations}
In this section we ponder on further complications one may encounter when performing inference on the detector data. As the APS method was the most successful one in the various experimental setups, the additional experiments and computations performed here will focus on this method for illustration.

\subsection{ Using inference to select theoretical models consistent with detector data}

In section~\ref{sec:exp2} we consider the matter potential coupling to be a generic function of radius. This is different from the setup used to generate the detector data which has a constant coupling $\tilde{V}_0$. The constant coupling is a special case for this section, and indeed, some of the optimization methods converge to it. When the detector data will be available to the community, one can not know for certain whether the theoretical model employed captures all the relevant aspects of neutrino flavor oscillations emitted from compact objects, and more probably, there will be more then one model to consider. 
To simulate this scenario, for the detector data generated based on table~\ref{table:Known}, we remove the neutrino self-interaction from Eq.~\ref{eq:model} and optimize for the matter coupling potential $V_0$. 
\begin{figure}[ht]
 \includegraphics[scale=0.5]{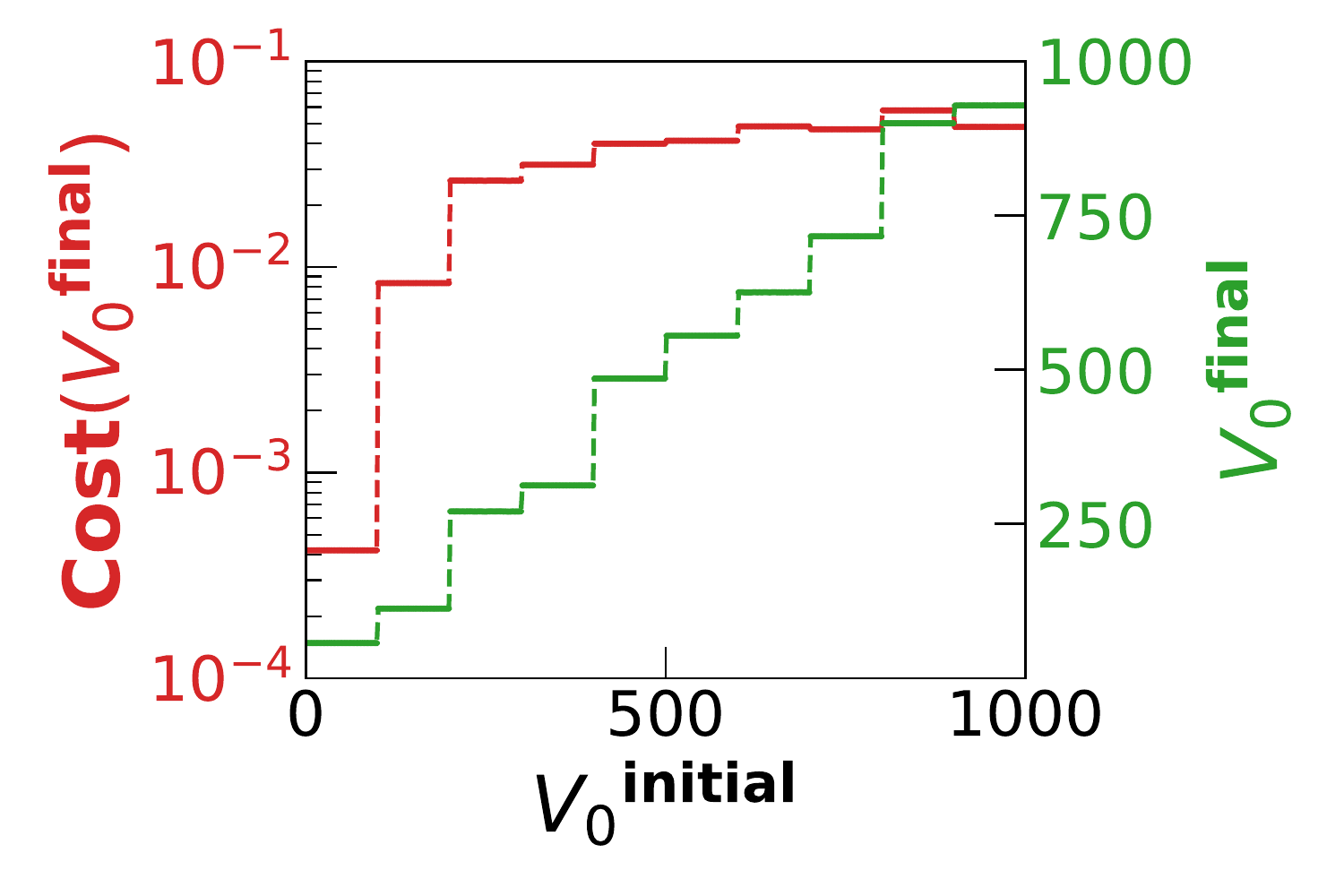}
 \caption{The final inferred value of $V_0$ (green) with the APS optimization method, and its  corresponding cost value (red), as function of the initial guess $V_0^{\text{initial}}$.  In this experiment, the optimization procedure was given a model with the neutrino-neutrino coupling $\mu_0=0$, even though the simulated \lq\lq detector\rq\rq\ data was generated using a model with $\mu_0 = 10$. The lack of a global convergence and the relatively high values of the cost function are indications that the model  with $\mu_0=0$ does not properly reproduce the detector measurements.}
 \label{fig:new_train}
\end{figure}
In Fig.~\ref{fig:new_train} we simultaneously plot the final inferred value of $V_0$ by the APS method, and its cost value, as function of the initial guess. To obtain this plots we used a grid of 500 uniform initial guesses in the range $[0,1000]$ and 1000 optimization iterations were employed per initial value. As the plot shows, there is no convergence to a global minimum, and the smallest cost value obtained is about five orders of magnitude higher than the one found by the same optimization method is Sec.~\ref{sec:exp1}. These are indications of how poorly the current model performs in reproducing the detector data. One could expect such differences to be used in discerning among the various proposed models when a detection is made. 

\subsection{ Inclusion of experimental uncertainties}

A second point that merits consideration is the inclusion of experimental uncertainties in inference. The results presented here can be interpreted as maximum likelihood estimates obtained from using the central value of measured neutrino polarization vectors. 
Assuming the measurement error to be tiny one can approximate, 
\begin{equation}
\frac{\bigtriangleup \text{Cost}}{\bigtriangleup V_0} \approx \frac{\partial \text{Cost}}{\partial V_0},    
\end{equation}
 and estimate the error in the inferred value of $V_0$. In the rare event the optimal solution is found, $\frac{\partial \text{Cost}}{\partial V_0}=0$, and the local gradient provides no useful information in estimating uncertainties.
  In practice, the optimization procedure converges close to the optimal solutions so local gradients are non-vanishing. However, detection errors may be small but not tiny and the outlined procedure might not provide robust estimates. Determining parameter uncertainties is, indeed, a rather complicated task and the presence of multiple minima with a cost value very close to the global optimum could lead to situations in which a small uncertainty in measurement could lead to large uncertainties in the parameter estimation. This is why one would need to probe large regions of parameter space through Monte Carlo sampling to properly represent uncertainties. When there is some information on the parameters of the model under consideration, one could perform a Bayesian analysis. These directions deserve to treated in depth and are outside the scope of this work. We plan to pursue them in the future.
 
\section{Conclusion}
\label{sec:conclusion}
 We have combined recent developments in deep learning with data assimilation, to examine what information is contained within a detected neutrino signal regarding complex astrophysical environments such as supernovae. 

By recasting the differential equations  of our model  as layers of a neural architecture, we accomplish two things. Firstly, we automatically satisfy physical constraints for the problem under investigation.  Secondly, we  demonstrate a model framework in which domain discretization is absent.  This framework has allowed us to focus on the prediction error (that is, the cost function),   by eliminating any free parameters associated with the grid resulting from domain discretization.

In addition, we have tested nine optimization algorithms that cover a wide range of techniques, and we have identified the ``Adaptive Particle Swarm Algorithm'' (APS) as best suited for our purposes in this paper. This algorithm, through its four evolutionary stages, is able to move out of local minima and thus has a high chance of finding a global minimum. 

The study conducted here has focused on a small system, primarily as a first testing ground for our framework.  We expect the computational complexity to inevitably increase with larger systems, and many degeneracies to be present in the parameter space. Thus, when the particle number is greatly increased, we might combine evolutionary algorithms such as APS for a wide parameter search, with a follow-up gradient- and Hessian-based method such as IPOPT as a secondary search within smaller optimal regions that are found by the first search.  We might also need to transition to distributed ordinary differential equation solvers, which can take advantage of computer clusters. 

 We intend to maintain a level of modeling complexity lower than that of three dimensional supernovae simulations (which take months for a single run to complete), and provide a computational service that is complementary to simulations and can function as a bridge to earth-based neutrino detection.  A more realistic setting, however, would require more than one affine parameter: temporal, in addition to spatial evolution. We may need to make away with rotational symmetry, leading to three affine spatial parameters. In this case, a combined approach, with the grid discretization of previous work applied to the spatial dimensions, and the new framework developed here applied to the temporal parameter, might prove useful.

\begin{acknowledgements}
We thank Anthony Mezzacappa for the useful discussions. GMF acknowledges NSF Grant No. PHY-1914242 at UCSD. 
ER,  AVP, and GMF  acknowledge the NSF N3AS Physics Frontier Center, NSF Grant No. PHY-2020275, and the Heising-Simons Foundation (2017-228). EA acknowledges an Institutional Support for Research and Creativity grant from New York Institute of Technology. 
Additionally, GMF acknowledges Department of Energy Scientific Discovery through Advanced Computing (SciDAC-4) grant register No. SN60152 (award number de-sc0018297). The work of AVP was also supported in part by the U.S. Department of Energy under contract number DE-AC02-76SF00515.
The authors acknowledge the Minnesota Supercomputing Institute (MSI) at the University of Minnesota for providing resources that contributed to the research results reported within this paper. See http://www.msi.umn.edu. In addition, HSL, a collection of Fortran codes for large scale scientific
 computation was used in conjunction with IPOPT. See http://www.hsl.rl.ac.uk.
\end{acknowledgements}
\newpage

\appendix

%

\begin{thebibliography}{98}%
\makeatletter
\providecommand \@ifxundefined [1]{%
 \@ifx{#1\undefined}
}%
\providecommand \@ifnum [1]{%
 \ifnum #1\expandafter \@firstoftwo
 \else \expandafter \@secondoftwo
 \fi
}%
\providecommand \@ifx [1]{%
 \ifx #1\expandafter \@firstoftwo
 \else \expandafter \@secondoftwo
 \fi
}%
\providecommand \natexlab [1]{#1}%
\providecommand \enquote  [1]{``#1''}%
\providecommand \bibnamefont  [1]{#1}%
\providecommand \bibfnamefont [1]{#1}%
\providecommand \citenamefont [1]{#1}%
\providecommand \href@noop [0]{\@secondoftwo}%
\providecommand \href [0]{\begingroup \@sanitize@url \@href}%
\providecommand \@href[1]{\@@startlink{#1}\@@href}%
\providecommand \@@href[1]{\endgroup#1\@@endlink}%
\providecommand \@sanitize@url [0]{\catcode `\\12\catcode `\$12\catcode
  `\&12\catcode `\#12\catcode `\^12\catcode `\_12\catcode `\%12\relax}%
\providecommand \@@startlink[1]{}%
\providecommand \@@endlink[0]{}%
\providecommand \url  [0]{\begingroup\@sanitize@url \@url }%
\providecommand \@url [1]{\endgroup\@href {#1}{\urlprefix }}%
\providecommand \urlprefix  [0]{URL }%
\providecommand \Eprint [0]{\href }%
\providecommand \doibase [0]{http://dx.doi.org/}%
\providecommand \selectlanguage [0]{\@gobble}%
\providecommand \bibinfo  [0]{\@secondoftwo}%
\providecommand \bibfield  [0]{\@secondoftwo}%
\providecommand \translation [1]{[#1]}%
\providecommand \BibitemOpen [0]{}%
\providecommand \bibitemStop [0]{}%
\providecommand \bibitemNoStop [0]{.\EOS\space}%
\providecommand \EOS [0]{\spacefactor3000\relax}%
\providecommand \BibitemShut  [1]{\csname bibitem#1\endcsname}%
\let\auto@bib@innerbib\@empty
\bibitem [{\citenamefont {Burrows}(1990)}]{Burrows1990}%
  \BibitemOpen
  \bibfield  {author} {\bibinfo {author} {\bibfnamefont {A}~\bibnamefont
  {Burrows}},\ }\bibfield  {title} {\enquote {\bibinfo {title} {Neutrinos from
  supernova explosions},}\ }\href {\doibase
  10.1146/annurev.ns.40.120190.001145} {\bibfield  {journal} {\bibinfo
  {journal} {Annual Review of Nuclear and Particle Science}\ }\textbf {\bibinfo
  {volume} {40}},\ \bibinfo {pages} {181--212} (\bibinfo {year}
  {1990})}\BibitemShut {NoStop}%
\bibitem [{\citenamefont {Mirizzi}\ \emph
  {et~al.}(2016{\natexlab{a}})\citenamefont {Mirizzi}, \citenamefont
  {Tamborra}, \citenamefont {Janka}, \citenamefont {Saviano}, \citenamefont
  {Scholberg}, \citenamefont {Bollig}, \citenamefont {Hudepohl},\ and\
  \citenamefont {Chakraborty}}]{Mirizzi:2015}%
  \BibitemOpen
  \bibfield  {author} {\bibinfo {author} {\bibfnamefont {Alessandro}\
  \bibnamefont {Mirizzi}}, \bibinfo {author} {\bibfnamefont {Irene}\
  \bibnamefont {Tamborra}}, \bibinfo {author} {\bibfnamefont {Hans-Thomas}\
  \bibnamefont {Janka}}, \bibinfo {author} {\bibfnamefont {Ninetta}\
  \bibnamefont {Saviano}}, \bibinfo {author} {\bibfnamefont {Kate}\
  \bibnamefont {Scholberg}}, \bibinfo {author} {\bibfnamefont {Robert}\
  \bibnamefont {Bollig}}, \bibinfo {author} {\bibfnamefont {Lorenz}\
  \bibnamefont {Hudepohl}}, \ and\ \bibinfo {author} {\bibfnamefont {Sovan}\
  \bibnamefont {Chakraborty}},\ }\bibfield  {title} {\enquote {\bibinfo {title}
  {{Supernova Neutrinos: Production, Oscillations and Detection}},}\ }\href
  {\doibase 10.1393/ncr/i2016-10120-8} {\bibfield  {journal} {\bibinfo
  {journal} {Riv. Nuovo Cim.}\ }\textbf {\bibinfo {volume} {39}},\ \bibinfo
  {pages} {1--112} (\bibinfo {year} {2016}{\natexlab{a}})}\BibitemShut
  {NoStop}%
\bibitem [{\citenamefont {Mart{\'i}nez-Pinedo}\ \emph
  {et~al.}(2017)\citenamefont {Mart{\'i}nez-Pinedo}, \citenamefont {Fischer},
  \citenamefont {Langanke}, \citenamefont {Lohs}, \citenamefont {Sieverding},\
  and\ \citenamefont {Wu}}]{Pinedo2017}%
  \BibitemOpen
  \bibfield  {author} {\bibinfo {author} {\bibfnamefont {Gabriel}\ \bibnamefont
  {Mart{\'i}nez-Pinedo}}, \bibinfo {author} {\bibfnamefont {Tobias}\
  \bibnamefont {Fischer}}, \bibinfo {author} {\bibfnamefont {Karlheinz}\
  \bibnamefont {Langanke}}, \bibinfo {author} {\bibfnamefont {Andreas}\
  \bibnamefont {Lohs}}, \bibinfo {author} {\bibfnamefont {Andre}\ \bibnamefont
  {Sieverding}}, \ and\ \bibinfo {author} {\bibfnamefont {Meng-Ru}\
  \bibnamefont {Wu}},\ }\enquote {\bibinfo {title} {Neutrinos and their impact
  on core-collapse supernova nucleosynthesis},}\ in\ \href {\doibase
  10.1007/978-3-319-21846-5_78} {\emph {\bibinfo {booktitle} {Handbook of
  Supernovae}}},\ \bibinfo {editor} {edited by\ \bibinfo {editor}
  {\bibfnamefont {Athem~W.}\ \bibnamefont {Alsabti}}\ and\ \bibinfo {editor}
  {\bibfnamefont {Paul}\ \bibnamefont {Murdin}}}\ (\bibinfo  {publisher}
  {Springer International Publishing},\ \bibinfo {address} {Cham},\ \bibinfo
  {year} {2017})\ pp.\ \bibinfo {pages} {1805--1841}\BibitemShut {NoStop}%
\bibitem [{\citenamefont {Müller}(2019)}]{Muller:2019}%
  \BibitemOpen
  \bibfield  {author} {\bibinfo {author} {\bibfnamefont {B.}~\bibnamefont
  {Müller}},\ }\bibfield  {title} {\enquote {\bibinfo {title} {Neutrino
  emission as diagnostics of core-collapse supernovae},}\ }\href {\doibase
  10.1146/annurev-nucl-101918-023434} {\bibfield  {journal} {\bibinfo
  {journal} {Annual Review of Nuclear and Particle Science}\ }\textbf {\bibinfo
  {volume} {69}},\ \bibinfo {pages} {253--278} (\bibinfo {year}
  {2019})}\BibitemShut {NoStop}%
\bibitem [{\citenamefont {Duan}\ \emph
  {et~al.}(2006{\natexlab{a}})\citenamefont {Duan}, \citenamefont {Fuller},
  \citenamefont {Carlson},\ and\ \citenamefont {Qian}}]{Duan:2006an}%
  \BibitemOpen
  \bibfield  {author} {\bibinfo {author} {\bibfnamefont {Huaiyu}\ \bibnamefont
  {Duan}}, \bibinfo {author} {\bibfnamefont {George~M.}\ \bibnamefont
  {Fuller}}, \bibinfo {author} {\bibfnamefont {J}~\bibnamefont {Carlson}}, \
  and\ \bibinfo {author} {\bibfnamefont {Yong-Zhong}\ \bibnamefont {Qian}},\
  }\bibfield  {title} {\enquote {\bibinfo {title} {{Simulation of Coherent
  Non-Linear Neutrino Flavor Transformation in the Supernova Environment. 1.
  Correlated Neutrino Trajectories}},}\ }\href {\doibase
  10.1103/PhysRevD.74.105014} {\bibfield  {journal} {\bibinfo  {journal} {Phys.
  Rev. D}\ }\textbf {\bibinfo {volume} {74}},\ \bibinfo {pages} {105014}
  (\bibinfo {year} {2006}{\natexlab{a}})},\ \Eprint
  {http://arxiv.org/abs/astro-ph/0606616} {arXiv:astro-ph/0606616} \BibitemShut
  {NoStop}%
\bibitem [{\citenamefont {Duan}\ \emph
  {et~al.}(2006{\natexlab{b}})\citenamefont {Duan}, \citenamefont {Fuller},\
  and\ \citenamefont {Qian}}]{Duan:2005cp}%
  \BibitemOpen
  \bibfield  {author} {\bibinfo {author} {\bibfnamefont {Huaiyu}\ \bibnamefont
  {Duan}}, \bibinfo {author} {\bibfnamefont {George~M.}\ \bibnamefont
  {Fuller}}, \ and\ \bibinfo {author} {\bibfnamefont {Yong-Zhong}\ \bibnamefont
  {Qian}},\ }\bibfield  {title} {\enquote {\bibinfo {title} {{Collective
  neutrino flavor transformation in supernovae}},}\ }\href {\doibase
  10.1103/PhysRevD.74.123004} {\bibfield  {journal} {\bibinfo  {journal} {Phys.
  Rev. D}\ }\textbf {\bibinfo {volume} {74}},\ \bibinfo {pages} {123004}
  (\bibinfo {year} {2006}{\natexlab{b}})},\ \Eprint
  {http://arxiv.org/abs/astro-ph/0511275} {arXiv:astro-ph/0511275} \BibitemShut
  {NoStop}%
\bibitem [{\citenamefont {Duan}\ \emph
  {et~al.}(2006{\natexlab{c}})\citenamefont {Duan}, \citenamefont {Fuller},
  \citenamefont {Carlson},\ and\ \citenamefont {Qian}}]{Duan:2006jv}%
  \BibitemOpen
  \bibfield  {author} {\bibinfo {author} {\bibfnamefont {Huaiyu}\ \bibnamefont
  {Duan}}, \bibinfo {author} {\bibfnamefont {George~M.}\ \bibnamefont
  {Fuller}}, \bibinfo {author} {\bibfnamefont {J.}~\bibnamefont {Carlson}}, \
  and\ \bibinfo {author} {\bibfnamefont {Yong-Zhong}\ \bibnamefont {Qian}},\
  }\bibfield  {title} {\enquote {\bibinfo {title} {{Coherent Development of
  Neutrino Flavor in the Supernova Environment}},}\ }\href {\doibase
  10.1103/PhysRevLett.97.241101} {\bibfield  {journal} {\bibinfo  {journal}
  {Phys. Rev. Lett.}\ }\textbf {\bibinfo {volume} {97}},\ \bibinfo {pages}
  {241101} (\bibinfo {year} {2006}{\natexlab{c}})},\ \Eprint
  {http://arxiv.org/abs/astro-ph/0608050} {arXiv:astro-ph/0608050} \BibitemShut
  {NoStop}%
\bibitem [{\citenamefont {Hannestad}\ \emph {et~al.}(2006)\citenamefont
  {Hannestad}, \citenamefont {Raffelt}, \citenamefont {Sigl},\ and\
  \citenamefont {Wong}}]{Hannestad:2006nj}%
  \BibitemOpen
  \bibfield  {author} {\bibinfo {author} {\bibfnamefont {Steen}\ \bibnamefont
  {Hannestad}}, \bibinfo {author} {\bibfnamefont {Georg~G.}\ \bibnamefont
  {Raffelt}}, \bibinfo {author} {\bibfnamefont {Gunter}\ \bibnamefont {Sigl}},
  \ and\ \bibinfo {author} {\bibfnamefont {Yvonne~Y.Y.}\ \bibnamefont {Wong}},\
  }\bibfield  {title} {\enquote {\bibinfo {title} {{Self-induced conversion in
  dense neutrino gases: Pendulum in flavour space}},}\ }\href {\doibase
  10.1103/PhysRevD.74.105010} {\bibfield  {journal} {\bibinfo  {journal} {Phys.
  Rev. D}\ }\textbf {\bibinfo {volume} {74}},\ \bibinfo {pages} {105010}
  (\bibinfo {year} {2006})},\ \bibinfo {note} {[Erratum: Phys.Rev.D 76, 029901
  (2007)]},\ \Eprint {http://arxiv.org/abs/astro-ph/0608695}
  {arXiv:astro-ph/0608695} \BibitemShut {NoStop}%
\bibitem [{\citenamefont {Raffelt}\ and\ \citenamefont
  {Smirnov}(2007{\natexlab{a}})}]{Raffelt:2007cb}%
  \BibitemOpen
  \bibfield  {author} {\bibinfo {author} {\bibfnamefont {Georg~G.}\
  \bibnamefont {Raffelt}}\ and\ \bibinfo {author} {\bibfnamefont {Alexei~Yu.}\
  \bibnamefont {Smirnov}},\ }\bibfield  {title} {\enquote {\bibinfo {title}
  {{Self-induced spectral splits in supernova neutrino fluxes}},}\ }\href
  {\doibase 10.1103/PhysRevD.76.081301} {\bibfield  {journal} {\bibinfo
  {journal} {Phys. Rev. D}\ }\textbf {\bibinfo {volume} {76}},\ \bibinfo
  {pages} {081301} (\bibinfo {year} {2007}{\natexlab{a}})},\ \bibinfo {note}
  {[Erratum: Phys.Rev.D 77, 029903 (2008)]},\ \Eprint
  {http://arxiv.org/abs/0705.1830} {arXiv:0705.1830 [hep-ph]} \BibitemShut
  {NoStop}%
\bibitem [{\citenamefont {Raffelt}\ and\ \citenamefont
  {Smirnov}(2007{\natexlab{b}})}]{Raffelt:2007xt}%
  \BibitemOpen
  \bibfield  {author} {\bibinfo {author} {\bibfnamefont {Georg~G.}\
  \bibnamefont {Raffelt}}\ and\ \bibinfo {author} {\bibfnamefont {Alexei~Yu.}\
  \bibnamefont {Smirnov}},\ }\bibfield  {title} {\enquote {\bibinfo {title}
  {{Adiabaticity and spectral splits in collective neutrino
  transformations}},}\ }\href {\doibase 10.1103/PhysRevD.76.125008} {\bibfield
  {journal} {\bibinfo  {journal} {Phys. Rev. D}\ }\textbf {\bibinfo {volume}
  {76}},\ \bibinfo {pages} {125008} (\bibinfo {year} {2007}{\natexlab{b}})},\
  \Eprint {http://arxiv.org/abs/0709.4641} {arXiv:0709.4641 [hep-ph]}
  \BibitemShut {NoStop}%
\bibitem [{\citenamefont {Dasgupta}\ and\ \citenamefont
  {Dighe}(2008)}]{Dasgupta:2007ws}%
  \BibitemOpen
  \bibfield  {author} {\bibinfo {author} {\bibfnamefont {Basudeb}\ \bibnamefont
  {Dasgupta}}\ and\ \bibinfo {author} {\bibfnamefont {Amol}\ \bibnamefont
  {Dighe}},\ }\bibfield  {title} {\enquote {\bibinfo {title} {{Collective
  three-flavor oscillations of supernova neutrinos}},}\ }\href {\doibase
  10.1103/PhysRevD.77.113002} {\bibfield  {journal} {\bibinfo  {journal} {Phys.
  Rev. D}\ }\textbf {\bibinfo {volume} {77}},\ \bibinfo {pages} {113002}
  (\bibinfo {year} {2008})},\ \Eprint {http://arxiv.org/abs/0712.3798}
  {arXiv:0712.3798 [hep-ph]} \BibitemShut {NoStop}%
\bibitem [{\citenamefont {Johns}\ and\ \citenamefont
  {Fuller}(2018)}]{Johns:2017oky}%
  \BibitemOpen
  \bibfield  {author} {\bibinfo {author} {\bibfnamefont {Lucas}\ \bibnamefont
  {Johns}}\ and\ \bibinfo {author} {\bibfnamefont {George~M.}\ \bibnamefont
  {Fuller}},\ }\bibfield  {title} {\enquote {\bibinfo {title} {{Strange
  mechanics of the neutrino flavor pendulum}},}\ }\href {\doibase
  10.1103/PhysRevD.97.023020} {\bibfield  {journal} {\bibinfo  {journal} {Phys.
  Rev. D}\ }\textbf {\bibinfo {volume} {97}},\ \bibinfo {pages} {023020}
  (\bibinfo {year} {2018})},\ \Eprint {http://arxiv.org/abs/1709.00518}
  {arXiv:1709.00518 [hep-ph]} \BibitemShut {NoStop}%
\bibitem [{\citenamefont {Cherry}\ \emph {et~al.}(2013)\citenamefont {Cherry},
  \citenamefont {Carlson}, \citenamefont {Friedland}, \citenamefont {Fuller},\
  and\ \citenamefont {Vlasenko}}]{Cherry:2013mv}%
  \BibitemOpen
  \bibfield  {author} {\bibinfo {author} {\bibfnamefont {John~F.}\ \bibnamefont
  {Cherry}}, \bibinfo {author} {\bibfnamefont {J.}~\bibnamefont {Carlson}},
  \bibinfo {author} {\bibfnamefont {Alexander}\ \bibnamefont {Friedland}},
  \bibinfo {author} {\bibfnamefont {George~M.}\ \bibnamefont {Fuller}}, \ and\
  \bibinfo {author} {\bibfnamefont {Alexey}\ \bibnamefont {Vlasenko}},\
  }\bibfield  {title} {\enquote {\bibinfo {title} {{Halo Modification of a
  Supernova Neutronization Neutrino Burst}},}\ }\href {\doibase
  10.1103/PhysRevD.87.085037} {\bibfield  {journal} {\bibinfo  {journal} {Phys.
  Rev. D}\ }\textbf {\bibinfo {volume} {87}},\ \bibinfo {pages} {085037}
  (\bibinfo {year} {2013})},\ \Eprint {http://arxiv.org/abs/1302.1159}
  {arXiv:1302.1159 [astro-ph.HE]} \BibitemShut {NoStop}%
\bibitem [{\citenamefont {Zaizen}\ \emph {et~al.}(2020)\citenamefont {Zaizen},
  \citenamefont {Cherry}, \citenamefont {Takiwaki}, \citenamefont {Horiuchi},
  \citenamefont {Kotake}, \citenamefont {Umeda},\ and\ \citenamefont
  {Yoshida}}]{Zaizen:2019ufj}%
  \BibitemOpen
  \bibfield  {author} {\bibinfo {author} {\bibfnamefont {Masamichi}\
  \bibnamefont {Zaizen}}, \bibinfo {author} {\bibfnamefont {John~F.}\
  \bibnamefont {Cherry}}, \bibinfo {author} {\bibfnamefont {Tomoya}\
  \bibnamefont {Takiwaki}}, \bibinfo {author} {\bibfnamefont {Shunsaku}\
  \bibnamefont {Horiuchi}}, \bibinfo {author} {\bibfnamefont {Kei}\
  \bibnamefont {Kotake}}, \bibinfo {author} {\bibfnamefont {Hideyuki}\
  \bibnamefont {Umeda}}, \ and\ \bibinfo {author} {\bibfnamefont {Takashi}\
  \bibnamefont {Yoshida}},\ }\bibfield  {title} {\enquote {\bibinfo {title}
  {{Neutrino halo effect on collective neutrino oscillation in iron
  core-collapse supernova model of a 9.6 $M_{\odot}$ star}},}\ }\href {\doibase
  10.1088/1475-7516/2020/06/011} {\bibfield  {journal} {\bibinfo  {journal}
  {JCAP}\ }\textbf {\bibinfo {volume} {06}},\ \bibinfo {pages} {011} (\bibinfo
  {year} {2020})},\ \Eprint {http://arxiv.org/abs/1908.10594} {arXiv:1908.10594
  [astro-ph.HE]} \BibitemShut {NoStop}%
\bibitem [{\citenamefont {Duan}\ and\ \citenamefont
  {Kneller}(2009)}]{Duan:2009cd}%
  \BibitemOpen
  \bibfield  {author} {\bibinfo {author} {\bibfnamefont {Huaiyu}\ \bibnamefont
  {Duan}}\ and\ \bibinfo {author} {\bibfnamefont {James~P}\ \bibnamefont
  {Kneller}},\ }\bibfield  {title} {\enquote {\bibinfo {title} {{Neutrino
  flavour transformation in supernovae}},}\ }\href {\doibase
  10.1088/0954-3899/36/11/113201} {\bibfield  {journal} {\bibinfo  {journal}
  {J. Phys. G}\ }\textbf {\bibinfo {volume} {36}},\ \bibinfo {pages} {113201}
  (\bibinfo {year} {2009})},\ \Eprint {http://arxiv.org/abs/0904.0974}
  {arXiv:0904.0974 [astro-ph.HE]} \BibitemShut {NoStop}%
\bibitem [{\citenamefont {Duan}\ \emph {et~al.}(2010)\citenamefont {Duan},
  \citenamefont {Fuller},\ and\ \citenamefont {Qian}}]{Duan:2010bg}%
  \BibitemOpen
  \bibfield  {author} {\bibinfo {author} {\bibfnamefont {Huaiyu}\ \bibnamefont
  {Duan}}, \bibinfo {author} {\bibfnamefont {George~M.}\ \bibnamefont
  {Fuller}}, \ and\ \bibinfo {author} {\bibfnamefont {Yong-Zhong}\ \bibnamefont
  {Qian}},\ }\bibfield  {title} {\enquote {\bibinfo {title} {{Collective
  Neutrino Oscillations}},}\ }\href {\doibase
  10.1146/annurev.nucl.012809.104524} {\bibfield  {journal} {\bibinfo
  {journal} {Ann. Rev. Nucl. Part. Sci.}\ }\textbf {\bibinfo {volume} {60}},\
  \bibinfo {pages} {569--594} (\bibinfo {year} {2010})},\ \Eprint
  {http://arxiv.org/abs/1001.2799} {arXiv:1001.2799 [hep-ph]} \BibitemShut
  {NoStop}%
\bibitem [{\citenamefont {Mirizzi}\ \emph
  {et~al.}(2016{\natexlab{b}})\citenamefont {Mirizzi}, \citenamefont
  {Tamborra}, \citenamefont {Janka}, \citenamefont {Saviano}, \citenamefont
  {Scholberg}, \citenamefont {Bollig}, \citenamefont {Hudepohl},\ and\
  \citenamefont {Chakraborty}}]{Mirizzi:2015eza}%
  \BibitemOpen
  \bibfield  {author} {\bibinfo {author} {\bibfnamefont {Alessandro}\
  \bibnamefont {Mirizzi}}, \bibinfo {author} {\bibfnamefont {Irene}\
  \bibnamefont {Tamborra}}, \bibinfo {author} {\bibfnamefont {Hans-Thomas}\
  \bibnamefont {Janka}}, \bibinfo {author} {\bibfnamefont {Ninetta}\
  \bibnamefont {Saviano}}, \bibinfo {author} {\bibfnamefont {Kate}\
  \bibnamefont {Scholberg}}, \bibinfo {author} {\bibfnamefont {Robert}\
  \bibnamefont {Bollig}}, \bibinfo {author} {\bibfnamefont {Lorenz}\
  \bibnamefont {Hudepohl}}, \ and\ \bibinfo {author} {\bibfnamefont {Sovan}\
  \bibnamefont {Chakraborty}},\ }\bibfield  {title} {\enquote {\bibinfo {title}
  {{Supernova Neutrinos: Production, Oscillations and Detection}},}\ }\href
  {\doibase 10.1393/ncr/i2016-10120-8} {\bibfield  {journal} {\bibinfo
  {journal} {Riv. Nuovo Cim.}\ }\textbf {\bibinfo {volume} {39}},\ \bibinfo
  {pages} {1--112} (\bibinfo {year} {2016}{\natexlab{b}})},\ \Eprint
  {http://arxiv.org/abs/1508.00785} {arXiv:1508.00785 [astro-ph.HE]}
  \BibitemShut {NoStop}%
\bibitem [{\citenamefont {Chakraborty}\ \emph {et~al.}(2016)\citenamefont
  {Chakraborty}, \citenamefont {Hansen}, \citenamefont {Izaguirre},\ and\
  \citenamefont {Raffelt}}]{Chakraborty:2016yeg}%
  \BibitemOpen
  \bibfield  {author} {\bibinfo {author} {\bibfnamefont {Sovan}\ \bibnamefont
  {Chakraborty}}, \bibinfo {author} {\bibfnamefont {Rasmus}\ \bibnamefont
  {Hansen}}, \bibinfo {author} {\bibfnamefont {Ignacio}\ \bibnamefont
  {Izaguirre}}, \ and\ \bibinfo {author} {\bibfnamefont {Georg}\ \bibnamefont
  {Raffelt}},\ }\bibfield  {title} {\enquote {\bibinfo {title} {{Collective
  neutrino flavor conversion: Recent developments}},}\ }\href {\doibase
  10.1016/j.nuclphysb.2016.02.012} {\bibfield  {journal} {\bibinfo  {journal}
  {Nucl. Phys. B}\ }\textbf {\bibinfo {volume} {908}},\ \bibinfo {pages}
  {366--381} (\bibinfo {year} {2016})},\ \Eprint
  {http://arxiv.org/abs/1602.02766} {arXiv:1602.02766 [hep-ph]} \BibitemShut
  {NoStop}%
\bibitem [{\citenamefont {{Banerjee}}\ \emph {et~al.}(2011)\citenamefont
  {{Banerjee}}, \citenamefont {{Dighe}},\ and\ \citenamefont
  {{Raffelt}}}]{2011PhRvD..84e3013B}%
  \BibitemOpen
  \bibfield  {author} {\bibinfo {author} {\bibfnamefont {Arka}\ \bibnamefont
  {{Banerjee}}}, \bibinfo {author} {\bibfnamefont {Amol}\ \bibnamefont
  {{Dighe}}}, \ and\ \bibinfo {author} {\bibfnamefont {Georg}\ \bibnamefont
  {{Raffelt}}},\ }\bibfield  {title} {\enquote {\bibinfo {title} {{Linearized
  flavor-stability analysis of dense neutrino streams}},}\ }\href {\doibase
  10.1103/PhysRevD.84.053013} {\bibfield  {journal} {\bibinfo  {journal}
  {\prd}\ }\textbf {\bibinfo {volume} {84}},\ \bibinfo {eid} {053013} (\bibinfo
  {year} {2011})},\ \Eprint {http://arxiv.org/abs/1107.2308} {arXiv:1107.2308
  [hep-ph]} \BibitemShut {NoStop}%
\bibitem [{\citenamefont {Raffelt}\ \emph {et~al.}(2013)\citenamefont
  {Raffelt}, \citenamefont {Sarikas},\ and\ \citenamefont
  {de~Sousa~Seixas}}]{Raffelt:2013rqa}%
  \BibitemOpen
  \bibfield  {author} {\bibinfo {author} {\bibfnamefont {Georg}\ \bibnamefont
  {Raffelt}}, \bibinfo {author} {\bibfnamefont {Srdjan}\ \bibnamefont
  {Sarikas}}, \ and\ \bibinfo {author} {\bibfnamefont {David}\ \bibnamefont
  {de~Sousa~Seixas}},\ }\bibfield  {title} {\enquote {\bibinfo {title} {{Axial
  Symmetry Breaking in Self-Induced Flavor Conversion of Supernova Neutrino
  Fluxes}},}\ }\href {\doibase 10.1103/PhysRevLett.111.091101} {\bibfield
  {journal} {\bibinfo  {journal} {Phys. Rev. Lett.}\ }\textbf {\bibinfo
  {volume} {111}},\ \bibinfo {pages} {091101} (\bibinfo {year} {2013})},\
  \bibinfo {note} {[Erratum: Phys.Rev.Lett. 113, 239903 (2014)]},\ \Eprint
  {http://arxiv.org/abs/1305.7140} {arXiv:1305.7140 [hep-ph]} \BibitemShut
  {NoStop}%
\bibitem [{\citenamefont {Chakraborty}\ and\ \citenamefont
  {Mirizzi}(2014)}]{Mirizzi:2013wda}%
  \BibitemOpen
  \bibfield  {author} {\bibinfo {author} {\bibfnamefont {Sovan}\ \bibnamefont
  {Chakraborty}}\ and\ \bibinfo {author} {\bibfnamefont {Alessandro}\
  \bibnamefont {Mirizzi}},\ }\bibfield  {title} {\enquote {\bibinfo {title}
  {{Multi-azimuthal-angle instability for different supernova neutrino
  fluxes}},}\ }\href {\doibase 10.1103/PhysRevD.90.033004} {\bibfield
  {journal} {\bibinfo  {journal} {Phys. Rev. D}\ }\textbf {\bibinfo {volume}
  {90}},\ \bibinfo {pages} {033004} (\bibinfo {year} {2014})},\ \Eprint
  {http://arxiv.org/abs/1308.5255} {arXiv:1308.5255 [hep-ph]} \BibitemShut
  {NoStop}%
\bibitem [{\citenamefont {Duan}\ and\ \citenamefont
  {Shalgar}(2015)}]{Duan:2014gfa}%
  \BibitemOpen
  \bibfield  {author} {\bibinfo {author} {\bibfnamefont {Huaiyu}\ \bibnamefont
  {Duan}}\ and\ \bibinfo {author} {\bibfnamefont {Shashank}\ \bibnamefont
  {Shalgar}},\ }\bibfield  {title} {\enquote {\bibinfo {title} {{Flavor
  instabilities in the neutrino line model}},}\ }\href {\doibase
  10.1016/j.physletb.2015.05.057} {\bibfield  {journal} {\bibinfo  {journal}
  {Phys. Lett. B}\ }\textbf {\bibinfo {volume} {747}},\ \bibinfo {pages}
  {139--143} (\bibinfo {year} {2015})},\ \Eprint
  {http://arxiv.org/abs/1412.7097} {arXiv:1412.7097 [hep-ph]} \BibitemShut
  {NoStop}%
\bibitem [{\citenamefont {{Abbar}}\ and\ \citenamefont
  {{Duan}}(2015)}]{2015PhLB..751...43A}%
  \BibitemOpen
  \bibfield  {author} {\bibinfo {author} {\bibfnamefont {Sajad}\ \bibnamefont
  {{Abbar}}}\ and\ \bibinfo {author} {\bibfnamefont {Huaiyu}\ \bibnamefont
  {{Duan}}},\ }\bibfield  {title} {\enquote {\bibinfo {title} {{Neutrino flavor
  instabilities in a time-dependent supernova model}},}\ }\href {\doibase
  10.1016/j.physletb.2015.10.019} {\bibfield  {journal} {\bibinfo  {journal}
  {Physics Letters B}\ }\textbf {\bibinfo {volume} {751}},\ \bibinfo {pages}
  {43--47} (\bibinfo {year} {2015})},\ \Eprint
  {http://arxiv.org/abs/1509.01538} {arXiv:1509.01538 [astro-ph.HE]}
  \BibitemShut {NoStop}%
\bibitem [{\citenamefont {{Chakraborty}}\ \emph {et~al.}(2016)\citenamefont
  {{Chakraborty}}, \citenamefont {{Hansen}}, \citenamefont {{Izaguirre}},\ and\
  \citenamefont {{Raffelt}}}]{2016JCAP...01..028C}%
  \BibitemOpen
  \bibfield  {author} {\bibinfo {author} {\bibfnamefont {S.}~\bibnamefont
  {{Chakraborty}}}, \bibinfo {author} {\bibfnamefont {R.~S.}\ \bibnamefont
  {{Hansen}}}, \bibinfo {author} {\bibfnamefont {I.}~\bibnamefont
  {{Izaguirre}}}, \ and\ \bibinfo {author} {\bibfnamefont {G.~G.}\ \bibnamefont
  {{Raffelt}}},\ }\bibfield  {title} {\enquote {\bibinfo {title} {{Self-induced
  flavor conversion of supernova neutrinos on small scales}},}\ }\href
  {\doibase 10.1088/1475-7516/2016/01/028} {\bibfield  {journal} {\bibinfo
  {journal} {JCAP}\ }\textbf {\bibinfo {volume} {2016}},\ \bibinfo {eid} {028}
  (\bibinfo {year} {2016})},\ \Eprint {http://arxiv.org/abs/1507.07569}
  {arXiv:1507.07569 [hep-ph]} \BibitemShut {NoStop}%
\bibitem [{\citenamefont {{Sawyer}}(2009)}]{2009PhRvD..79j5003S}%
  \BibitemOpen
  \bibfield  {author} {\bibinfo {author} {\bibfnamefont {R.~F.}\ \bibnamefont
  {{Sawyer}}},\ }\bibfield  {title} {\enquote {\bibinfo {title} {{Multiangle
  instability in dense neutrino systems}},}\ }\href {\doibase
  10.1103/PhysRevD.79.105003} {\bibfield  {journal} {\bibinfo  {journal}
  {\prd}\ }\textbf {\bibinfo {volume} {79}},\ \bibinfo {eid} {105003} (\bibinfo
  {year} {2009})},\ \Eprint {http://arxiv.org/abs/0803.4319} {arXiv:0803.4319
  [astro-ph]} \BibitemShut {NoStop}%
\bibitem [{\citenamefont {{Sawyer}}(2016)}]{2016PhRvL.116h1101S}%
  \BibitemOpen
  \bibfield  {author} {\bibinfo {author} {\bibfnamefont {R.~F.}\ \bibnamefont
  {{Sawyer}}},\ }\bibfield  {title} {\enquote {\bibinfo {title} {{Neutrino
  Cloud Instabilities Just above the Neutrino Sphere of a Supernova}},}\ }\href
  {\doibase 10.1103/PhysRevLett.116.081101} {\bibfield  {journal} {\bibinfo
  {journal} {\prl}\ }\textbf {\bibinfo {volume} {116}},\ \bibinfo {eid}
  {081101} (\bibinfo {year} {2016})},\ \Eprint
  {http://arxiv.org/abs/1509.03323} {arXiv:1509.03323 [astro-ph.HE]}
  \BibitemShut {NoStop}%
\bibitem [{\citenamefont {{Dasgupta}}\ and\ \citenamefont
  {{Mirizzi}}(2015)}]{2015PhRvD..92l5030D}%
  \BibitemOpen
  \bibfield  {author} {\bibinfo {author} {\bibfnamefont {Basudeb}\ \bibnamefont
  {{Dasgupta}}}\ and\ \bibinfo {author} {\bibfnamefont {Alessandro}\
  \bibnamefont {{Mirizzi}}},\ }\bibfield  {title} {\enquote {\bibinfo {title}
  {{Temporal instability enables neutrino flavor conversions deep inside
  supernovae}},}\ }\href {\doibase 10.1103/PhysRevD.92.125030} {\bibfield
  {journal} {\bibinfo  {journal} {\prd}\ }\textbf {\bibinfo {volume} {92}},\
  \bibinfo {eid} {125030} (\bibinfo {year} {2015})},\ \Eprint
  {http://arxiv.org/abs/1509.03171} {arXiv:1509.03171 [hep-ph]} \BibitemShut
  {NoStop}%
\bibitem [{\citenamefont {Izaguirre}\ \emph {et~al.}(2017)\citenamefont
  {Izaguirre}, \citenamefont {Raffelt},\ and\ \citenamefont
  {Tamborra}}]{Izaguirre:2016gsx}%
  \BibitemOpen
  \bibfield  {author} {\bibinfo {author} {\bibfnamefont {Ignacio}\ \bibnamefont
  {Izaguirre}}, \bibinfo {author} {\bibfnamefont {Georg}\ \bibnamefont
  {Raffelt}}, \ and\ \bibinfo {author} {\bibfnamefont {Irene}\ \bibnamefont
  {Tamborra}},\ }\bibfield  {title} {\enquote {\bibinfo {title} {{Fast Pairwise
  Conversion of Supernova Neutrinos: A Dispersion-Relation Approach}},}\ }\href
  {\doibase 10.1103/PhysRevLett.118.021101} {\bibfield  {journal} {\bibinfo
  {journal} {Phys. Rev. Lett.}\ }\textbf {\bibinfo {volume} {118}},\ \bibinfo
  {pages} {021101} (\bibinfo {year} {2017})},\ \Eprint
  {http://arxiv.org/abs/1610.01612} {arXiv:1610.01612 [hep-ph]} \BibitemShut
  {NoStop}%
\bibitem [{\citenamefont {Dasgupta}\ \emph {et~al.}(2017)\citenamefont
  {Dasgupta}, \citenamefont {Mirizzi},\ and\ \citenamefont
  {Sen}}]{Dasgupta:2016dbv}%
  \BibitemOpen
  \bibfield  {author} {\bibinfo {author} {\bibfnamefont {Basudeb}\ \bibnamefont
  {Dasgupta}}, \bibinfo {author} {\bibfnamefont {Alessandro}\ \bibnamefont
  {Mirizzi}}, \ and\ \bibinfo {author} {\bibfnamefont {Manibrata}\ \bibnamefont
  {Sen}},\ }\bibfield  {title} {\enquote {\bibinfo {title} {{Fast neutrino
  flavor conversions near the supernova core with realistic flavor-dependent
  angular distributions}},}\ }\href {\doibase 10.1088/1475-7516/2017/02/019}
  {\bibfield  {journal} {\bibinfo  {journal} {JCAP}\ }\textbf {\bibinfo
  {volume} {02}},\ \bibinfo {pages} {019} (\bibinfo {year} {2017})},\ \Eprint
  {http://arxiv.org/abs/1609.00528} {arXiv:1609.00528 [hep-ph]} \BibitemShut
  {NoStop}%
\bibitem [{\citenamefont {Capozzi}\ \emph {et~al.}(2017)\citenamefont
  {Capozzi}, \citenamefont {Dasgupta}, \citenamefont {Lisi}, \citenamefont
  {Marrone},\ and\ \citenamefont {Mirizzi}}]{Capozzi:2017gqd}%
  \BibitemOpen
  \bibfield  {author} {\bibinfo {author} {\bibfnamefont {Francesco}\
  \bibnamefont {Capozzi}}, \bibinfo {author} {\bibfnamefont {Basudeb}\
  \bibnamefont {Dasgupta}}, \bibinfo {author} {\bibfnamefont {Eligio}\
  \bibnamefont {Lisi}}, \bibinfo {author} {\bibfnamefont {Antonio}\
  \bibnamefont {Marrone}}, \ and\ \bibinfo {author} {\bibfnamefont
  {Alessandro}\ \bibnamefont {Mirizzi}},\ }\bibfield  {title} {\enquote
  {\bibinfo {title} {{Fast flavor conversions of supernova neutrinos:
  Classifying instabilities via dispersion relations}},}\ }\href {\doibase
  10.1103/PhysRevD.96.043016} {\bibfield  {journal} {\bibinfo  {journal} {Phys.
  Rev. D}\ }\textbf {\bibinfo {volume} {96}},\ \bibinfo {pages} {043016}
  (\bibinfo {year} {2017})},\ \Eprint {http://arxiv.org/abs/1706.03360}
  {arXiv:1706.03360 [hep-ph]} \BibitemShut {NoStop}%
\bibitem [{\citenamefont {{Dasgupta}}\ and\ \citenamefont
  {{Sen}}(2018)}]{2018PhRvD..97b3017D}%
  \BibitemOpen
  \bibfield  {author} {\bibinfo {author} {\bibfnamefont {Basudeb}\ \bibnamefont
  {{Dasgupta}}}\ and\ \bibinfo {author} {\bibfnamefont {Manibrata}\
  \bibnamefont {{Sen}}},\ }\bibfield  {title} {\enquote {\bibinfo {title}
  {{Fast neutrino flavor conversion as oscillations in a quartic potential}},}\
  }\href {\doibase 10.1103/PhysRevD.97.023017} {\bibfield  {journal} {\bibinfo
  {journal} {\prd}\ }\textbf {\bibinfo {volume} {97}},\ \bibinfo {eid} {023017}
  (\bibinfo {year} {2018})},\ \Eprint {http://arxiv.org/abs/1709.08671}
  {arXiv:1709.08671 [hep-ph]} \BibitemShut {NoStop}%
\bibitem [{\citenamefont {{Abbar}}\ and\ \citenamefont
  {{Volpe}}(2019)}]{2019PhLB..790..545A}%
  \BibitemOpen
  \bibfield  {author} {\bibinfo {author} {\bibfnamefont {Sajad}\ \bibnamefont
  {{Abbar}}}\ and\ \bibinfo {author} {\bibfnamefont {Maria~Cristina}\
  \bibnamefont {{Volpe}}},\ }\bibfield  {title} {\enquote {\bibinfo {title}
  {{On fast neutrino flavor conversion modes in the nonlinear regime}},}\
  }\href {\doibase 10.1016/j.physletb.2019.02.002} {\bibfield  {journal}
  {\bibinfo  {journal} {Physics Letters B}\ }\textbf {\bibinfo {volume}
  {790}},\ \bibinfo {pages} {545--550} (\bibinfo {year} {2019})},\ \Eprint
  {http://arxiv.org/abs/1811.04215} {arXiv:1811.04215 [astro-ph.HE]}
  \BibitemShut {NoStop}%
\bibitem [{\citenamefont {{Shalgar}}\ and\ \citenamefont
  {{Tamborra}}(2019)}]{2019ApJ...883...80S}%
  \BibitemOpen
  \bibfield  {author} {\bibinfo {author} {\bibfnamefont {Shashank}\
  \bibnamefont {{Shalgar}}}\ and\ \bibinfo {author} {\bibfnamefont {Irene}\
  \bibnamefont {{Tamborra}}},\ }\bibfield  {title} {\enquote {\bibinfo {title}
  {{On the Occurrence of Crossings between the Angular Distributions of
  Electron Neutrinos and Antineutrinos in the Supernova Core}},}\ }\href
  {\doibase 10.3847/1538-4357/ab38ba} {\bibfield  {journal} {\bibinfo
  {journal} {\apj}\ }\textbf {\bibinfo {volume} {883}},\ \bibinfo {eid} {80}
  (\bibinfo {year} {2019})},\ \Eprint {http://arxiv.org/abs/1904.07236}
  {arXiv:1904.07236 [astro-ph.HE]} \BibitemShut {NoStop}%
\bibitem [{\citenamefont {{Capozzi}}\ \emph {et~al.}(2019)\citenamefont
  {{Capozzi}}, \citenamefont {{Dasgupta}}, \citenamefont {{Mirizzi}},
  \citenamefont {{Sen}},\ and\ \citenamefont {{Sigl}}}]{2019PhRvL.122i1101C}%
  \BibitemOpen
  \bibfield  {author} {\bibinfo {author} {\bibfnamefont {Francesco}\
  \bibnamefont {{Capozzi}}}, \bibinfo {author} {\bibfnamefont {Basudeb}\
  \bibnamefont {{Dasgupta}}}, \bibinfo {author} {\bibfnamefont {Alessandro}\
  \bibnamefont {{Mirizzi}}}, \bibinfo {author} {\bibfnamefont {Manibrata}\
  \bibnamefont {{Sen}}}, \ and\ \bibinfo {author} {\bibfnamefont {G{\"u}nter}\
  \bibnamefont {{Sigl}}},\ }\bibfield  {title} {\enquote {\bibinfo {title}
  {{Collisional Triggering of Fast Flavor Conversions of Supernova
  Neutrinos}},}\ }\href {\doibase 10.1103/PhysRevLett.122.091101} {\bibfield
  {journal} {\bibinfo  {journal} {\prl}\ }\textbf {\bibinfo {volume} {122}},\
  \bibinfo {eid} {091101} (\bibinfo {year} {2019})},\ \Eprint
  {http://arxiv.org/abs/1808.06618} {arXiv:1808.06618 [hep-ph]} \BibitemShut
  {NoStop}%
\bibitem [{\citenamefont {{Delfan Azari}}\ \emph {et~al.}(2019)\citenamefont
  {{Delfan Azari}}, \citenamefont {{Yamada}}, \citenamefont {{Morinaga}},
  \citenamefont {{Iwakami}}, \citenamefont {{Okawa}}, \citenamefont
  {{Nagakura}},\ and\ \citenamefont {{Sumiyoshi}}}]{2019PhRvD..99j3011D}%
  \BibitemOpen
  \bibfield  {author} {\bibinfo {author} {\bibfnamefont {Milad}\ \bibnamefont
  {{Delfan Azari}}}, \bibinfo {author} {\bibfnamefont {Shoichi}\ \bibnamefont
  {{Yamada}}}, \bibinfo {author} {\bibfnamefont {Taiki}\ \bibnamefont
  {{Morinaga}}}, \bibinfo {author} {\bibfnamefont {Wakana}\ \bibnamefont
  {{Iwakami}}}, \bibinfo {author} {\bibfnamefont {Hirotada}\ \bibnamefont
  {{Okawa}}}, \bibinfo {author} {\bibfnamefont {Hiroki}\ \bibnamefont
  {{Nagakura}}}, \ and\ \bibinfo {author} {\bibfnamefont {Kohsuke}\
  \bibnamefont {{Sumiyoshi}}},\ }\bibfield  {title} {\enquote {\bibinfo {title}
  {{Linear analysis of fast-pairwise collective neutrino oscillations in
  core-collapse supernovae based on the results of Boltzmann simulations}},}\
  }\href {\doibase 10.1103/PhysRevD.99.103011} {\bibfield  {journal} {\bibinfo
  {journal} {\prd}\ }\textbf {\bibinfo {volume} {99}},\ \bibinfo {eid} {103011}
  (\bibinfo {year} {2019})},\ \Eprint {http://arxiv.org/abs/1902.07467}
  {arXiv:1902.07467 [astro-ph.HE]} \BibitemShut {NoStop}%
\bibitem [{\citenamefont {{Nagakura}}\ \emph {et~al.}(2019)\citenamefont
  {{Nagakura}}, \citenamefont {{Morinaga}}, \citenamefont {{Kato}},\ and\
  \citenamefont {{Yamada}}}]{2019ApJ...886..139N}%
  \BibitemOpen
  \bibfield  {author} {\bibinfo {author} {\bibfnamefont {Hiroki}\ \bibnamefont
  {{Nagakura}}}, \bibinfo {author} {\bibfnamefont {Taiki}\ \bibnamefont
  {{Morinaga}}}, \bibinfo {author} {\bibfnamefont {Chinami}\ \bibnamefont
  {{Kato}}}, \ and\ \bibinfo {author} {\bibfnamefont {Shoichi}\ \bibnamefont
  {{Yamada}}},\ }\bibfield  {title} {\enquote {\bibinfo {title} {{Fast-pairwise
  Collective Neutrino Oscillations Associated with Asymmetric Neutrino
  Emissions in Core-collapse Supernovae}},}\ }\href {\doibase
  10.3847/1538-4357/ab4cf2} {\bibfield  {journal} {\bibinfo  {journal} {\apj}\
  }\textbf {\bibinfo {volume} {886}},\ \bibinfo {eid} {139} (\bibinfo {year}
  {2019})},\ \Eprint {http://arxiv.org/abs/1910.04288} {arXiv:1910.04288
  [astro-ph.HE]} \BibitemShut {NoStop}%
\bibitem [{\citenamefont {Johns}\ \emph {et~al.}(2020)\citenamefont {Johns},
  \citenamefont {Nagakura}, \citenamefont {Fuller},\ and\ \citenamefont
  {Burrows}}]{Johns:2019izj}%
  \BibitemOpen
  \bibfield  {author} {\bibinfo {author} {\bibfnamefont {Lucas}\ \bibnamefont
  {Johns}}, \bibinfo {author} {\bibfnamefont {Hiroki}\ \bibnamefont
  {Nagakura}}, \bibinfo {author} {\bibfnamefont {George~M.}\ \bibnamefont
  {Fuller}}, \ and\ \bibinfo {author} {\bibfnamefont {Adam}\ \bibnamefont
  {Burrows}},\ }\bibfield  {title} {\enquote {\bibinfo {title} {{Neutrino
  oscillations in supernovae: angular moments and fast instabilities}},}\
  }\href {\doibase 10.1103/PhysRevD.101.043009} {\bibfield  {journal} {\bibinfo
   {journal} {Phys. Rev. D}\ }\textbf {\bibinfo {volume} {101}},\ \bibinfo
  {pages} {043009} (\bibinfo {year} {2020})},\ \Eprint
  {http://arxiv.org/abs/1910.05682} {arXiv:1910.05682 [hep-ph]} \BibitemShut
  {NoStop}%
\bibitem [{\citenamefont {Chakraborty}\ and\ \citenamefont
  {Chakraborty}(2020)}]{Chakraborty:2019wxe}%
  \BibitemOpen
  \bibfield  {author} {\bibinfo {author} {\bibfnamefont {Madhurima}\
  \bibnamefont {Chakraborty}}\ and\ \bibinfo {author} {\bibfnamefont {Sovan}\
  \bibnamefont {Chakraborty}},\ }\bibfield  {title} {\enquote {\bibinfo {title}
  {{Three flavor neutrino conversions in supernovae: slow \& fast
  instabilities}},}\ }\href {\doibase 10.1088/1475-7516/2020/01/005} {\bibfield
   {journal} {\bibinfo  {journal} {JCAP}\ }\textbf {\bibinfo {volume} {01}},\
  \bibinfo {pages} {005} (\bibinfo {year} {2020})},\ \Eprint
  {http://arxiv.org/abs/1909.10420} {arXiv:1909.10420 [hep-ph]} \BibitemShut
  {NoStop}%
\bibitem [{\citenamefont {Morinaga}\ and\ \citenamefont
  {Yamada}(2018)}]{Morinaga:2018aug}%
  \BibitemOpen
  \bibfield  {author} {\bibinfo {author} {\bibfnamefont {Taiki}\ \bibnamefont
  {Morinaga}}\ and\ \bibinfo {author} {\bibfnamefont {Shoichi}\ \bibnamefont
  {Yamada}},\ }\bibfield  {title} {\enquote {\bibinfo {title} {{Linear
  stability analysis of collective neutrino oscillations without spurious
  modes}},}\ }\href {\doibase 10.1103/PhysRevD.97.023024} {\bibfield  {journal}
  {\bibinfo  {journal} {Phys. Rev. D}\ }\textbf {\bibinfo {volume} {97}},\
  \bibinfo {pages} {023024} (\bibinfo {year} {2018})},\ \Eprint
  {http://arxiv.org/abs/1803.05913} {arXiv:1803.05913 [hep-ph]} \BibitemShut
  {NoStop}%
\bibitem [{\citenamefont {Abbar}\ \emph {et~al.}(2019)\citenamefont {Abbar},
  \citenamefont {Duan}, \citenamefont {Sumiyoshi}, \citenamefont {Takiwaki},\
  and\ \citenamefont {Volpe}}]{Abbar:2018shq}%
  \BibitemOpen
  \bibfield  {author} {\bibinfo {author} {\bibfnamefont {Sajad}\ \bibnamefont
  {Abbar}}, \bibinfo {author} {\bibfnamefont {Huaiyu}\ \bibnamefont {Duan}},
  \bibinfo {author} {\bibfnamefont {Kohsuke}\ \bibnamefont {Sumiyoshi}},
  \bibinfo {author} {\bibfnamefont {Tomoya}\ \bibnamefont {Takiwaki}}, \ and\
  \bibinfo {author} {\bibfnamefont {Maria~Cristina}\ \bibnamefont {Volpe}},\
  }\bibfield  {title} {\enquote {\bibinfo {title} {{On the occurrence of fast
  neutrino flavor conversions in multidimensional supernova models}},}\ }\href
  {\doibase 10.1103/PhysRevD.100.043004} {\bibfield  {journal} {\bibinfo
  {journal} {Phys. Rev. D}\ }\textbf {\bibinfo {volume} {100}},\ \bibinfo
  {pages} {043004} (\bibinfo {year} {2019})},\ \Eprint
  {http://arxiv.org/abs/1812.06883} {arXiv:1812.06883 [astro-ph.HE]}
  \BibitemShut {NoStop}%
\bibitem [{\citenamefont {Cherry}\ \emph {et~al.}(2019)\citenamefont {Cherry},
  \citenamefont {Fuller}, \citenamefont {Horiuchi}, \citenamefont {Kotake},
  \citenamefont {Takiwaki},\ and\ \citenamefont {Fischer}}]{Cherry:2019vkv}%
  \BibitemOpen
  \bibfield  {author} {\bibinfo {author} {\bibfnamefont {John~F.}\ \bibnamefont
  {Cherry}}, \bibinfo {author} {\bibfnamefont {George~M.}\ \bibnamefont
  {Fuller}}, \bibinfo {author} {\bibfnamefont {Shunsaku}\ \bibnamefont
  {Horiuchi}}, \bibinfo {author} {\bibfnamefont {Kei}\ \bibnamefont {Kotake}},
  \bibinfo {author} {\bibfnamefont {Tomoya}\ \bibnamefont {Takiwaki}}, \ and\
  \bibinfo {author} {\bibfnamefont {Tobias}\ \bibnamefont {Fischer}},\
  }\bibfield  {title} {\enquote {\bibinfo {title} {{Time of Flight and
  Supernova Progenitor Effects on the Neutrino Halo}},}\ }\href@noop {} {\
  (\bibinfo {year} {2019})},\ \Eprint {http://arxiv.org/abs/1912.11489}
  {arXiv:1912.11489 [astro-ph.HE]} \BibitemShut {NoStop}%
\bibitem [{\citenamefont {Morinaga}\ \emph {et~al.}(2020)\citenamefont
  {Morinaga}, \citenamefont {Nagakura}, \citenamefont {Kato},\ and\
  \citenamefont {Yamada}}]{Morinaga:2019wsv}%
  \BibitemOpen
  \bibfield  {author} {\bibinfo {author} {\bibfnamefont {Taiki}\ \bibnamefont
  {Morinaga}}, \bibinfo {author} {\bibfnamefont {Hiroki}\ \bibnamefont
  {Nagakura}}, \bibinfo {author} {\bibfnamefont {Chinami}\ \bibnamefont
  {Kato}}, \ and\ \bibinfo {author} {\bibfnamefont {Shoichi}\ \bibnamefont
  {Yamada}},\ }\bibfield  {title} {\enquote {\bibinfo {title} {{Fast
  neutrino-flavor conversion in the preshock region of core-collapse
  supernovae}},}\ }\href {\doibase 10.1103/PhysRevResearch.2.012046} {\bibfield
   {journal} {\bibinfo  {journal} {Phys. Rev. Res.}\ }\textbf {\bibinfo
  {volume} {2}},\ \bibinfo {pages} {012046} (\bibinfo {year} {2020})},\ \Eprint
  {http://arxiv.org/abs/1909.13131} {arXiv:1909.13131 [astro-ph.HE]}
  \BibitemShut {NoStop}%
\bibitem [{\citenamefont {Bhattacharyya}\ and\ \citenamefont
  {Dasgupta}(2020)}]{Bhattacharyya:2020dhu}%
  \BibitemOpen
  \bibfield  {author} {\bibinfo {author} {\bibfnamefont {Soumya}\ \bibnamefont
  {Bhattacharyya}}\ and\ \bibinfo {author} {\bibfnamefont {Basudeb}\
  \bibnamefont {Dasgupta}},\ }\bibfield  {title} {\enquote {\bibinfo {title}
  {{Late-time behavior of fast neutrino oscillations}},}\ }\href {\doibase
  10.1103/PhysRevD.102.063018} {\bibfield  {journal} {\bibinfo  {journal}
  {Phys. Rev. D}\ }\textbf {\bibinfo {volume} {102}},\ \bibinfo {pages}
  {063018} (\bibinfo {year} {2020})},\ \Eprint
  {http://arxiv.org/abs/2005.00459} {arXiv:2005.00459 [hep-ph]} \BibitemShut
  {NoStop}%
\bibitem [{\citenamefont {Abbar}(2020)}]{Abbar:2020fcl}%
  \BibitemOpen
  \bibfield  {author} {\bibinfo {author} {\bibfnamefont {Sajad}\ \bibnamefont
  {Abbar}},\ }\bibfield  {title} {\enquote {\bibinfo {title} {{Searching for
  Fast Neutrino Flavor Conversion Modes in Core-collapse Supernova
  Simulations}},}\ }\href {\doibase 10.1088/1475-7516/2020/05/027} {\bibfield
  {journal} {\bibinfo  {journal} {JCAP}\ }\textbf {\bibinfo {volume} {05}},\
  \bibinfo {pages} {027} (\bibinfo {year} {2020})},\ \Eprint
  {http://arxiv.org/abs/2003.00969} {arXiv:2003.00969 [astro-ph.HE]}
  \BibitemShut {NoStop}%
\bibitem [{\citenamefont {Capozzi}\ \emph {et~al.}(2020)\citenamefont
  {Capozzi}, \citenamefont {Abbar}, \citenamefont {Bollig},\ and\ \citenamefont
  {Janka}}]{Capozzi:2020syn}%
  \BibitemOpen
  \bibfield  {author} {\bibinfo {author} {\bibfnamefont {Francesco}\
  \bibnamefont {Capozzi}}, \bibinfo {author} {\bibfnamefont {Sajad}\
  \bibnamefont {Abbar}}, \bibinfo {author} {\bibfnamefont {Robert}\
  \bibnamefont {Bollig}}, \ and\ \bibinfo {author} {\bibfnamefont {H.~Thomas}\
  \bibnamefont {Janka}},\ }\bibfield  {title} {\enquote {\bibinfo {title}
  {{Fast neutrino flavor conversions in one-dimensional core-collapse supernova
  models with and without muon creation}},}\ }\href@noop {} {\  (\bibinfo
  {year} {2020})},\ \Eprint {http://arxiv.org/abs/2012.08525} {arXiv:2012.08525
  [astro-ph.HE]} \BibitemShut {NoStop}%
\bibitem [{\citenamefont {Tamborra}\ and\ \citenamefont
  {Shalgar}(2020)}]{Tamborra:2020cul}%
  \BibitemOpen
  \bibfield  {author} {\bibinfo {author} {\bibfnamefont {Irene}\ \bibnamefont
  {Tamborra}}\ and\ \bibinfo {author} {\bibfnamefont {Shashank}\ \bibnamefont
  {Shalgar}},\ }\bibfield  {title} {\enquote {\bibinfo {title} {{New
  Developments in Flavor Evolution of a Dense Neutrino Gas}},}\ }\href
  {\doibase 10.1146/annurev-nucl-102920-050505} {\  (\bibinfo {year} {2020}),\
  10.1146/annurev-nucl-102920-050505},\ \Eprint
  {http://arxiv.org/abs/2011.01948} {arXiv:2011.01948 [astro-ph.HE]}
  \BibitemShut {NoStop}%
\bibitem [{\citenamefont {Abi}\ \emph {et~al.}(2020)\citenamefont {Abi} \emph
  {et~al.}}]{Abi:2020}%
  \BibitemOpen
  \bibfield  {author} {\bibinfo {author} {\bibfnamefont {Babak}\ \bibnamefont
  {Abi}} \emph {et~al.} (\bibinfo {collaboration} {DUNE}),\ }\bibfield  {title}
  {\enquote {\bibinfo {title} {{Deep Underground Neutrino Experiment (DUNE),
  Far Detector Technical Design Report, Volume II DUNE Physics}},}\ }\href@noop
  {} {\  (\bibinfo {year} {2020})},\ \Eprint {http://arxiv.org/abs/2002.03005}
  {arXiv:2002.03005 [hep-ex]} \BibitemShut {NoStop}%
\bibitem [{\citenamefont {Abe}\ \emph {et~al.}(2018)\citenamefont {Abe} \emph
  {et~al.}}]{Abe:2018}%
  \BibitemOpen
  \bibfield  {author} {\bibinfo {author} {\bibfnamefont {K.}~\bibnamefont
  {Abe}} \emph {et~al.} (\bibinfo {collaboration} {Hyper-Kamiokande}),\
  }\bibfield  {title} {\enquote {\bibinfo {title} {{Hyper-Kamiokande Design
  Report}},}\ }\href@noop {} {\  (\bibinfo {year} {2018})},\ \Eprint
  {http://arxiv.org/abs/1805.04163} {arXiv:1805.04163 [physics.ins-det]}
  \BibitemShut {NoStop}%
\bibitem [{\citenamefont {He}\ \emph {et~al.}(2016)\citenamefont {He},
  \citenamefont {Zhang}, \citenamefont {Ren},\ and\ \citenamefont
  {Sun}}]{He2016}%
  \BibitemOpen
  \bibfield  {author} {\bibinfo {author} {\bibfnamefont {Kaiming}\ \bibnamefont
  {He}}, \bibinfo {author} {\bibfnamefont {Xiangyu}\ \bibnamefont {Zhang}},
  \bibinfo {author} {\bibfnamefont {Shaoqing}\ \bibnamefont {Ren}}, \ and\
  \bibinfo {author} {\bibfnamefont {Jian}\ \bibnamefont {Sun}},\ }\bibfield
  {title} {\enquote {\bibinfo {title} {Deep residual learning for image
  recognition},}\ }in\ \href@noop {} {\emph {\bibinfo {booktitle} {The IEEE
  Conference on Computer Vision and Pattern Recognition (CVPR)}}}\ (\bibinfo
  {year} {2016})\BibitemShut {NoStop}%
\bibitem [{\citenamefont {Zoph}\ \emph {et~al.}(2018)\citenamefont {Zoph},
  \citenamefont {Vasudevan}, \citenamefont {Shlens},\ and\ \citenamefont
  {Le}}]{Zoph2018}%
  \BibitemOpen
  \bibfield  {author} {\bibinfo {author} {\bibfnamefont {Barret}\ \bibnamefont
  {Zoph}}, \bibinfo {author} {\bibfnamefont {Vijay}\ \bibnamefont {Vasudevan}},
  \bibinfo {author} {\bibfnamefont {Jonathon}\ \bibnamefont {Shlens}}, \ and\
  \bibinfo {author} {\bibfnamefont {Quoc~V.}\ \bibnamefont {Le}},\ }\bibfield
  {title} {\enquote {\bibinfo {title} {Learning transferable architectures for
  scalable image recognition},}\ }in\ \href@noop {} {\emph {\bibinfo
  {booktitle} {The IEEE Conference on Computer Vision and Pattern Recognition
  (CVPR)}}}\ (\bibinfo {year} {2018})\BibitemShut {NoStop}%
\bibitem [{\citenamefont {Traore}\ \emph {et~al.}(2018)\citenamefont {Traore},
  \citenamefont {Kamsu-Foguem},\ and\ \citenamefont {Tangara}}]{Traore2018}%
  \BibitemOpen
  \bibfield  {author} {\bibinfo {author} {\bibfnamefont {Boukaye~Boubacar}\
  \bibnamefont {Traore}}, \bibinfo {author} {\bibfnamefont {Bernard}\
  \bibnamefont {Kamsu-Foguem}}, \ and\ \bibinfo {author} {\bibfnamefont {Fana}\
  \bibnamefont {Tangara}},\ }\bibfield  {title} {\enquote {\bibinfo {title}
  {Deep convolution neural network for image recognition},}\ }\href@noop {}
  {\bibfield  {journal} {\bibinfo  {journal} {Ecol. Informatics}\ }\textbf
  {\bibinfo {volume} {48}},\ \bibinfo {pages} {257--268} (\bibinfo {year}
  {2018})}\BibitemShut {NoStop}%
\bibitem [{\citenamefont {Young}\ \emph {et~al.}(2017)\citenamefont {Young},
  \citenamefont {Hazarika}, \citenamefont {Poria},\ and\ \citenamefont
  {Cambria}}]{Yong2018}%
  \BibitemOpen
  \bibfield  {author} {\bibinfo {author} {\bibfnamefont {Tom}\ \bibnamefont
  {Young}}, \bibinfo {author} {\bibfnamefont {Devamanyu}\ \bibnamefont
  {Hazarika}}, \bibinfo {author} {\bibfnamefont {Soujanya}\ \bibnamefont
  {Poria}}, \ and\ \bibinfo {author} {\bibfnamefont {Erik}\ \bibnamefont
  {Cambria}},\ }\bibfield  {title} {\enquote {\bibinfo {title} {Recent trends
  in deep learning based natural language processing},}\ }\href@noop {}
  {\bibfield  {journal} {\bibinfo  {journal} {CoRR}\ }\textbf {\bibinfo
  {volume} {abs/1708.02709}} (\bibinfo {year} {2017})},\ \Eprint
  {http://arxiv.org/abs/1708.02709} {1708.02709} \BibitemShut {NoStop}%
\bibitem [{\citenamefont {Otter}\ \emph {et~al.}(2018)\citenamefont {Otter},
  \citenamefont {Medina},\ and\ \citenamefont {Kalita}}]{Otter2018}%
  \BibitemOpen
  \bibfield  {author} {\bibinfo {author} {\bibfnamefont {Daniel~W.}\
  \bibnamefont {Otter}}, \bibinfo {author} {\bibfnamefont {Julian~R.}\
  \bibnamefont {Medina}}, \ and\ \bibinfo {author} {\bibfnamefont {Jugal~K.}\
  \bibnamefont {Kalita}},\ }\bibfield  {title} {\enquote {\bibinfo {title} {A
  survey of the usages of deep learning in natural language processing},}\
  }\href@noop {} {\bibfield  {journal} {\bibinfo  {journal} {CoRR}\ }\textbf
  {\bibinfo {volume} {abs/1807.10854}} (\bibinfo {year} {2018})}\BibitemShut
  {NoStop}%
\bibitem [{\citenamefont {Tang}\ \emph {et~al.}(2019)\citenamefont {Tang},
  \citenamefont {Pan}, \citenamefont {Yin},\ and\ \citenamefont
  {Khateeb}}]{Tang2019}%
  \BibitemOpen
  \bibfield  {author} {\bibinfo {author} {\bibfnamefont {Binhua}\ \bibnamefont
  {Tang}}, \bibinfo {author} {\bibfnamefont {Zixiang}\ \bibnamefont {Pan}},
  \bibinfo {author} {\bibfnamefont {Kang}\ \bibnamefont {Yin}}, \ and\ \bibinfo
  {author} {\bibfnamefont {Asif}\ \bibnamefont {Khateeb}},\ }\bibfield  {title}
  {\enquote {\bibinfo {title} {Recent advances of deep learning in
  bioinformatics and computational biology},}\ }\href@noop {} {\bibfield
  {journal} {\bibinfo  {journal} {Frontiers in genetics}\ }\textbf {\bibinfo
  {volume} {10}},\ \bibinfo {pages} {214--214} (\bibinfo {year}
  {2019})}\BibitemShut {NoStop}%
\bibitem [{\citenamefont {Zemouri}\ \emph {et~al.}(2019)\citenamefont
  {Zemouri}, \citenamefont {Zerhouni},\ and\ \citenamefont
  {Racoceanu}}]{Zemouri2019}%
  \BibitemOpen
  \bibfield  {author} {\bibinfo {author} {\bibfnamefont {Ryad}\ \bibnamefont
  {Zemouri}}, \bibinfo {author} {\bibfnamefont {Noureddine}\ \bibnamefont
  {Zerhouni}}, \ and\ \bibinfo {author} {\bibfnamefont {Daniel}\ \bibnamefont
  {Racoceanu}},\ }\bibfield  {title} {\enquote {\bibinfo {title} {Deep learning
  in the biomedical applications: Recent and future status},}\ }\href {\doibase
  10.3390/app9081526} {\bibfield  {journal} {\bibinfo  {journal} {Applied
  Sciences}\ }\textbf {\bibinfo {volume} {9}},\ \bibinfo {pages} {1526}
  (\bibinfo {year} {2019})}\BibitemShut {NoStop}%
\bibitem [{\citenamefont {Armstrong}\ \emph {et~al.}(2017)\citenamefont
  {Armstrong}, \citenamefont {Patwardhan}, \citenamefont {Johns}, \citenamefont
  {Kishimoto}, \citenamefont {Abarbanel},\ and\ \citenamefont
  {Fuller}}]{Armstrong:2017}%
  \BibitemOpen
  \bibfield  {author} {\bibinfo {author} {\bibfnamefont {Eve}\ \bibnamefont
  {Armstrong}}, \bibinfo {author} {\bibfnamefont {Amol~V.}\ \bibnamefont
  {Patwardhan}}, \bibinfo {author} {\bibfnamefont {Lucas}\ \bibnamefont
  {Johns}}, \bibinfo {author} {\bibfnamefont {Chad~T.}\ \bibnamefont
  {Kishimoto}}, \bibinfo {author} {\bibfnamefont {Henry D.~I.}\ \bibnamefont
  {Abarbanel}}, \ and\ \bibinfo {author} {\bibfnamefont {George~M.}\
  \bibnamefont {Fuller}},\ }\bibfield  {title} {\enquote {\bibinfo {title} {An
  optimization-based approach to calculating neutrino flavor evolution},}\
  }\href {\doibase 10.1103/PhysRevD.96.083008} {\bibfield  {journal} {\bibinfo
  {journal} {Phys. Rev. D}\ }\textbf {\bibinfo {volume} {96}},\ \bibinfo
  {pages} {083008} (\bibinfo {year} {2017})}\BibitemShut {NoStop}%
\bibitem [{\citenamefont {Armstrong}\ \emph {et~al.}(2020)\citenamefont
  {Armstrong}, \citenamefont {Patwardhan}, \citenamefont {Rrapaj},
  \citenamefont {Ardizi},\ and\ \citenamefont {Fuller}}]{Armstrong:2020}%
  \BibitemOpen
  \bibfield  {author} {\bibinfo {author} {\bibfnamefont {Eve}\ \bibnamefont
  {Armstrong}}, \bibinfo {author} {\bibfnamefont {Amol~V.}\ \bibnamefont
  {Patwardhan}}, \bibinfo {author} {\bibfnamefont {Ermal}\ \bibnamefont
  {Rrapaj}}, \bibinfo {author} {\bibfnamefont {Sina~Fallah}\ \bibnamefont
  {Ardizi}}, \ and\ \bibinfo {author} {\bibfnamefont {George~M.}\ \bibnamefont
  {Fuller}},\ }\bibfield  {title} {\enquote {\bibinfo {title} {Inference offers
  a metric to constrain dynamical models of neutrino flavor transformation},}\
  }\href {\doibase 10.1103/PhysRevD.102.043013} {\bibfield  {journal} {\bibinfo
   {journal} {Phys. Rev. D}\ }\textbf {\bibinfo {volume} {102}},\ \bibinfo
  {pages} {043013} (\bibinfo {year} {2020})}\BibitemShut {NoStop}%
\bibitem [{\citenamefont {Chen}\ \emph {et~al.}(2018)\citenamefont {Chen},
  \citenamefont {Rubanova}, \citenamefont {Bettencourt},\ and\ \citenamefont
  {Duvenaud}}]{Chen2018}%
  \BibitemOpen
  \bibfield  {author} {\bibinfo {author} {\bibfnamefont {Ricky T.~Q.}\
  \bibnamefont {Chen}}, \bibinfo {author} {\bibfnamefont {Yulia}\ \bibnamefont
  {Rubanova}}, \bibinfo {author} {\bibfnamefont {Jesse}\ \bibnamefont
  {Bettencourt}}, \ and\ \bibinfo {author} {\bibfnamefont {David~K}\
  \bibnamefont {Duvenaud}},\ }\bibfield  {title} {\enquote {\bibinfo {title}
  {Neural ordinary differential equations},}\ }in\ \href@noop {} {\emph
  {\bibinfo {booktitle} {Advances in Neural Information Processing Systems
  31}}},\ \bibinfo {editor} {edited by\ \bibinfo {editor} {\bibfnamefont
  {S.}~\bibnamefont {Bengio}}, \bibinfo {editor} {\bibfnamefont
  {H.}~\bibnamefont {Wallach}}, \bibinfo {editor} {\bibfnamefont
  {H.}~\bibnamefont {Larochelle}}, \bibinfo {editor} {\bibfnamefont
  {K.}~\bibnamefont {Grauman}}, \bibinfo {editor} {\bibfnamefont
  {N.}~\bibnamefont {Cesa-Bianchi}}, \ and\ \bibinfo {editor} {\bibfnamefont
  {R.}~\bibnamefont {Garnett}}}\ (\bibinfo  {publisher} {Curran Associates,
  Inc.},\ \bibinfo {year} {2018})\ pp.\ \bibinfo {pages}
  {6571--6583}\BibitemShut {NoStop}%
\bibitem [{\citenamefont {Hornik}\ \emph {et~al.}(1989)\citenamefont {Hornik},
  \citenamefont {Stinchcombe},\ and\ \citenamefont {White}}]{HORNIK1989}%
  \BibitemOpen
  \bibfield  {author} {\bibinfo {author} {\bibfnamefont {Kurt}\ \bibnamefont
  {Hornik}}, \bibinfo {author} {\bibfnamefont {Maxwell}\ \bibnamefont
  {Stinchcombe}}, \ and\ \bibinfo {author} {\bibfnamefont {Halbert}\
  \bibnamefont {White}},\ }\bibfield  {title} {\enquote {\bibinfo {title}
  {Multilayer feedforward networks are universal approximators},}\ }\href
  {\doibase https://doi.org/10.1016/0893-6080(89)90020-8} {\bibfield  {journal}
  {\bibinfo  {journal} {Neural Networks}\ }\textbf {\bibinfo {volume} {2}},\
  \bibinfo {pages} {359 -- 366} (\bibinfo {year} {1989})}\BibitemShut {NoStop}%
\bibitem [{\citenamefont {Keister}(2015)}]{Keister:2014ufa}%
  \BibitemOpen
  \bibfield  {author} {\bibinfo {author} {\bibfnamefont {B.D.}\ \bibnamefont
  {Keister}},\ }\bibfield  {title} {\enquote {\bibinfo {title} {{Numerical and
  Physical Stability of Supernova Neutrino Flavor Evolution}},}\ }\href
  {\doibase 10.1088/0031-8949/90/8/088008} {\bibfield  {journal} {\bibinfo
  {journal} {Phys. Scripta}\ }\textbf {\bibinfo {volume} {90}},\ \bibinfo
  {pages} {088008} (\bibinfo {year} {2015})},\ \Eprint
  {http://arxiv.org/abs/1408.4729} {arXiv:1408.4729 [astro-ph.HE]} \BibitemShut
  {NoStop}%
\bibitem [{\citenamefont {Tarantola}(2005)}]{Tarantola2005}%
  \BibitemOpen
  \bibfield  {author} {\bibinfo {author} {\bibfnamefont {A.}~\bibnamefont
  {Tarantola}},\ }\href@noop {} {\emph {\bibinfo {title} {Inverse Problem
  Theory and Methods for Model Parameter Estimation}}},\ Other Titles in
  Applied Mathematics\ (\bibinfo  {publisher} {Society for Industrial and
  Applied Mathematics},\ \bibinfo {year} {2005})\BibitemShut {NoStop}%
\bibitem [{\citenamefont {vanMilligen}\ \emph {et~al.}(1995)\citenamefont
  {vanMilligen}, \citenamefont {Tribaldos},\ and\ \citenamefont
  {Jim{\'e}nez}}]{vanMilligen1995}%
  \BibitemOpen
  \bibfield  {author} {\bibinfo {author} {\bibfnamefont {BP}~\bibnamefont
  {vanMilligen}}, \bibinfo {author} {\bibnamefont {Tribaldos}}, \ and\ \bibinfo
  {author} {\bibnamefont {Jim{\'e}nez}},\ }\bibfield  {title} {\enquote
  {\bibinfo {title} {Neural network differential equation and plasma
  equilibrium solver.}}\ }\href@noop {} {\bibfield  {journal} {\bibinfo
  {journal} {Physical review letters}\ }\textbf {\bibinfo {volume} {75 20}},\
  \bibinfo {pages} {3594--3597} (\bibinfo {year} {1995})}\BibitemShut {NoStop}%
\bibitem [{\citenamefont {{Lagaris}}\ \emph {et~al.}(1998)\citenamefont
  {{Lagaris}}, \citenamefont {{Likas}},\ and\ \citenamefont
  {{Fotiadis}}}]{Lagaris1998}%
  \BibitemOpen
  \bibfield  {author} {\bibinfo {author} {\bibfnamefont {I.~E.}\ \bibnamefont
  {{Lagaris}}}, \bibinfo {author} {\bibfnamefont {A.}~\bibnamefont {{Likas}}},
  \ and\ \bibinfo {author} {\bibfnamefont {D.~I.}\ \bibnamefont {{Fotiadis}}},\
  }\bibfield  {title} {\enquote {\bibinfo {title} {Artificial neural networks
  for solving ordinary and partial differential equations},}\ }\href@noop {}
  {\bibfield  {journal} {\bibinfo  {journal} {IEEE Transactions on Neural
  Networks}\ }\textbf {\bibinfo {volume} {9}},\ \bibinfo {pages} {987--1000}
  (\bibinfo {year} {1998})}\BibitemShut {NoStop}%
\bibitem [{\citenamefont {Berg}\ and\ \citenamefont
  {Nystr{\"o}m}(2018)}]{Berg2018}%
  \BibitemOpen
  \bibfield  {author} {\bibinfo {author} {\bibfnamefont {Jens}\ \bibnamefont
  {Berg}}\ and\ \bibinfo {author} {\bibfnamefont {Kaj}\ \bibnamefont
  {Nystr{\"o}m}},\ }\bibfield  {title} {\enquote {\bibinfo {title} {A unified
  deep artificial neural network approach to partial differential equations in
  complex geometries},}\ }\href@noop {} {\bibfield  {journal} {\bibinfo
  {journal} {Neurocomputing}\ }\textbf {\bibinfo {volume} {317}},\ \bibinfo
  {pages} {28--41} (\bibinfo {year} {2018})}\BibitemShut {NoStop}%
\bibitem [{\citenamefont {Magill}\ \emph {et~al.}(2018)\citenamefont {Magill},
  \citenamefont {Qureshi},\ and\ \citenamefont {de~Haan}}]{Magill2018}%
  \BibitemOpen
  \bibfield  {author} {\bibinfo {author} {\bibfnamefont {Martin}\ \bibnamefont
  {Magill}}, \bibinfo {author} {\bibfnamefont {Faisal}\ \bibnamefont
  {Qureshi}}, \ and\ \bibinfo {author} {\bibfnamefont {Hendrick}\ \bibnamefont
  {de~Haan}},\ }\bibfield  {title} {\enquote {\bibinfo {title} {Neural networks
  trained to solve differential equations learn general representations},}\
  }in\ \href@noop {} {\emph {\bibinfo {booktitle} {Advances in Neural
  Information Processing Systems 31}}},\ \bibinfo {editor} {edited by\ \bibinfo
  {editor} {\bibfnamefont {S.}~\bibnamefont {Bengio}}, \bibinfo {editor}
  {\bibfnamefont {H.}~\bibnamefont {Wallach}}, \bibinfo {editor} {\bibfnamefont
  {H.}~\bibnamefont {Larochelle}}, \bibinfo {editor} {\bibfnamefont
  {K.}~\bibnamefont {Grauman}}, \bibinfo {editor} {\bibfnamefont
  {N.}~\bibnamefont {Cesa-Bianchi}}, \ and\ \bibinfo {editor} {\bibfnamefont
  {R.}~\bibnamefont {Garnett}}}\ (\bibinfo  {publisher} {Curran Associates,
  Inc.},\ \bibinfo {year} {2018})\ pp.\ \bibinfo {pages}
  {4071--4081}\BibitemShut {NoStop}%
\bibitem [{\citenamefont {Raissi}\ \emph {et~al.}(2019)\citenamefont {Raissi},
  \citenamefont {Perdikaris},\ and\ \citenamefont {Karniadakis}}]{RAISSI2019}%
  \BibitemOpen
  \bibfield  {author} {\bibinfo {author} {\bibfnamefont {M.}~\bibnamefont
  {Raissi}}, \bibinfo {author} {\bibfnamefont {P.}~\bibnamefont {Perdikaris}},
  \ and\ \bibinfo {author} {\bibfnamefont {G.E.}\ \bibnamefont {Karniadakis}},\
  }\bibfield  {title} {\enquote {\bibinfo {title} {Physics-informed neural
  networks: A deep learning framework for solving forward and inverse problems
  involving nonlinear partial differential equations},}\ }\href@noop {}
  {\bibfield  {journal} {\bibinfo  {journal} {Journal of Computational
  Physics}\ }\textbf {\bibinfo {volume} {378}},\ \bibinfo {pages} {686 -- 707}
  (\bibinfo {year} {2019})}\BibitemShut {NoStop}%
\bibitem [{\citenamefont {Hairer}\ and\ \citenamefont
  {Wanner}(1999)}]{Hairier1999}%
  \BibitemOpen
  \bibfield  {author} {\bibinfo {author} {\bibfnamefont {Ernst}\ \bibnamefont
  {Hairer}}\ and\ \bibinfo {author} {\bibfnamefont {Gerhard}\ \bibnamefont
  {Wanner}},\ }\bibfield  {title} {\enquote {\bibinfo {title} {Stiff
  differential equations solved by radau methods},}\ }\href@noop {} {\bibfield
  {journal} {\bibinfo  {journal} {Journal of Computational and Applied
  Mathematics}\ }\textbf {\bibinfo {volume} {111}},\ \bibinfo {pages} {93 --
  111} (\bibinfo {year} {1999})}\BibitemShut {NoStop}%
\bibitem [{\citenamefont {{Revels}}\ \emph {et~al.}(2016)\citenamefont
  {{Revels}}, \citenamefont {{Lubin}},\ and\ \citenamefont
  {{Papamarkou}}}]{Revels2016}%
  \BibitemOpen
  \bibfield  {author} {\bibinfo {author} {\bibfnamefont {J.}~\bibnamefont
  {{Revels}}}, \bibinfo {author} {\bibfnamefont {M.}~\bibnamefont {{Lubin}}}, \
  and\ \bibinfo {author} {\bibfnamefont {T.}~\bibnamefont {{Papamarkou}}},\
  }\bibfield  {title} {\enquote {\bibinfo {title} {Forward-mode automatic
  differentiation in {J}ulia},}\ }\href {https://arxiv.org/abs/1607.07892}
  {\bibfield  {journal} {\bibinfo  {journal} {arXiv:1607.07892 [cs.MS]}\ }
  (\bibinfo {year} {2016})}\BibitemShut {NoStop}%
\bibitem [{\citenamefont {Baydin}\ \emph {et~al.}(2018)\citenamefont {Baydin},
  \citenamefont {Pearlmutter}, \citenamefont {Radul},\ and\ \citenamefont
  {Siskind}}]{Baydin2018}%
  \BibitemOpen
  \bibfield  {author} {\bibinfo {author} {\bibfnamefont {Atilim~Gunes}\
  \bibnamefont {Baydin}}, \bibinfo {author} {\bibfnamefont {Barak~A.}\
  \bibnamefont {Pearlmutter}}, \bibinfo {author} {\bibfnamefont
  {Alexey~Andreyevich}\ \bibnamefont {Radul}}, \ and\ \bibinfo {author}
  {\bibfnamefont {Jeffrey~Mark}\ \bibnamefont {Siskind}},\ }\bibfield  {title}
  {\enquote {\bibinfo {title} {Automatic differentiation in machine learning: a
  survey},}\ }\href@noop {} {\bibfield  {journal} {\bibinfo  {journal} {Journal
  of Machine Learning Research}\ }\textbf {\bibinfo {volume} {18}},\ \bibinfo
  {pages} {1--43} (\bibinfo {year} {2018})}\BibitemShut {NoStop}%
\bibitem [{\citenamefont {Sandu}\ \emph {et~al.}(2003)\citenamefont {Sandu},
  \citenamefont {Daescu},\ and\ \citenamefont {Carmichael}}]{SANDU2003}%
  \BibitemOpen
  \bibfield  {author} {\bibinfo {author} {\bibfnamefont {Adrian}\ \bibnamefont
  {Sandu}}, \bibinfo {author} {\bibfnamefont {Dacian~N.}\ \bibnamefont
  {Daescu}}, \ and\ \bibinfo {author} {\bibfnamefont {Gregory~R.}\ \bibnamefont
  {Carmichael}},\ }\bibfield  {title} {\enquote {\bibinfo {title} {Direct and
  adjoint sensitivity analysis of chemical kinetic systems with kpp: Part
  i—theory and software tools},}\ }\href@noop {} {\bibfield  {journal}
  {\bibinfo  {journal} {Atmospheric Environment}\ }\textbf {\bibinfo {volume}
  {37}},\ \bibinfo {pages} {5083 -- 5096} (\bibinfo {year} {2003})}\BibitemShut
  {NoStop}%
\bibitem [{\citenamefont {Cao}\ \emph {et~al.}(2003)\citenamefont {Cao},
  \citenamefont {Li}, \citenamefont {Petzold},\ and\ \citenamefont
  {Serban}}]{Ca02006}%
  \BibitemOpen
  \bibfield  {author} {\bibinfo {author} {\bibfnamefont {Yang}\ \bibnamefont
  {Cao}}, \bibinfo {author} {\bibfnamefont {Shengtai}\ \bibnamefont {Li}},
  \bibinfo {author} {\bibfnamefont {Linda}\ \bibnamefont {Petzold}}, \ and\
  \bibinfo {author} {\bibfnamefont {Radu}\ \bibnamefont {Serban}},\ }\bibfield
  {title} {\enquote {\bibinfo {title} {Adjoint sensitivity analysis for
  differential-algebraic equations: The adjoint dae system and its numerical
  solution},}\ }\href {\doibase 10.1137/S1064827501380630} {\bibfield
  {journal} {\bibinfo  {journal} {SIAM Journal on Scientific Computing}\
  }\textbf {\bibinfo {volume} {24}},\ \bibinfo {pages} {1076--1089} (\bibinfo
  {year} {2003})}\BibitemShut {NoStop}%
\bibitem [{\citenamefont {Bezanson}\ \emph {et~al.}(2017)\citenamefont
  {Bezanson}, \citenamefont {Edelman}, \citenamefont {Karpinski},\ and\
  \citenamefont {Shah}}]{Bezanson2017}%
  \BibitemOpen
  \bibfield  {author} {\bibinfo {author} {\bibfnamefont {Jeff}\ \bibnamefont
  {Bezanson}}, \bibinfo {author} {\bibfnamefont {Alan}\ \bibnamefont
  {Edelman}}, \bibinfo {author} {\bibfnamefont {Stefan}\ \bibnamefont
  {Karpinski}}, \ and\ \bibinfo {author} {\bibfnamefont {Viral~B.}\
  \bibnamefont {Shah}},\ }\bibfield  {title} {\enquote {\bibinfo {title}
  {Julia: A fresh approach to numerical computing},}\ }\href {\doibase
  10.1137/141000671} {\bibfield  {journal} {\bibinfo  {journal} {SIAM Review}\
  }\textbf {\bibinfo {volume} {59}},\ \bibinfo {pages} {65--98} (\bibinfo
  {year} {2017})}\BibitemShut {NoStop}%
\bibitem [{\citenamefont {Rackauckas}\ and\ \citenamefont
  {Nie}(2017)}]{Rackauckas2017}%
  \BibitemOpen
  \bibfield  {author} {\bibinfo {author} {\bibfnamefont {Christopher}\
  \bibnamefont {Rackauckas}}\ and\ \bibinfo {author} {\bibfnamefont
  {Q}~\bibnamefont {Nie}},\ }\bibfield  {title} {\enquote {\bibinfo {title}
  {Differentialequations.jl – a performant and feature-rich ecosystem for
  solving differential equations in julia},}\ }\href@noop {} {\bibfield
  {journal} {\bibinfo  {journal} {Journal of Open Research Software}\ }\textbf
  {\bibinfo {volume} {5}},\ \bibinfo {pages} {15} (\bibinfo {year}
  {2017})}\BibitemShut {NoStop}%
\bibitem [{\citenamefont {Innes}\ \emph {et~al.}(2018)\citenamefont {Innes},
  \citenamefont {Saba}, \citenamefont {Fischer}, \citenamefont {Gandhi},
  \citenamefont {Rudilosso}, \citenamefont {Joy}, \citenamefont {Karmali},
  \citenamefont {Pal},\ and\ \citenamefont {Shah}}]{Innes2018}%
  \BibitemOpen
  \bibfield  {author} {\bibinfo {author} {\bibfnamefont {Michael}\ \bibnamefont
  {Innes}}, \bibinfo {author} {\bibfnamefont {Elliot}\ \bibnamefont {Saba}},
  \bibinfo {author} {\bibfnamefont {Keno}\ \bibnamefont {Fischer}}, \bibinfo
  {author} {\bibfnamefont {Dhairya}\ \bibnamefont {Gandhi}}, \bibinfo {author}
  {\bibfnamefont {Marco~Concetto}\ \bibnamefont {Rudilosso}}, \bibinfo {author}
  {\bibfnamefont {Neethu~Mariya}\ \bibnamefont {Joy}}, \bibinfo {author}
  {\bibfnamefont {Tejan}\ \bibnamefont {Karmali}}, \bibinfo {author}
  {\bibfnamefont {Avik}\ \bibnamefont {Pal}}, \ and\ \bibinfo {author}
  {\bibfnamefont {Viral}\ \bibnamefont {Shah}},\ }\bibfield  {title} {\enquote
  {\bibinfo {title} {Fashionable modelling with flux},}\ }\href@noop {}
  {\bibfield  {journal} {\bibinfo  {journal} {CoRR}\ }\textbf {\bibinfo
  {volume} {abs/1811.01457}} (\bibinfo {year} {2018})}\BibitemShut {NoStop}%
\bibitem [{\citenamefont {Rackauckas}\ \emph {et~al.}(2019)\citenamefont
  {Rackauckas}, \citenamefont {Innes}, \citenamefont {Ma}, \citenamefont
  {Bettencourt}, \citenamefont {White},\ and\ \citenamefont
  {Dixit}}]{Rackauckas2019}%
  \BibitemOpen
  \bibfield  {author} {\bibinfo {author} {\bibfnamefont {Christopher}\
  \bibnamefont {Rackauckas}}, \bibinfo {author} {\bibfnamefont {Mike}\
  \bibnamefont {Innes}}, \bibinfo {author} {\bibfnamefont {Yingbo}\
  \bibnamefont {Ma}}, \bibinfo {author} {\bibfnamefont {Jesse}\ \bibnamefont
  {Bettencourt}}, \bibinfo {author} {\bibfnamefont {Lyndon}\ \bibnamefont
  {White}}, \ and\ \bibinfo {author} {\bibfnamefont {Vaibhav}\ \bibnamefont
  {Dixit}},\ }\bibfield  {title} {\enquote {\bibinfo {title} {Diffeqflux.jl -
  {A} julia library for neural differential equations},}\ }\href@noop {}
  {\bibfield  {journal} {\bibinfo  {journal} {CoRR}\ }\textbf {\bibinfo
  {volume} {abs/1902.02376}} (\bibinfo {year} {2019})}\BibitemShut {NoStop}%
\bibitem [{\citenamefont {Bottou}(1998)}]{Bottou1998}%
  \BibitemOpen
  \bibfield  {author} {\bibinfo {author} {\bibfnamefont {L\'{e}on}\
  \bibnamefont {Bottou}},\ }\bibfield  {title} {\enquote {\bibinfo {title}
  {Online algorithms and stochastic approximations},}\ }in\ \href@noop {}
  {\emph {\bibinfo {booktitle} {Online Learning and Neural Networks}}},\
  \bibinfo {editor} {edited by\ \bibinfo {editor} {\bibfnamefont {David}\
  \bibnamefont {Saad}}}\ (\bibinfo  {publisher} {Cambridge University Press},\
  \bibinfo {address} {Cambridge, UK},\ \bibinfo {year} {1998})\BibitemShut
  {NoStop}%
\bibitem [{\citenamefont {Duchi}\ \emph {et~al.}(2011)\citenamefont {Duchi},
  \citenamefont {Hazan},\ and\ \citenamefont {Singer}}]{Duchi2011}%
  \BibitemOpen
  \bibfield  {author} {\bibinfo {author} {\bibfnamefont {John}\ \bibnamefont
  {Duchi}}, \bibinfo {author} {\bibfnamefont {Elad}\ \bibnamefont {Hazan}}, \
  and\ \bibinfo {author} {\bibfnamefont {Yoram}\ \bibnamefont {Singer}},\
  }\bibfield  {title} {\enquote {\bibinfo {title} {Adaptive subgradient methods
  for online learning and stochastic optimization},}\ }\href@noop {} {\bibfield
   {journal} {\bibinfo  {journal} {J. Mach. Learn. Res.}\ }\textbf {\bibinfo
  {volume} {12}},\ \bibinfo {pages} {2121–2159} (\bibinfo {year}
  {2011})}\BibitemShut {NoStop}%
\bibitem [{\citenamefont {Kingma}\ and\ \citenamefont {Ba}(2015)}]{Kingma2015}%
  \BibitemOpen
  \bibfield  {author} {\bibinfo {author} {\bibfnamefont {Diederik~P}\
  \bibnamefont {Kingma}}\ and\ \bibinfo {author} {\bibfnamefont {Jimmy~Lei}\
  \bibnamefont {Ba}},\ }\bibfield  {title} {\enquote {\bibinfo {title} {Adam: A
  method for stochastic gradient descent},}\ }in\ \href@noop {} {\emph
  {\bibinfo {booktitle} {ICLR: International Conference on Learning
  Representations}}}\ (\bibinfo {year} {2015})\BibitemShut {NoStop}%
\bibitem [{\citenamefont {Malouf}(2002)}]{Robert2002}%
  \BibitemOpen
  \bibfield  {author} {\bibinfo {author} {\bibfnamefont {Robert}\ \bibnamefont
  {Malouf}},\ }\bibfield  {title} {\enquote {\bibinfo {title} {A comparison of
  algorithms for maximum entropy parameter estimation},}\ }in\ \href {\doibase
  10.3115/1118853.1118871} {\emph {\bibinfo {booktitle} {Proceedings of the 6th
  Conference on Natural Language Learning - Volume 20}}},\ \bibinfo {series and
  number} {COLING-02}\ (\bibinfo  {publisher} {Association for Computational
  Linguistics},\ \bibinfo {address} {USA},\ \bibinfo {year} {2002})\ p.\
  \bibinfo {pages} {1–7}\BibitemShut {NoStop}%
\bibitem [{\citenamefont {Goffe}\ \emph {et~al.}(1994)\citenamefont {Goffe},
  \citenamefont {Ferrier},\ and\ \citenamefont {Rogers}}]{Goffe1994}%
  \BibitemOpen
  \bibfield  {author} {\bibinfo {author} {\bibfnamefont {William}\ \bibnamefont
  {Goffe}}, \bibinfo {author} {\bibfnamefont {Gary}\ \bibnamefont {Ferrier}}, \
  and\ \bibinfo {author} {\bibfnamefont {John}\ \bibnamefont {Rogers}},\
  }\bibfield  {title} {\enquote {\bibinfo {title} {Global optimization of
  statistical functions with simulated annealing},}\ }\href@noop {} {\bibfield
  {journal} {\bibinfo  {journal} {Journal of Econometrics}\ }\textbf {\bibinfo
  {volume} {60}},\ \bibinfo {pages} {65--99} (\bibinfo {year}
  {1994})}\BibitemShut {NoStop}%
\bibitem [{\citenamefont {L.}(1996)}]{Goffe1996}%
  \BibitemOpen
  \bibfield  {author} {\bibinfo {author} {\bibfnamefont {Goffe~William}\
  \bibnamefont {L.}},\ }\bibfield  {title} {\enquote {\bibinfo {title} {Simann:
  A global optimization algorithm using simulated annealing},}\ }\href@noop {}
  {\bibfield  {journal} {\bibinfo  {journal} {Studies in Nonlinear Dynamics \&
  Econometrics}\ }\textbf {\bibinfo {volume} {1}},\ \bibinfo {pages} {1--9}
  (\bibinfo {year} {1996})}\BibitemShut {NoStop}%
\bibitem [{\citenamefont {{Runarsson}}\ and\ \citenamefont {{Xin
  Yao}}(2000)}]{Runarsson2000}%
  \BibitemOpen
  \bibfield  {author} {\bibinfo {author} {\bibfnamefont {T.~P.}\ \bibnamefont
  {{Runarsson}}}\ and\ \bibinfo {author} {\bibnamefont {{Xin Yao}}},\
  }\bibfield  {title} {\enquote {\bibinfo {title} {Stochastic ranking for
  constrained evolutionary optimization},}\ }\href@noop {} {\bibfield
  {journal} {\bibinfo  {journal} {IEEE Transactions on Evolutionary
  Computation}\ }\textbf {\bibinfo {volume} {4}},\ \bibinfo {pages} {284--294}
  (\bibinfo {year} {2000})}\BibitemShut {NoStop}%
\bibitem [{\citenamefont {Runarsson}\ and\ \citenamefont
  {Yao}(2005)}]{Runarsson2005}%
  \BibitemOpen
  \bibfield  {author} {\bibinfo {author} {\bibfnamefont {T.~P.}\ \bibnamefont
  {Runarsson}}\ and\ \bibinfo {author} {\bibfnamefont {Xin}\ \bibnamefont
  {Yao}},\ }\bibfield  {title} {\enquote {\bibinfo {title} {Search biases in
  constrained evolutionary optimization},}\ }\href {\doibase
  10.1109/TSMCC.2004.841906} {\bibfield  {journal} {\bibinfo  {journal} {Trans.
  Sys. Man Cyber Part C}\ }\textbf {\bibinfo {volume} {35}},\ \bibinfo {pages}
  {233–243} (\bibinfo {year} {2005})}\BibitemShut {NoStop}%
\bibitem [{\citenamefont {Nelder}\ and\ \citenamefont
  {Mead}(1965)}]{Nelder1965}%
  \BibitemOpen
  \bibfield  {author} {\bibinfo {author} {\bibfnamefont {John~A.}\ \bibnamefont
  {Nelder}}\ and\ \bibinfo {author} {\bibfnamefont {Roger}\ \bibnamefont
  {Mead}},\ }\bibfield  {title} {\enquote {\bibinfo {title} {A simplex method
  for function minimization},}\ }\href@noop {} {\bibfield  {journal} {\bibinfo
  {journal} {Computer Journal}\ }\textbf {\bibinfo {volume} {7}},\ \bibinfo
  {pages} {308--313} (\bibinfo {year} {1965})}\BibitemShut {NoStop}%
\bibitem [{\citenamefont {{Zhan}}\ \emph {et~al.}(2009)\citenamefont {{Zhan}},
  \citenamefont {{Zhang}}, \citenamefont {{Li}},\ and\ \citenamefont
  {{Chung}}}]{Zhan2009}%
  \BibitemOpen
  \bibfield  {author} {\bibinfo {author} {\bibfnamefont {Z.}~\bibnamefont
  {{Zhan}}}, \bibinfo {author} {\bibfnamefont {J.}~\bibnamefont {{Zhang}}},
  \bibinfo {author} {\bibfnamefont {Y.}~\bibnamefont {{Li}}}, \ and\ \bibinfo
  {author} {\bibfnamefont {H.~S.}\ \bibnamefont {{Chung}}},\ }\bibfield
  {title} {\enquote {\bibinfo {title} {Adaptive particle swarm optimization},}\
  }\href@noop {} {\bibfield  {journal} {\bibinfo  {journal} {IEEE Transactions
  on Systems, Man, and Cybernetics, Part B (Cybernetics)}\ }\textbf {\bibinfo
  {volume} {39}},\ \bibinfo {pages} {1362--1381} (\bibinfo {year}
  {2009})}\BibitemShut {NoStop}%
\bibitem [{\citenamefont {W\"{a}chter}\ and\ \citenamefont
  {Biegler}(2006)}]{Wachter2006}%
  \BibitemOpen
  \bibfield  {author} {\bibinfo {author} {\bibfnamefont {Andreas}\ \bibnamefont
  {W\"{a}chter}}\ and\ \bibinfo {author} {\bibfnamefont {Lorenz~T.}\
  \bibnamefont {Biegler}},\ }\bibfield  {title} {\enquote {\bibinfo {title} {On
  the implementation of an interior-point filter line-search algorithm for
  large-scale nonlinear programming},}\ }\href@noop {} {\bibfield  {journal}
  {\bibinfo  {journal} {Math. Program.}\ }\textbf {\bibinfo {volume} {106}},\
  \bibinfo {pages} {25–57} (\bibinfo {year} {2006})}\BibitemShut {NoStop}%
\bibitem [{\citenamefont {Nocedal}\ and\ \citenamefont
  {Wright}(2006)}]{NoceWrig06}%
  \BibitemOpen
  \bibfield  {author} {\bibinfo {author} {\bibfnamefont {Jorge}\ \bibnamefont
  {Nocedal}}\ and\ \bibinfo {author} {\bibfnamefont {Stephen~J.}\ \bibnamefont
  {Wright}},\ }\href@noop {} {\emph {\bibinfo {title} {Numerical
  Optimization}}},\ \bibinfo {edition} {2nd}\ ed.\ (\bibinfo  {publisher}
  {Springer},\ \bibinfo {address} {New York, NY, USA},\ \bibinfo {year}
  {2006})\BibitemShut {NoStop}%
\bibitem [{\citenamefont {Madsen}\ and\ \citenamefont
  {{\v{Z}}ilinskas}(2002)}]{Madsen2002}%
  \BibitemOpen
  \bibfield  {author} {\bibinfo {author} {\bibfnamefont {Kaj}\ \bibnamefont
  {Madsen}}\ and\ \bibinfo {author} {\bibfnamefont {Julius}\ \bibnamefont
  {{\v{Z}}ilinskas}},\ }\enquote {\bibinfo {title} {Parallel branch-and-bound
  attraction based methods for global optimzation},}\ in\ \href {\doibase
  10.1007/0-306-47648-7_10} {\emph {\bibinfo {booktitle} {Stochastic and Global
  Optimization}}},\ \bibinfo {editor} {edited by\ \bibinfo {editor}
  {\bibfnamefont {Gintautas}\ \bibnamefont {Dzemyda}}, \bibinfo {editor}
  {\bibfnamefont {Vyd{\={u}}nas}\ \bibnamefont {{\v{S}}altenis}}, \ and\
  \bibinfo {editor} {\bibfnamefont {Antanas}\ \bibnamefont {{\v{Z}}ilinskas}}}\
  (\bibinfo  {publisher} {Springer US},\ \bibinfo {address} {Boston, MA},\
  \bibinfo {year} {2002})\ pp.\ \bibinfo {pages} {175--187}\BibitemShut
  {NoStop}%
\bibitem [{\citenamefont {Rinnooy~Kan}\ and\ \citenamefont
  {Timmer}(1987)}]{Kan1987}%
  \BibitemOpen
  \bibfield  {author} {\bibinfo {author} {\bibfnamefont {A.~H.~G.}\
  \bibnamefont {Rinnooy~Kan}}\ and\ \bibinfo {author} {\bibfnamefont {G.~T.}\
  \bibnamefont {Timmer}},\ }\bibfield  {title} {\enquote {\bibinfo {title}
  {Stochastic global optimization methods part i: Clustering methods},}\ }\href
  {\doibase 10.1007/BF02592070} {\bibfield  {journal} {\bibinfo  {journal}
  {Mathematical Programming}\ }\textbf {\bibinfo {volume} {39}},\ \bibinfo
  {pages} {27--56} (\bibinfo {year} {1987})}\BibitemShut {NoStop}%
\bibitem [{\citenamefont {Kucherenko}\ and\ \citenamefont
  {Sytsko}(2005)}]{Kucherenko2005}%
  \BibitemOpen
  \bibfield  {author} {\bibinfo {author} {\bibfnamefont {Sergei}\ \bibnamefont
  {Kucherenko}}\ and\ \bibinfo {author} {\bibfnamefont {Yury}\ \bibnamefont
  {Sytsko}},\ }\bibfield  {title} {\enquote {\bibinfo {title} {Application of
  deterministic low-discrepancy sequences in global optimization},}\ }\href
  {\doibase 10.1007/s10589-005-4615-1} {\bibfield  {journal} {\bibinfo
  {journal} {Computational Optimization and Applications}\ }\textbf {\bibinfo
  {volume} {30}},\ \bibinfo {pages} {297--318} (\bibinfo {year}
  {2005})}\BibitemShut {NoStop}%
\bibitem [{\citenamefont {Powell}(2009)}]{Powell2009}%
  \BibitemOpen
  \bibfield  {author} {\bibinfo {author} {\bibfnamefont {M.~J.~D.}\
  \bibnamefont {Powell}},\ }\bibfield  {title} {\enquote {\bibinfo {title} {The
  bobyqa algorithm for bound constrained optimization without derivatives},}\
  }\href@noop {} {\bibfield  {journal} {\bibinfo  {journal} {technical report
  NA2009/06}\ } (\bibinfo {year} {2009})}\BibitemShut {NoStop}%
\bibitem [{\citenamefont {Svanberg}(2002)}]{Svanberg2002}%
  \BibitemOpen
  \bibfield  {author} {\bibinfo {author} {\bibfnamefont {Krister}\ \bibnamefont
  {Svanberg}},\ }\bibfield  {title} {\enquote {\bibinfo {title} {A class of
  globally convergent optimization methods based on conservative convex
  separable approximations},}\ }\href@noop {} {\bibfield  {journal} {\bibinfo
  {journal} {SIAM Journal on Optimization}\ }\textbf {\bibinfo {volume} {12}},\
  \bibinfo {pages} {555--573} (\bibinfo {year} {2002})}\BibitemShut {NoStop}%
\bibitem [{\citenamefont {Nocedal}(1980)}]{Nocedal1980}%
  \BibitemOpen
  \bibfield  {author} {\bibinfo {author} {\bibfnamefont {Jorge}\ \bibnamefont
  {Nocedal}},\ }\bibfield  {title} {\enquote {\bibinfo {title} {Updating
  quasi-newton matrices with limited storage},}\ }\href@noop {} {\bibfield
  {journal} {\bibinfo  {journal} {Mathematics of Computation}\ }\textbf
  {\bibinfo {volume} {35}},\ \bibinfo {pages} {773–782} (\bibinfo {year}
  {1980})}\BibitemShut {NoStop}%
\bibitem [{\citenamefont {Liu}\ and\ \citenamefont {Nocedal}(1989)}]{Liu1989}%
  \BibitemOpen
  \bibfield  {author} {\bibinfo {author} {\bibfnamefont {Dong~C.}\ \bibnamefont
  {Liu}}\ and\ \bibinfo {author} {\bibfnamefont {Jorge}\ \bibnamefont
  {Nocedal}},\ }\bibfield  {title} {\enquote {\bibinfo {title} {On the limited
  memory bfgs method for large scale optimization},}\ }\href {\doibase
  10.1007/BF01589116} {\bibfield  {journal} {\bibinfo  {journal} {Mathematical
  Programming}\ }\textbf {\bibinfo {volume} {45}},\ \bibinfo {pages} {503--528}
  (\bibinfo {year} {1989})}\BibitemShut {NoStop}%
\bibitem [{\citenamefont {Luenberger}\ and\ \citenamefont
  {Ye}(2015)}]{Luenberger2015}%
  \BibitemOpen
  \bibfield  {author} {\bibinfo {author} {\bibfnamefont {David~G.}\
  \bibnamefont {Luenberger}}\ and\ \bibinfo {author} {\bibfnamefont {Yinyu}\
  \bibnamefont {Ye}},\ }\href@noop {} {\emph {\bibinfo {title} {Linear and
  Nonlinear Programming}}}\ (\bibinfo  {publisher} {Springer Publishing
  Company, Incorporated},\ \bibinfo {year} {2015})\BibitemShut {NoStop}%
\bibitem [{\citenamefont {{Vikhar}}(2016)}]{Vikhar2016}%
  \BibitemOpen
  \bibfield  {author} {\bibinfo {author} {\bibfnamefont {P.~A.}\ \bibnamefont
  {{Vikhar}}},\ }\bibfield  {title} {\enquote {\bibinfo {title} {Evolutionary
  algorithms: A critical review and its future prospects},}\ }in\ \href@noop {}
  {\emph {\bibinfo {booktitle} {2016 International Conference on Global Trends
  in Signal Processing, Information Computing and Communication (ICGTSPICC)}}}\
  (\bibinfo {year} {2016})\ pp.\ \bibinfo {pages} {261--265}\BibitemShut
  {NoStop}%
\bibitem [{\citenamefont {Johnson}()}]{Johnson}%
  \BibitemOpen
  \bibfield  {author} {\bibinfo {author} {\bibfnamefont {Steven~G.}\
  \bibnamefont {Johnson}},\ }\bibfield  {title} {\enquote {\bibinfo {title}
  {The nlopt nonlinear optimization package},}\ }\href {\doibase
  ab-initio.mit.edu/nlopt} {\ ab-initio.mit.edu/nlopt}\BibitemShut {NoStop}%
\bibitem [{\citenamefont {Mogensen}\ and\ \citenamefont
  {Riseth}(2018)}]{mogensen2018optim}%
  \BibitemOpen
  \bibfield  {author} {\bibinfo {author} {\bibfnamefont {Patrick~Kofod}\
  \bibnamefont {Mogensen}}\ and\ \bibinfo {author} {\bibfnamefont
  {Asbj{\o}rn~Nilsen}\ \bibnamefont {Riseth}},\ }\bibfield  {title} {\enquote
  {\bibinfo {title} {Optim: A mathematical optimization package for {Julia}},}\
  }\href {\doibase 10.21105/joss.00615} {\bibfield  {journal} {\bibinfo
  {journal} {Journal of Open Source Software}\ }\textbf {\bibinfo {volume}
  {3}},\ \bibinfo {pages} {615} (\bibinfo {year} {2018})}\BibitemShut {NoStop}%
\end{thebibliography}%
\end{document}